%% file: MEGPhysicsPaper.tex
\documentclass[epjc3,twocolumn,a4paper,fleqn]{svjour3} % epjc
\pdfoutput=1
\usepackage[fleqn]{amsmath}
\usepackage{graphicx}
\usepackage{epstopdf}
\usepackage{fix-cm}
\usepackage{import}
\usepackage{txfonts}
\usepackage{epsfig}
\usepackage{listings}
\usepackage{lineno}
\usepackage{mdwlist}
\usepackage{subfigure}
\usepackage{pennames}
\usepackage{hyperref}
\usepackage[british]{babel}
\usepackage{color}

\RequirePackage[normalem]{ulem} %DIF PREAMBLE
\RequirePackage{color}\definecolor{RED}{rgb}{1,0,0}\definecolor{BLUE}{rgb}{0,0,1} %DIF PREAMBLE
\RequirePackage[normalem]{ulem} %DIF PREAMBLE
\RequirePackage{color}\definecolor{RED}{rgb}{1,0,0}\definecolor{BLUE}{rgb}{0,0,1} %DIF PREAMBLE
\graphicspath{{./}{../00_introductionLFV/}{../01_detector/}{../02_run/}{../03_reconstruction/}{../04_analysis/}{../05_results/}{../06_conclusions/}}
\DeclareGraphicsExtensions{.eps,.pdf,.jpeg,.jpg,.png}

\def\vector#1{\mbox{\boldmath $#1$}}
\newcommand{\meg}{\ifmmode\mu^+ \to {\rm e}^+ \gamma\else$\mu^+ \to \mathrm{e}^+ \gamma$\fi}
\newcommand{\michel}{\ifmmode\mu^+ \to {\rm e}^+ \nu\bar{\nu}\else$\mu^+ \to \mathrm{e}^+ \nu\bar{\nu}$\fi}
\newcommand{\radiative}{\ifmmode\mu^+ \to {\rm e}^+\gamma \nu\bar{\nu}\else$\mu^+ \to \mathrm{e}^+\gamma \nu\bar{\nu}$\fi}
\newcommand{\conv}{\ifmmode\mu^- \to {\rm e}^-\else$\mu^- \to \mathrm{e}^-$\fi}
\newcommand{\mute}{\ifmmode\mu^+ \to 3{\rm e}\else $\mu^+ \to 3\mathrm{e}$\fi}
\newcommand{\aif}{\ifmmode\mathrm{e}^+ \mathrm{e}^- \to \gamma\gamma \else$\mathrm{e}^+ \mathrm{e}^- \to \gamma \gamma$\fi}

\newcommand*{\megsign}        {\mathrm{\mu}^+ \to \mathrm{e}^+ \mathrm{\gamma}}

\newcommand*{\egammapair}     {\mathrm{e}^+\gamma}
\newcommand*{\egamma}         {E_{\mathrm{\gamma}}}
\newcommand*{\epositron}      {E_\mathrm{e^+}}
\newcommand*{\tpositron}      {t_\mathrm{e^+}}
\newcommand*{\tgamma}         {t_{\mathrm{\gamma}}}
\newcommand*{\tegamma}        {t_{\mathrm{e^+ \gamma}}}
\newcommand*{\Thetaegamma}    {\Theta_{\mathrm{e^+ \gamma}}}
\newcommand*{\cosThetaegamma} {\cos{\Thetaegamma}}
\newcommand*{\thetaegamma}    {\theta_{\mathrm{e^+ \gamma}}}
\newcommand*{\phiegamma}      {\phi_{\mathrm{e^+ \gamma}}}
\newcommand*{\thetae}         {\theta_\mathrm{e^+}}
\newcommand*{\phie}           {\phi_\mathrm{e^+}}
\newcommand*{\thetagamma}     {\theta_\mathrm{\gamma}}
\newcommand*{\phigamma}       {\phi_\mathrm{\gamma}}

\newcommand*{\posAIF}         {\mathrm{e^+_{AIF}}}
\newcommand*{\pos}            {\mathrm{e^+}}

\newcommand*{\BR}     { {\cal B} }
\newcommand*{\xpos}          {x_\mathrm{e^+}}
\newcommand*{\ypos}          {y_\mathrm{e^+}}
\newcommand*{\zpos}          {z_\mathrm{e^+}}

\newcommand*{\posgamma}      {\vector{r}_{\gamma}}
\newcommand*{\qualitye}      {\vector{q}_{\rm e^+}}

\newcommand*{\ugamma}         {u_{\gamma}}
\newcommand*{\vgamma}         {v_{\gamma}}
\newcommand*{\wgamma}         {w_{\gamma}}

\newcommand*{\nsig}           {N_{\rm sig}}
\newcommand*{\nrd}            {N_{\rm RMD}}

\newcommand*{\nacc}            {N_{\rm ACC}}
\newcommand*{\nobs}           {N_{\rm obs}}

\newcommand*{\sens}     { {\cal S}_{90}}
\newcommand*{\ul}     { {\cal B}_{90}}
\newcommand*{\bestfit}     { {\cal B}_\mathrm{fit}}

\newcommand*{\mathtentative}{}
\def\mathtentative#1#{\mathcoloraux{#1}}
\newcommand*{\mathcoloraux}[3]{%
  \protect\leavevmode
  \begingroup
    \color#1{#2}#3%
  \endgroup
}
\journalname{Eur. Phys. J. C} % epjc

\begin{document}

%\begin{frontmatter} %nima

\title{Search for the Lepton Flavour Violating Decay \meg\ with the Full Dataset of the MEG Experiment}

\include{author-meg-epjc} % epjc

\thankstext[*]{e1}{Corresponding author: fabrizio.cei@pi.infn.it} % epjc
\thankstext[$\dagger$]{e2}{Deceased } % epjc
\maketitle % epjc
 
\begin{abstract}

The final results of the search for the lepton flavour violating decay 
$\megsign$ based on the full dataset collected by the MEG experiment 
at the Paul Scherrer Institut in the period 2009--2013 and 
totalling $7.5\times 10^{14}$ stopped muons on target are presented. 
No significant excess of events is observed in the dataset with respect to the expected background 
and a new upper limit on the branching ratio of this decay of $\BR(\meg) < 4.2 \times 10^{-13}$ (90\%\ confidence level) is established,
which represents the most stringent limit on the existence of this decay to date.
\end{abstract}

%\begin{keyword} % nima
 % epjc
%\end{keyword} % nima
%\end{frontmatter} %nima

\tableofcontents 

%\linenumbers

% --------------------------- 00_introduction -------------------
\subimport{../00_introduction/}{introduction.tex}

% --------------------------- 01_detector -----------------------
\subimport{../01_detector/}{detector.tex}

% --------------------------- 02_run ----------------------------
%\subimport{../02_run/}{run.tex}

% --------------------------- 03_reconstructio -------------------------------
\subimport{../03_reconstruction/}{reconstruction.tex}

% --------------------------- 04_analysis -------------------------------
\subimport{../04_analysisresults/}{analysisresults.tex}

% --------------------------- 05_results------------------------
%%%%%\subimport{../05_results/}{results.tex}

% --------------------------- 06_conclusion ---------------------------
\subimport{../06_conclusions/}{conclusions.tex}

\section*{Acknowledgments}

We are grateful for the support and co-operation provided 
by PSI as the host laboratory and to the technical and 
engineering staff of our institutes. This work is
supported by DOE DEFG02-91ER40679 (USA), INFN
(Italy), MEXT KAKENHI 22000004 and 26000004 (Japan),
Schweizerischer Nationalfonds (SNF) Grant 200021\_137738,
the Russian Federation Ministry of Education and Science and Russian Fund
for Basic Research grant RFBR-14-22-03071.

\bibliographystyle{my}
\bibliography{MEG}
\end{document}

%% file: author-meg-epjc.tex
\input{institute}

\date{Received: date / Accepted: date}

\author{
        A.~M.~Baldini~\thanksref{addr4}$^a$ \and
        Y.~Bao~\thanksref{addr1} \and
        E.~Baracchini~\thanksref{addr3,addr17} \and
        C.~Bemporad~\thanksref{addr4}$^{ab}$ \and
        F.~Berg~\thanksref{addr1,addr2} \and
        M.~Biasotti~\thanksref{addr5}$^{ab}$ \and
        G.~Boca~\thanksref{addr7}$^{ab}$ \and
        M.~Cascella~\thanksref{addr18,addr6}$^{ab}$ \and
        P.~W.~Cattaneo~\thanksref{addr7}$^{a}$  \and
        G.~Cavoto~\thanksref{addr8}$^{a}$ \and
        F.~Cei~\thanksref{e1,addr4}$^{ab}$ \and
        C.~Cerri~\thanksref{addr4}$^{a}$ \and
        G.~Chiarello~\thanksref{addr6}$^{ab}$ \and
        C.~Chiri~\thanksref{addr6}$^{ab}$ \and
        A.~Corvaglia~\thanksref{addr6}$^{ab}$ \and
        A.~de Bari~\thanksref{addr7}$^{ab}$ \and
        M.~De Gerone~\thanksref{addr5}$^{a}$ \and
        T.~Doke~\thanksref{e2,addr10} \and
        A.~D'Onofrio~\thanksref{addr4}$^{ab}$ \and
        S.~Dussoni~\thanksref{addr4}$^{a}$\and
        J.~Egger~\thanksref{addr1} \and
        Y.~Fujii~\thanksref{addr3}  \and
        L.~Galli~\thanksref{addr4}$^{a}$ \and
        F.~Gatti~\thanksref{addr5}$^{ab}$ \and
        F.~Grancagnolo~\thanksref{addr6}$^{a}$ \and
        M.~Grassi~\thanksref{addr4}$^{a}$ \and
        A.~Graziosi~\thanksref{addr8}$^{ab}$ \and
        D.~N.~Grigoriev~\thanksref{addr12,addr14,addr15} \and
        T.~Haruyama~\thanksref{addr9} \and
        M.~Hildebrandt~\thanksref{addr1} \and
        Z.~Hodge~\thanksref{addr1,addr2} \and
        K.~Ieki~\thanksref{addr3}  \and
        F.~Ignatov~\thanksref{addr12,addr15} \and
        T.~Iwamoto~\thanksref{addr3}  \and
        D.~Kaneko~\thanksref{addr3}  \and
        T.~I.~Kang~\thanksref{addr11}  \and
        P.-R.~Kettle~\thanksref{addr1} \and
        B.~I.~Khazin~\thanksref{e2,addr12,addr15} \and
        N.~Khomutov~\thanksref{addr13} \and
        A.~Korenchenko~\thanksref{e2,addr13}  \and
        N.~Kravchuk~\thanksref{addr13}  \and
        G.~M.~A.~Lim~\thanksref{addr11} \and
        A.~Maki~\thanksref{addr9}  \and
        S.~Mihara~\thanksref{addr9}  \and
        W.~Molzon~\thanksref{addr11} \and
        Toshinori~Mori~\thanksref{addr3}  \and
        F.~Morsani~\thanksref{addr4}$^{a}$ \and
        A.~Mtchedilishvili~\thanksref{addr1}  \and
        D.~Mzavia~\thanksref{e2,addr13}  \and 
        S.~Nakaura~\thanksref{addr3}  \and 
        R.~Nard\`o~\thanksref{addr7}$^{ab}$ \and
        D.~Nicol\`o~\thanksref{addr4}$^{ab}$ \and
        H.~Nishiguchi~\thanksref{addr9}  \and
        M.~Nishimura~\thanksref{addr3}  \and 
        S.~Ogawa~\thanksref{addr3}  \and
        W.~Ootani~\thanksref{addr3}  \and
        S.~Orito~\thanksref{e2,addr3}  \and
        M.~Panareo~\thanksref{addr6}$^{ab}$ \and
        A.~Papa~\thanksref{addr1} \and
        R.~Pazzi~\thanksref{e2,addr4} \and
        A.~Pepino~\thanksref{addr6}$^{ab}$ \and
        G.~Piredda~\thanksref{e2,addr8}$^{a}$ \and
        G.~Pizzigoni~\thanksref{addr5}$^{ab}$ \and
        A.~Popov~\thanksref{addr12,addr15} \and
        F.~Raffaelli~\thanksref{addr4}$^{a}$ \and
        F.~Renga~\thanksref{addr1,addr8}$^{ab}$ \and
        E.~Ripiccini~\thanksref{addr8}$^{ab}$ \and
        S.~Ritt~\thanksref{addr1} \and
        M.~Rossella~\thanksref{addr7}$^{a}$ \and
        G.~Rutar~\thanksref{addr1,addr2} \and
        R.~Sawada~\thanksref{addr3}  \and
        F.~Sergiampietri~\thanksref{addr4}$^{a}$ \and
        G.~Signorelli~\thanksref{addr4}$^{a}$ \and
        M.~Simonetta~\thanksref{addr7}$^{ab}$  \and
        G.~F.~Tassielli~\thanksref{addr6}$^{a}$ \and
        F.~Tenchini~\thanksref{addr4}$^{ab}$ \and
        Y.~Uchiyama~\thanksref{addr3} \and
        M.~Venturini~\thanksref{addr4}$^{ac}$ \and
        C.~Voena~\thanksref{addr8}$^{a}$ \and
        A.~Yamamoto~\thanksref{addr9} \and
        K.~Yoshida~\thanksref{addr3} \and
        Z.~You~\thanksref{addr11} \and
        Yu.V.~Yudin~\thanksref{addr12,addr15} \and
        D.~Zanello~\thanksref{addr8}
}

\institute{\PSI \label{addr1} 
           \and
              \ETHZ \label{addr2}
%             \email{fauthor@example.com}           %  \\
           \and
              \ICEPP \label{addr3}
           \and
             \INFNPi \label{addr4}
           \and
             \UCI    \label{addr11}
           \and
             \INFNPv \label{addr7}
           \and
             \INFNRm \label{addr8}
           \and
             \INFNGe \label{addr5}
           \and
             \Waseda \label{addr10}
           \and
             \BINP   \label{addr12}
           \and
             \KEK    \label{addr9}
           \and
             \JINR   \label{addr13}
           \and
             \INFNLe \label{addr6} 
           \and
             \NOVST  \label{addr14}
           \and
             \NOVS   \label{addr15}
%             \ScuolaPi \label{addr16}   
           \and  
             \INFNLNF \label{addr17}   
           \and  
             \LONDON \label{addr18}   
}

%% file: institute.tex
\newcommand*{\INFNPi}{INFN Sezione di Pisa$^{a}$; Dipartimento di Fisica$^{b}$ dell'Universit\`a, Largo B.~Pontecorvo~3, 56127 Pisa; Scuola Normale Superiore$^{c}$, Piazza dei Cavalieri, 56127 Pisa Italy}
\newcommand*{\INFNGe}{INFN Sezione di Genova$^{a}$; Dipartimento di Fisica$^{b}$ dell'Universit\`a, Via Dodecaneso 33, 16146 Genova, Italy}
\newcommand*{\INFNPv}{INFN Sezione di Pavia$^{a}$; Dipartimento di Fisica$^{b}$ dell'Universit\`a, Via Bassi 6, 27100 Pavia, Italy}
\newcommand*{\INFNRm}{INFN Sezione di Roma$^{a}$; Dipartimento di Fisica$^{b}$ dell'Universit\`a ``Sapienza'', Piazzale A.~Moro, 00185 Roma, Italy}
\newcommand*{\INFNLe}{INFN Sezione di Lecce$^{a}$; Dipartimento di Matematica e Fisica$^{b}$ dell'Universit\`a del Salento, Via per Arnesano, 73100 Lecce, Italy}
\newcommand*{\ICEPP} {ICEPP, The University of Tokyo, 7-3-1 Hongo, Bunkyo-ku, Tokyo 113-0033, Japan }
\newcommand*{\UCI}   {University of California, Irvine, CA 92697, USA}
\newcommand*{\KEK}   {KEK, High Energy Accelerator Research Organization 1-1 Oho, Tsukuba, Ibaraki, 305-0801, Japan}
\newcommand*{\PSI}   {Paul Scherrer Institut PSI, 5232, Villigen, Switzerland}
\newcommand*{\Waseda}{Research Institute for Science and Engineering, Waseda~University, 3-4-1 Okubo, Shinjuku-ku, Tokyo 169-8555, Japan}
\newcommand*{\BINP}  {Budker Institute of Nuclear Physics of Siberian Branch of Russian Academy of Sciences, 630090, Novosibirsk, Russia}
\newcommand*{\JINR}  {Joint Institute for Nuclear Research, 141980 Dubna, Russia}
\newcommand*{\ETHZ}  {Swiss Federal Institute of Technology ETH, CH-8093 Z\" urich, Switzerland}
\newcommand*{\NOVS}  {Novosibirsk State University, 630090 Novosibirsk, Russia}
\newcommand*{\NOVST} {Novosibirsk State Technical University, 630092 Novosibirsk, Russia}
\newcommand*{\INFNLNF}{\textit{Present Address}: INFN, Laboratori Nazionali di Frascati, Via 
E. Fermi, 40-00044 Frascati, Rome, Italy}
\newcommand*{\LONDON}{\textit{Present Address}: Department of Physics \& 
Astronomy, University College London, Gower Street, London, WC1E 6BT, UK}

%% file: introduction.tex
\section{Introduction}
\label{sec:introduction}
%
%{Section coordinator: P.W. Cattaneo \\
%Text:  1.5 \\
%Figure: 2
%}

The Standard Model (SM) of particle physics allows \linebreak charged lepton flavour violating (CLFV) processes with only extremely small branching 
ratios ($\ll 10^{-50}$) even when accounting for measured neutrino mass 
differences and mixing angles. 
Therefore, such decays are free from SM physics backgrounds associated with processes involving, either directly or indirectly, 
hadronic states and are ideal laboratories for searching for new physics beyond the SM. A positive signal 
would be an unambiguous evidence for physics beyond the SM.

The existence of such decays at measurable rates not far below current upper limits is suggested
by many SM extensions, such as supersymmetry \cite{barbieri}. An extensive 
review of the theoretical expectations for CLFV is provided in \cite{Mihara:2013zna}.
CLFV searches with improved sensitivity probe 
new regions of the parameter spaces of SM extensions, and CLFV decay \meg\ is particularly sensitive to new physics.
The MEG collaboration has searched for \meg\ decay
at the Paul Scherrer Institut (PSI) in Switzerland in the period 2008-2013.  A 
detailed report of the experiment motivation, design criteria, and goals is available in reference 
\cite{mori_1999,meg1infn} and references therein.  We have previously reported \cite{meg2009,meg2010,meg2013} results of partial datasets including a limit on the branching ratio for this decay $\BR < 5.7\times 10^{-13}$ at 90\%\ C.L.

\begin{figure*}[t]
\centering
  \includegraphics[width=\textwidth,angle=0] {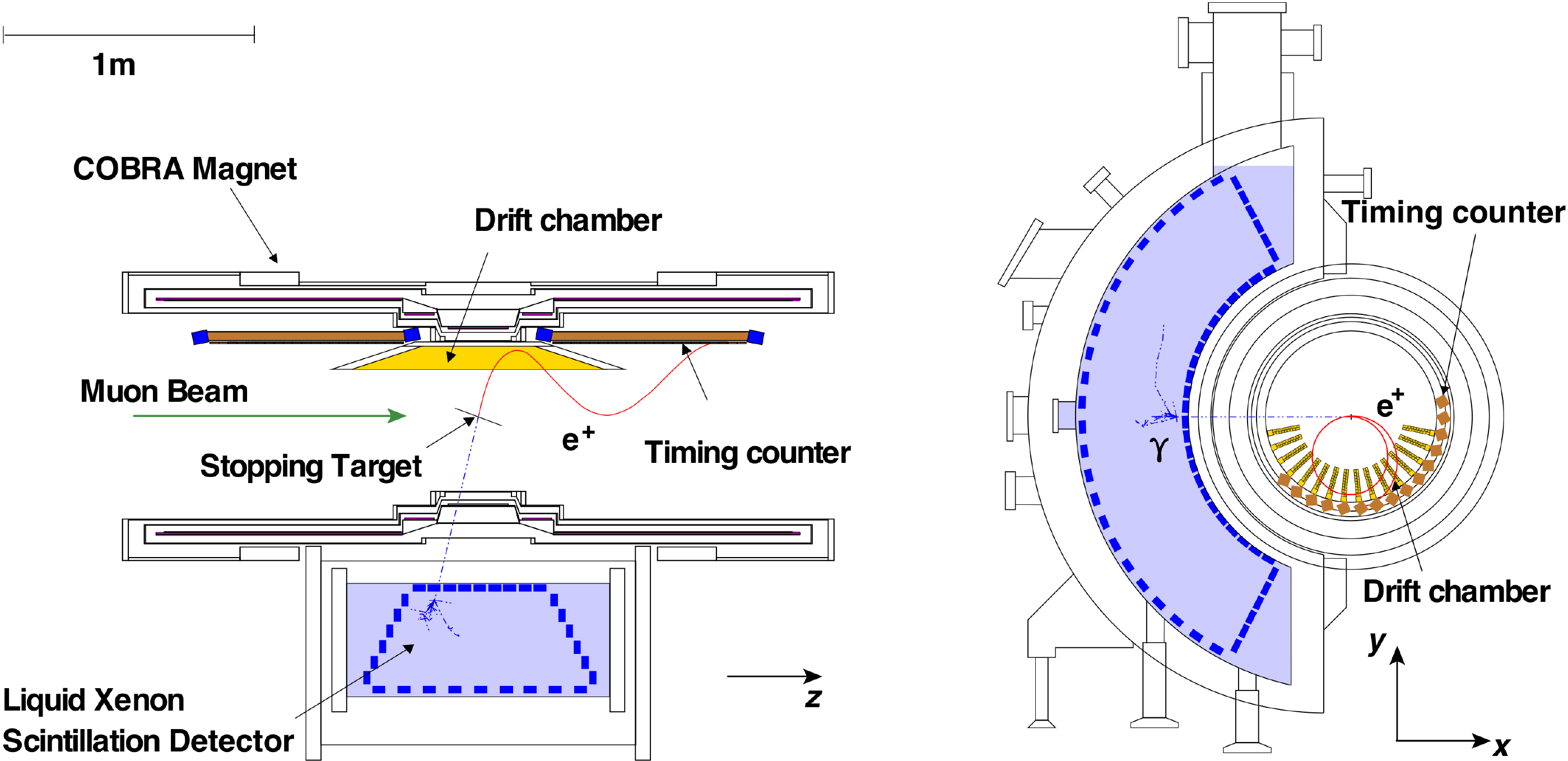}
 \caption{A schematic view of the MEG detector showing a simulated event. 
}
 \label{introduction:megdet}
\end{figure*}

The signal consists of a positron and a photon back-to-back,
each with energy of $52.83$~MeV (half of the muon mass), and with a common origin in space and time.
Figure~\ref{introduction:megdet} shows cut schematic views of the MEG apparatus. 
Positive muons are 
stopped in a thin plastic target at the centre of a spectrometer based on a superconducting solenoid. 
The decay positron's trajectory is measured in a
magnetic field by a set of low-mass drift chambers and 
a scintillation counter array is used to measure its time.
The photon momentum vector, interaction point and timing are measured 
by a homogeneous liquid xenon calorimeter located outside the magnet
and covering the angular region opposite to the acceptance of the spectrometer. 
The total geometrical acceptance of the detector for the signal is $\approx 11$\%.

The signal can be mimicked by various processes, with the positron and photon originating either from a single radiative muon decay (RMD) (\radiative) or from the accidental coincidence of a positron and a photon from different processes. In the latter case, the photon can be produced by radiative muon decay or by Bremsstrahlung or positron annihilation-in-flight (AIF) (\aif).
Accidental coincidences between a positron and a photon from different processes, each close 
in energy to their kinematic limit and with origin, direction and timing coincident within the detector resolutions
are the dominant source of background. 

Since the rate of accidental coincidences is proportional to the square 
of the $\mu^+$ decay rate, 
% while the rate of signal events is linearly proportional,
the signal to background ratio and data collection efficiency are optimised by using a 
direct-current rather than pulsed beam. 
% rather than a pulsed beam. 
Hence, the high intensity continuous surface $\mu^+$ beam
(see Sect.~\ref{sec:beam}) at PSI is the ideal facility for such a search.

The remainder of this paper is organised as follows. 
After a brief introduction to the detector and 
to the data acquisition system (Sect.~\ref{sec:detector}), the 
reconstruction algorithms are presented in detail (Sect.~\ref{sec:reconstruction}), followed by
an in-depth discussion of the analysis of the full MEG dataset 
and of the results (Sect.~\ref{sec:analysis}).
Finally, in the conclusions, some prospects for future improvements are outlined (Sect.~\ref{sec:conclusions}).

%% file: detector.tex
\section{MEG detector }
\label{sec:detector}
%
%{\it Editor's comments: \\
%Section coordinator: Paolo Walter Cattaneo \\
%Text:  5.\\
%Figure: 5.
%}

The MEG detector is briefly presented in the following, emphasising the aspects relevant to the analysis;
a detailed description is available in \cite{megdet}. Briefly, it consists of the $\mu^+$ beam, a thin stopping target, a thin-walled,  superconducting magnet,
a drift chamber array (DCH), scintillating timing counters (TC), and a liquid xenon calorimeter (LXe detector). 

In this paper we adopt a cylindrical coordinate system ${\it (r,\phi,z)}$ with origin
 at the centre of the magnet (see Fig.~\ref{introduction:megdet}).
The {\it z}-axis is parallel to the magnet axis and directed along the 
$\mu^+$ beam.
The axis defining $\phi=90^\circ$ (the {\it y}-axis of the corresponding Cartesian coordinate system)
is directed upwards and, as a consequence, the {\it x}-axis 
is directed opposite to the centre of the LXe detector.
Positrons move along trajectories with decreasing $\phi$-coordinate.
When required, the polar angle $\theta$ with respect to the {\it z}-axis is also used.
The region with ${\it z}<0$ is referred to as upstream, that with ${\it z}>0$ as downstream.

\subsection{Muon beam}
\label{sec:beam}

The requirement to stop a large number of $\mu^+$ in a thin target of small transverse size drives the beam requirements: high flux, small transverse size, 
small momentum spread and small contamination, e.g. from positrons.  
These goals are met by the 2.2 mA PSI proton cyclotron and $\pi$E5 channel in combination with the MEG beam line, which produces one of the 
world\textsc{\char13}s most intense continuous $\mu^+$ beams. It is a {\it surface muon beam} produced by $\pi^+$ decay near the surface of the 
production target. It can deliver more than $10^8$~$\mu^+$/s at 28~MeV/c in a momentum bite of 5-7\%. 
To maximise the experiment's sensitivity, the beam is tuned to a $\mu^+$ 
stopping rate of 3$\times 10^7$, limited by the rate 
capabilities of the tracking system and the rate of accidental backgrounds, given the MEG detector resolutions. 
The ratio of $\pos$ to $\mu^+$ flux in the beam is $\approx 8$, and the positrons are efficiently removed by a combination of a Wien filter 
and collimator system. The muon momentum distribution at the target is optimised by a degrader system 
comprised of a 300~$\mu$m thick mylar\textsuperscript{\textregistered} foil and the He-air atmosphere inside the 
spectrometer in front of the target. The round, Gaussian beam-spot profile has $\sigma_{x,y} \approx 10$~mm. 

The muons at the production target are produced fully polarized ($P_{\mu^+}=-1$) and they reach the stopping target with a residual polarization 
$P_{\mu^+} = -0.86 \pm 0.02 ~ {\rm (stat)} ~ { }^{+ 0.05}_{-0.06} ~ {\rm (syst)}$ consistent with the expectations \cite{Baldini:2015lwl}.

Other beam tunes are used for calibration purposes, including a $\pi^-$ 
tune at 70.5~MeV/c used to produce monochromatic photons via pion charge exchange and a 
53~MeV/c positron beam tune to produce Mott-scattered positrons close to the energy of a signal positron (Sect.~\ref{sec:calibration}).

\subsection{Muon stopping target}

\begin{figure}[t]
\centering
  \includegraphics[width=0.475\textwidth,angle=0] {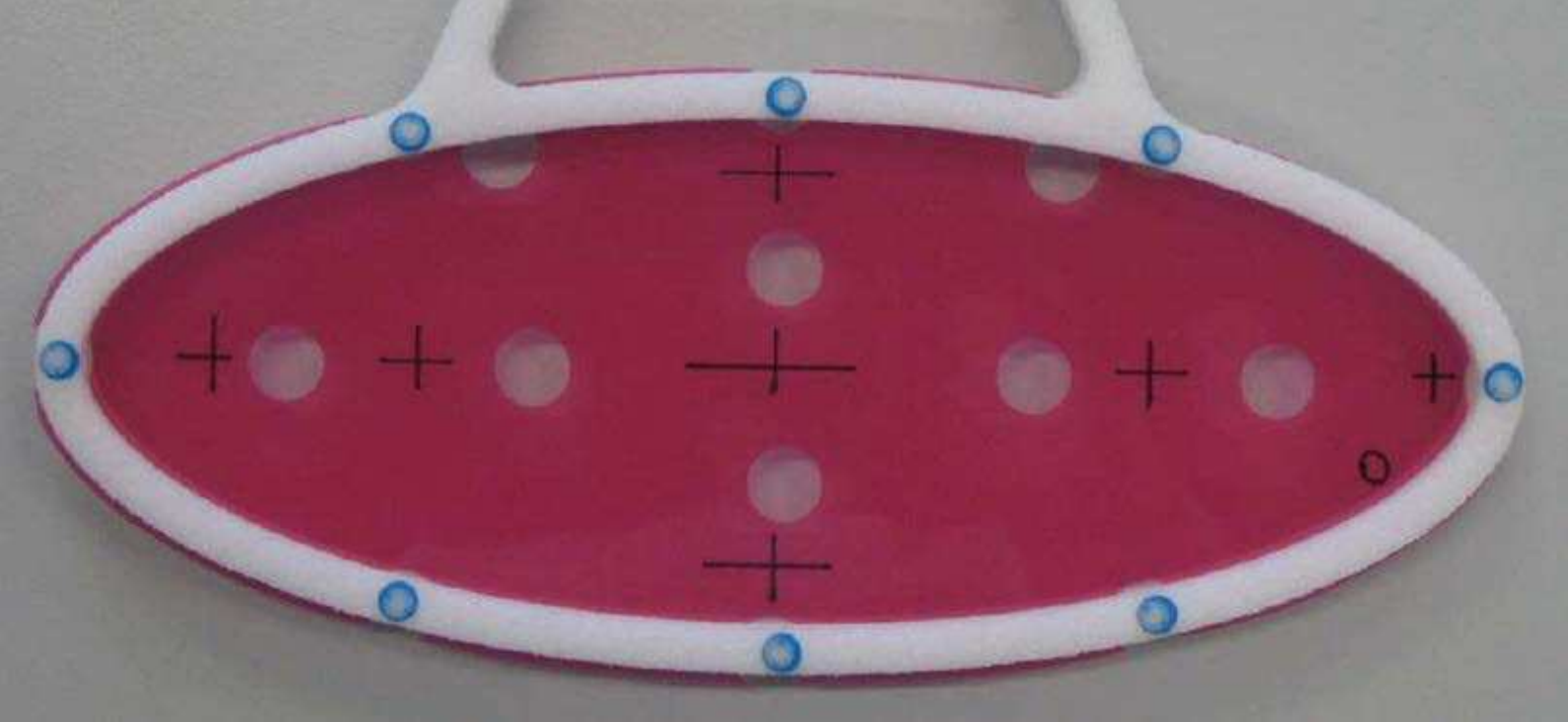}
 \caption{The thin muon stopping target mounted in a Rohacell frame.}
 \label{fig:target}
\end{figure}

Positive muons are stopped in a thin target at the centre of the spectrometer, where they decay at rest. 
The target is optimised to satisfy conflicting goals of maximising stopping efficiency ($\approx 80\%$) while 
minimising multiple scattering, Bremsstrahlung and AIF of positrons from muon decays. 
The target is composed of a 205~$\mu$m thick layer of polyethylene and 
polyester (density $0.895$~g/cm$^3$) with an elliptical shape with semi-major and semi-minor axes 
of 10~cm and 4~cm. The target foil is equipped with seven cross marks and eight holes 
of radius 0.5~cm, used for optical survey and for software alignment purposes. The foil is mounted in a Rohacell\textsuperscript{\textregistered} 
frame, which is attached to the tracking system support frame and positioned with the target normal vector in the horizontal plane and at an angle $\theta$ 
 $\approx\!70^\circ$. The target before installation in the detector is shown in Fig.~\ref{fig:target}.

\subsection{COBRA magnet }

\begin{figure}[t]
\centering
 \includegraphics[width=0.48\textwidth,angle=0] {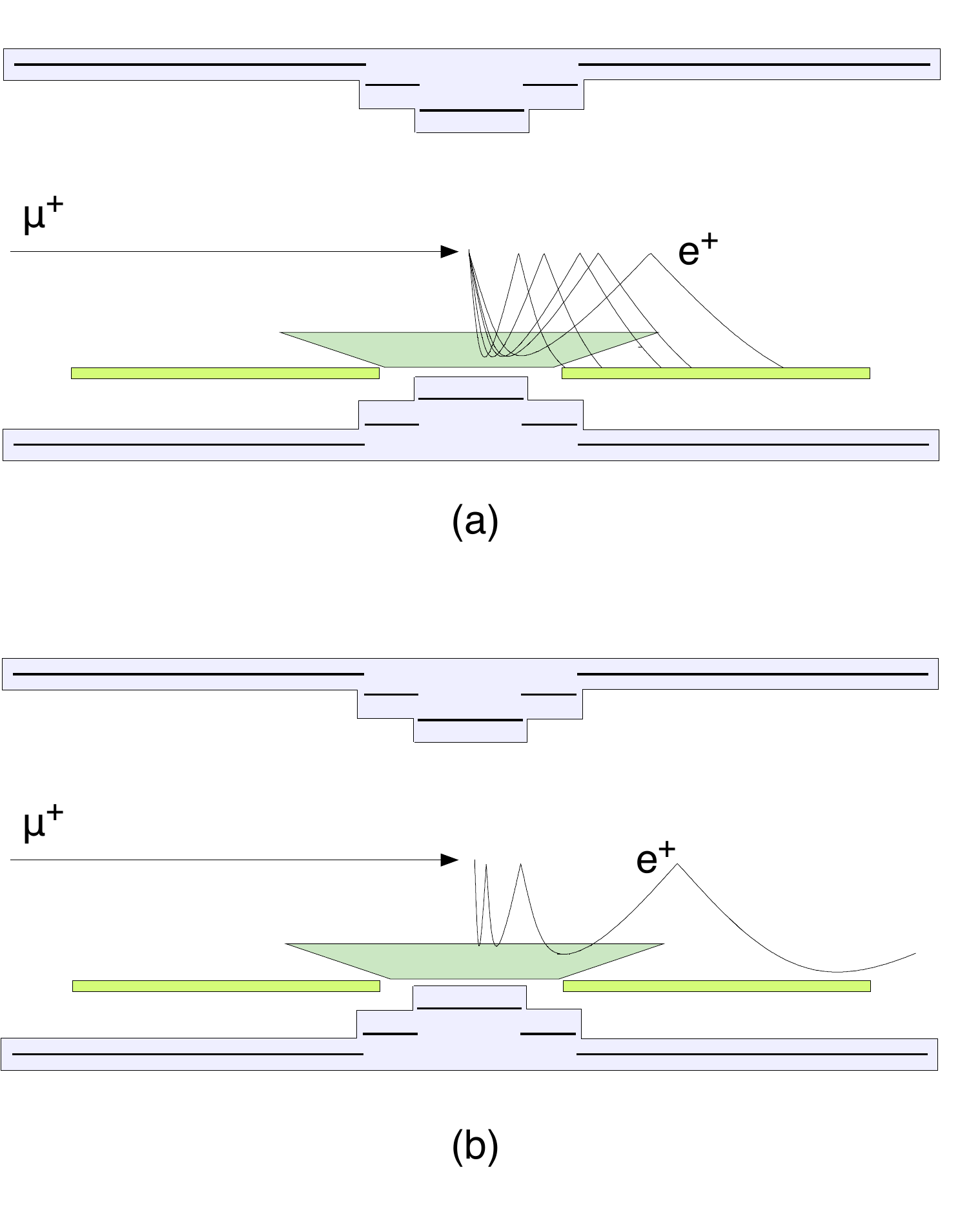}
 \caption{
  Concept of the gradient magnetic field of COBRA.
  The positrons follow trajectories at constant bending radius weakly dependent on the emission angle $\theta_{\pos}$ 
  (a) and those emitted from the target with small longitudinal momentum 
  ($\theta_{\pos}\approx 90^{\circ}$) are quickly swept away from the central region (b).
}
 \label{fig:COBRA}
\end{figure}

The COBRA (constant bending radius) magnet \cite{ootani_2004} is a thin-walled, superconducting magnet with an axially graded magnetic field,
ranging from 1.27~T at the centre to 0.49~T at either end of the magnet cryostat.
The graded field has the advantage with respect to a uniform solenoidal field that particles produced with small longitudinal momentum 
have a much shorter latency time in the spectrometer, allowing stable operation in a high-rate environment.
Additionally, the graded magnetic field is designed so that positrons emitted from the target
follow a trajectory with almost constant projected bending radius, only weakly dependent on the
emission polar angle $\theta_{\pos}$ (see Fig.~\ref{fig:COBRA}(a)), even for positrons emitted with substantial longitudinal momentum. 

The central part of the coil and cryostat accounts for $0.197\,{X}_0$, thereby maintaining high transmission of signal photons to the 
LXe detector outside the COBRA cryostat. 
The COBRA magnet is also equipped with a pair of compensation coils to reduce the stray field to the level necessary  
to operate the photomultiplier tubes (PMTs) in the LXe detector. 

The COBRA magnetic field was measured with a commercial Hall probe 
mounted on a wagon moving along $z$, $r$ and $\phi$ 
in the ranges 
$|z|<110\,\mathrm{cm}$,
$0^{\circ}<\phi<360^{\circ}$ and
$0 < r < 29\,\mathrm{cm}$, covering most of the positron tracking volume.
The probe contained three Hall sensors orthogonally aligned to measure $B_z, B_r$ and $B_\phi$ individually.
Because the main (axial) field component  is much larger than the others, even small angular misalignments of the other probes could cause large errors in $B_r$ and $B_\phi$.
Therefore, only the measured values of $B_z$ are used in the analysis and the secondary components 
$B_r$ and $B_\phi$ are reconstructed from the measured $B_z$ using Maxwell's equations as

\begin{eqnarray}
   B_\phi(z, r, \phi) &=& B_\phi(z_\mathrm{B}, r, \phi) + \frac{1}{r}\int_{z_\mathrm{B}}^{z}\frac{\partial B_z(z', r, \phi)}{\partial \phi}dz' \label{eq:reconstructed Bphi} \nonumber \\
   B_r(z, r, \phi) &=& B_r(z_\mathrm{B}, r, \phi) + \int_{z_\mathrm{B}}^{z}\frac{\partial B_z(z', r, \phi)}{\partial r}dz'. \label{eq:reconstructed Br} \nonumber
\end{eqnarray}

The measured values of $B_r$ and $B_\phi$ are required 
only at $z_\mathrm{B}=1\,\mathrm{mm}$ near the symmetry plane of the magnet where 
the measured value of $B_r$ is minimised ($|B_r(z_\mathrm{B}, r, \phi)|<2\times 10^{-3}$~T) as expected. 
The effect of the misalignment of the $B_\phi$-measuring sensor 
on $B_\phi(z_\mathrm{B}, r, \phi)$ is estimated by checking the consistency 
of the reconstructed $B_r$ and $B_\phi$ with Maxwell's equations.

The continuous magnetic field map used in the analysis is obtained 
by interpolating the reconstructed magnetic field 
at the measurement grid points by a B-spline fit \cite{Splines_DeBoor}.

\subsection{Drift chamber system}
\label{sec:dch-system}

The DCH system \cite{Hildebrandt2010111}
is designed to ensure precise measurement of the trajectory and momentum of positrons from \meg\ decays.
It is designed to satisfy several requirements:
operate at high rates, primarily from positrons from $\mu^+$ 
decays in the target;
have low mass to improve kinematic resolution (dominated by scattering) 
and to minimise production of photons by positron AIF; 
and provide excellent resolution in the measurement of the radial and longitudinal coordinates.

\begin{figure}[t]
\centering
 \includegraphics[width=0.48\textwidth,angle=0] {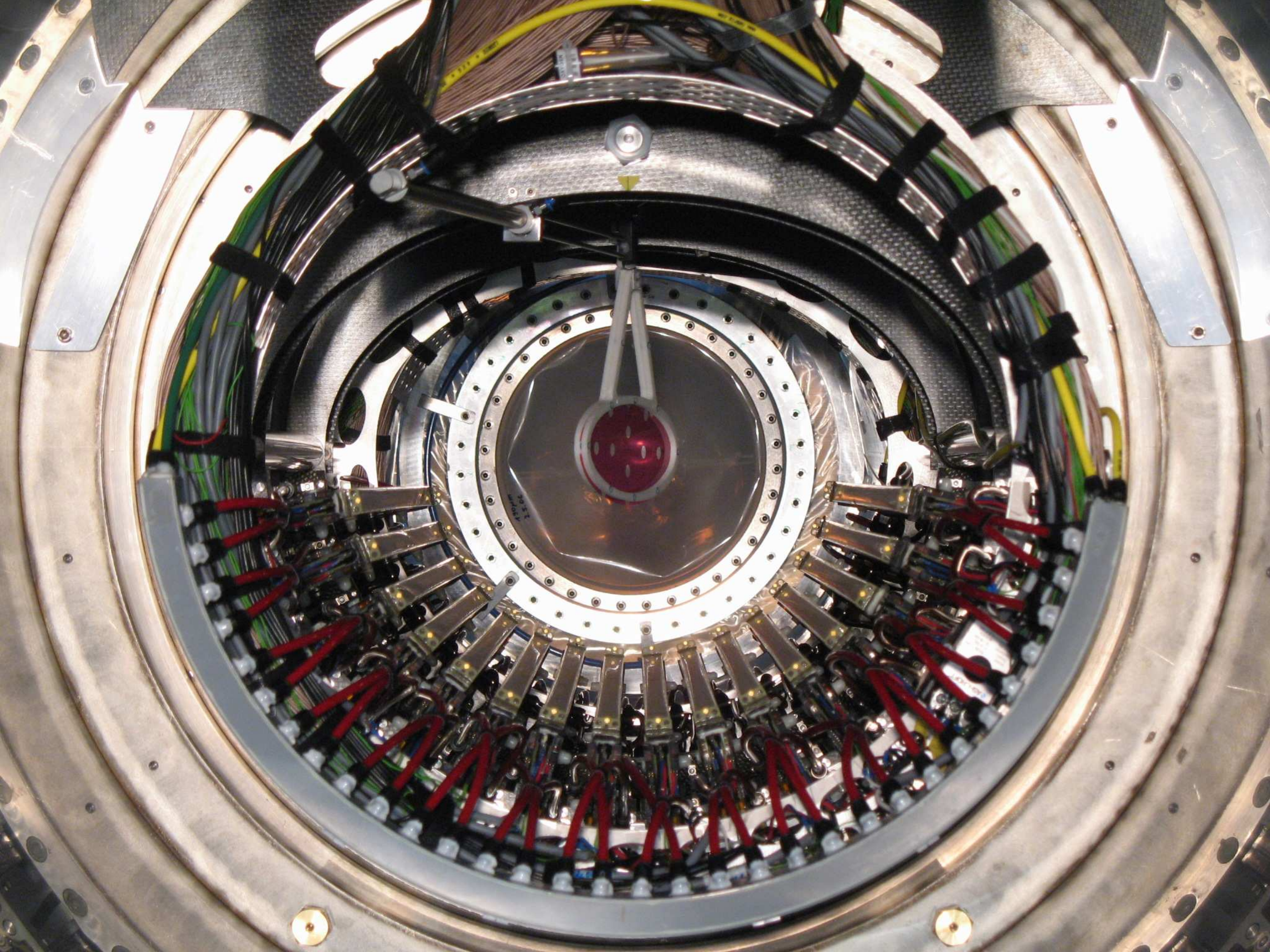}
 \caption{
View of the DCH system from the downstream side of the MEG detector. The muon 
stopping target is placed in the centre and the 16 DCH modules are mounted in a semi-circular array.
}
 \label{fig:dctar}
\end{figure}

\begin{figure}[t]
\centering
  \includegraphics[width=0.48\textwidth,angle=0] {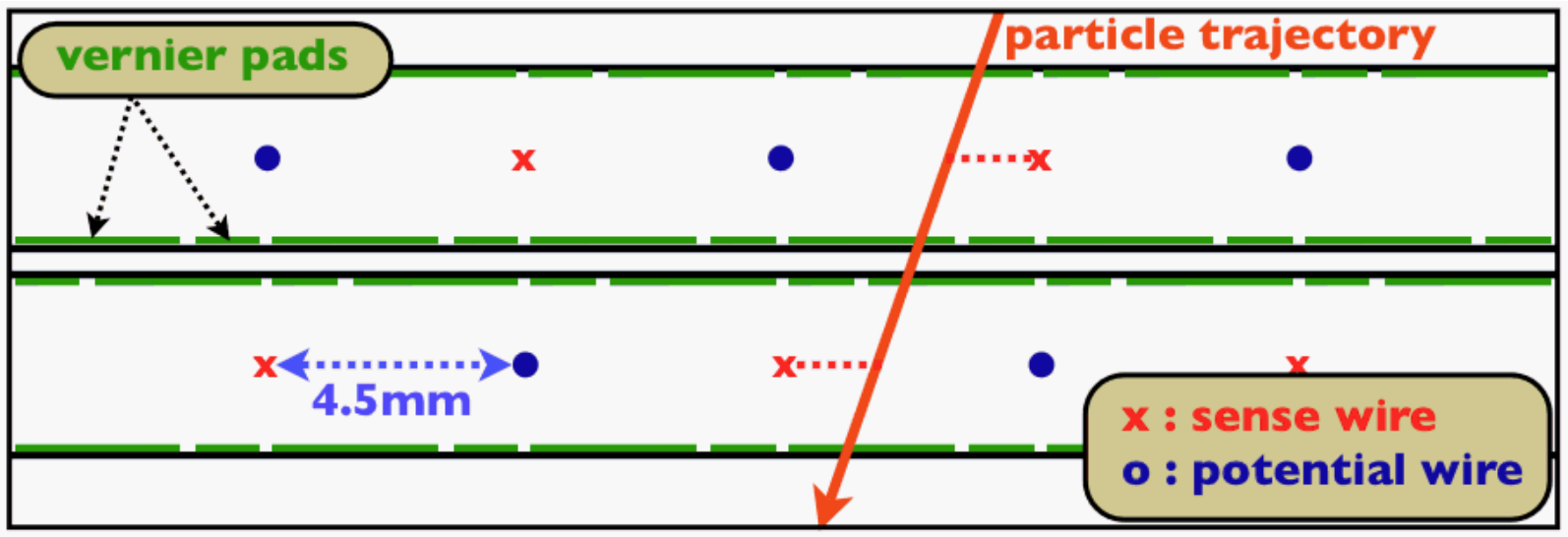}
 \caption{
Schematic view of the cell structure of a DCH plane.
}
 \label{fig:cell}
\end{figure}

\begin{figure}[t]
\centering
  \includegraphics[width=0.48\textwidth,angle=0] {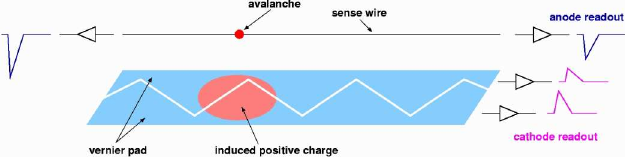}
 \caption{
Schematic view of the Vernier pad method showing the pad shape and offsets. Only one of the two cathode pads in each cell is shown. 
}
 \label{fig:vernier}
\end{figure}

The DCH system consists of 16 identical, independent modules placed inside COBRA,
aligned in a semi-circle with 10.5$^{\circ}$ spacing, and covering
the azimuthal region 
between $191.25^\circ$ and $348.75^\circ$ and the
radial region between 19.3~cm and 27.9~cm (see Fig.~\ref{fig:dctar}). 
Each module has a trapezoidal shape with base lengths of 40~cm and 104~cm,
without supporting structure on the long (inner) side 
to reduce the amount of material intercepted by signal positrons.
A module consists of two independent detector planes, 
each consisting of two cathode foils (12.5~$\mu$m-thick aluminised polyamide) separated by 7~mm 
and filled with a 50:50 mixture of He:C$_2$H$_6$. 
A plane of alternating axial anode and potential wires is situated midway between the cathode foils 
with a pitch of 4.5~mm.
The two planes of cells are separated by 3~mm and 
the two wire arrays in the same module are staggered by
half a drift cell to help resolve left-right position ambiguities (see Fig.~\ref{fig:cell}).
A double wedge pad
structure is etched on both cathodes with a Vernier pattern of cycle $\lambda = 5$ cm as shown in Fig.~\ref{fig:vernier}.
The pad geometry is designed to allow a precise measurement of the axial coordinate of the hit by comparing the signals induced on the four pads in each cell. 
The average amount of material intercepted by a positron track in a DCH module is
\mbox{$2.6\times 10^{-4}\,X_0$}, with the total material along a typical signal positron track of 
\mbox{$2.0\times 10^{-3}\,X_0$}.

% following not really necessary since it is seen in the figure. 
%The modules are mounted with the long side in the inner part of the spectrometer
%(small radius) and the short one positioned on the central coil of the magnet (large radius) 
%(for a sketch see Fig.~\ref{introduction:megdet}).
%
\subsection{Timing counter }
%added requirement to measure position
The TC \cite{DeGerone:2011te,DeGerone:2011zz}
is designed to measure precisely the impact time and position of signal positrons
and to infer the muon decay time
by correcting for the track length from the target to the TC obtained from the DCH information.

The main requirements of the TC are:
\begin{itemize}
\item provide full acceptance for signal positrons in the DCH acceptance
matching the tight mechanical constraints dictated by the DCH system and COBRA;
\item ability to operate at high rate in a high and non-uniform magnetic field;
\item fast and approximate ($\approx 5$~cm resolution) determination of the positron 
impact point for the online trigger;
\item good ($\approx 1$~cm) positron impact point position resolution in the offline event analysis;
\item excellent ($\approx 50$~ps) time resolution of the positron impact point.
\end{itemize}

The system consists of an upstream and a downstream sector, as shown in Fig.~\ref{introduction:megdet}.

\begin{figure}[t]
\centering
  \includegraphics[width=0.48\textwidth,angle=0] {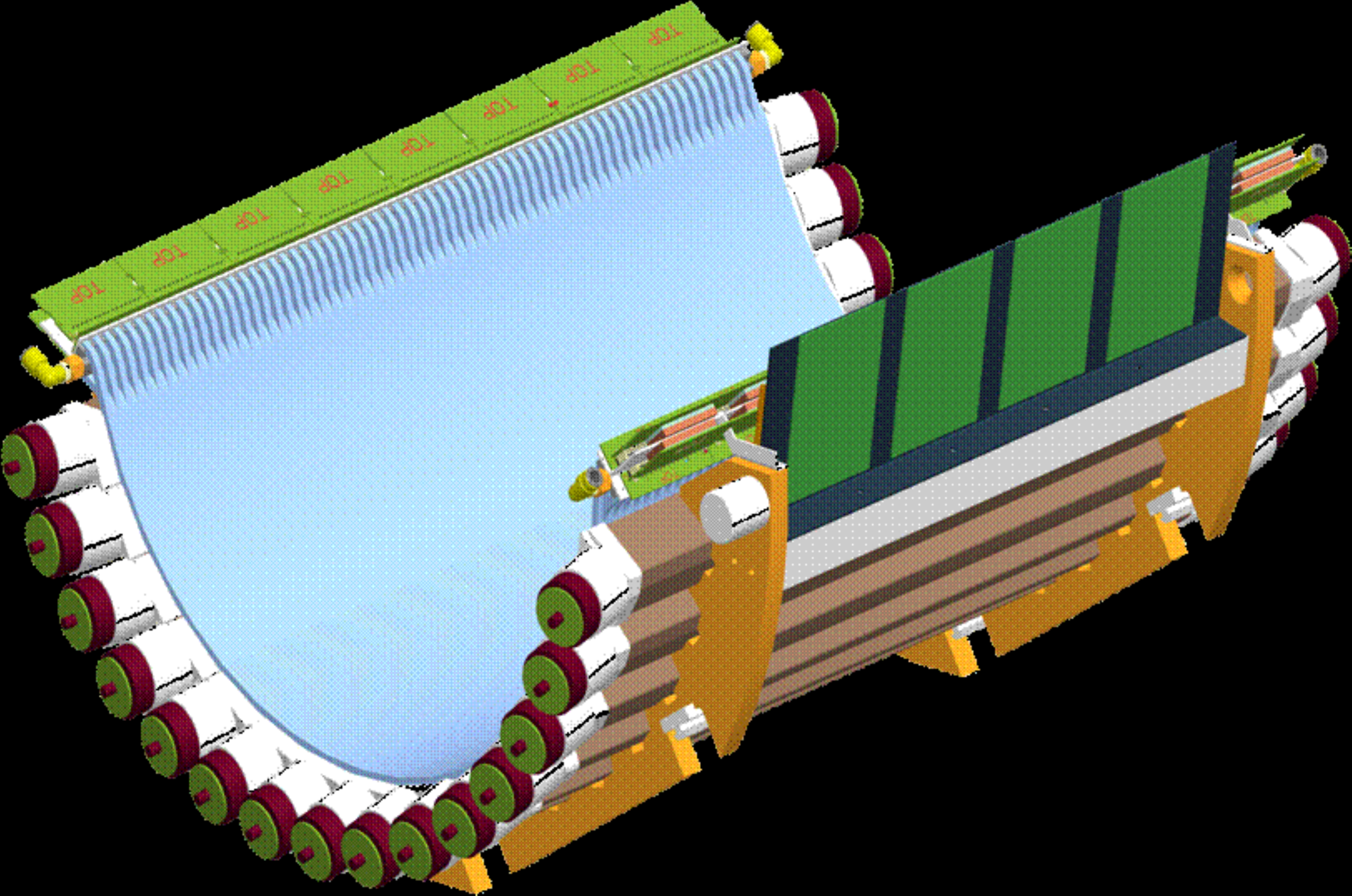}
 \caption{
Schematic picture of a TC sector. Scintillator bars are read out by a PMT at each end.
}
 \label{fig:tcall}
\end{figure}

Each sector (see Fig.~\ref{fig:tcall}) is barrel shaped with full angular coverage 
for signal positrons within the photon and positron acceptance of the LXe detector and DCH.
It consists of an array of 15 scintillating bars with
a $10.5^\circ$ pitch between adjacent bars. 
Each bar has an
approximate square cross-section of size $4.0\times 4.0\times 79.6\,\mathrm{cm}^3$ and 
is read out by a
fine-mesh, magnetic field tolerant, 2"~PMT at each end.
The inner radius of a sector is 29.5~cm, such that only positrons with a momentum close to that of 
signal positrons hit the TC.

\subsection{Liquid xenon detector}

The LXe photon detector \cite{Sawada2010258,baldini_2005_nim} requires excellent
position, time and energy resolutions to minimise
the number of accidental coincidences between photons and positrons from different muon decays,
which comprise the dominant background process (see Sect.~\ref{sec:accbackg}).

It is a homogeneous calorimeter able to
contain fully the shower induced by a 52.83~MeV photon and 
measure the photon interaction vertex, interaction time
and energy with high efficiency. The photon direction is not directly measured
in the LXe detector, rather it is inferred by the direction of a line between 
the photon interaction vertex in the LXe detector and 
the intercept of the positron trajectory at the stopping target. 

Liquid xenon, with its high density and short radiation
length, is an efficient detection medium for photons; optimal 
resolution is achieved, at least at low energies, if both the ionisation and scintillation 
signals are detected. 
In the high rate MEG environment, only the scintillation light with its very fast 
signal, is detected.

\begin{figure}[t]
\centering
  \includegraphics[width=0.48\textwidth,angle=0] {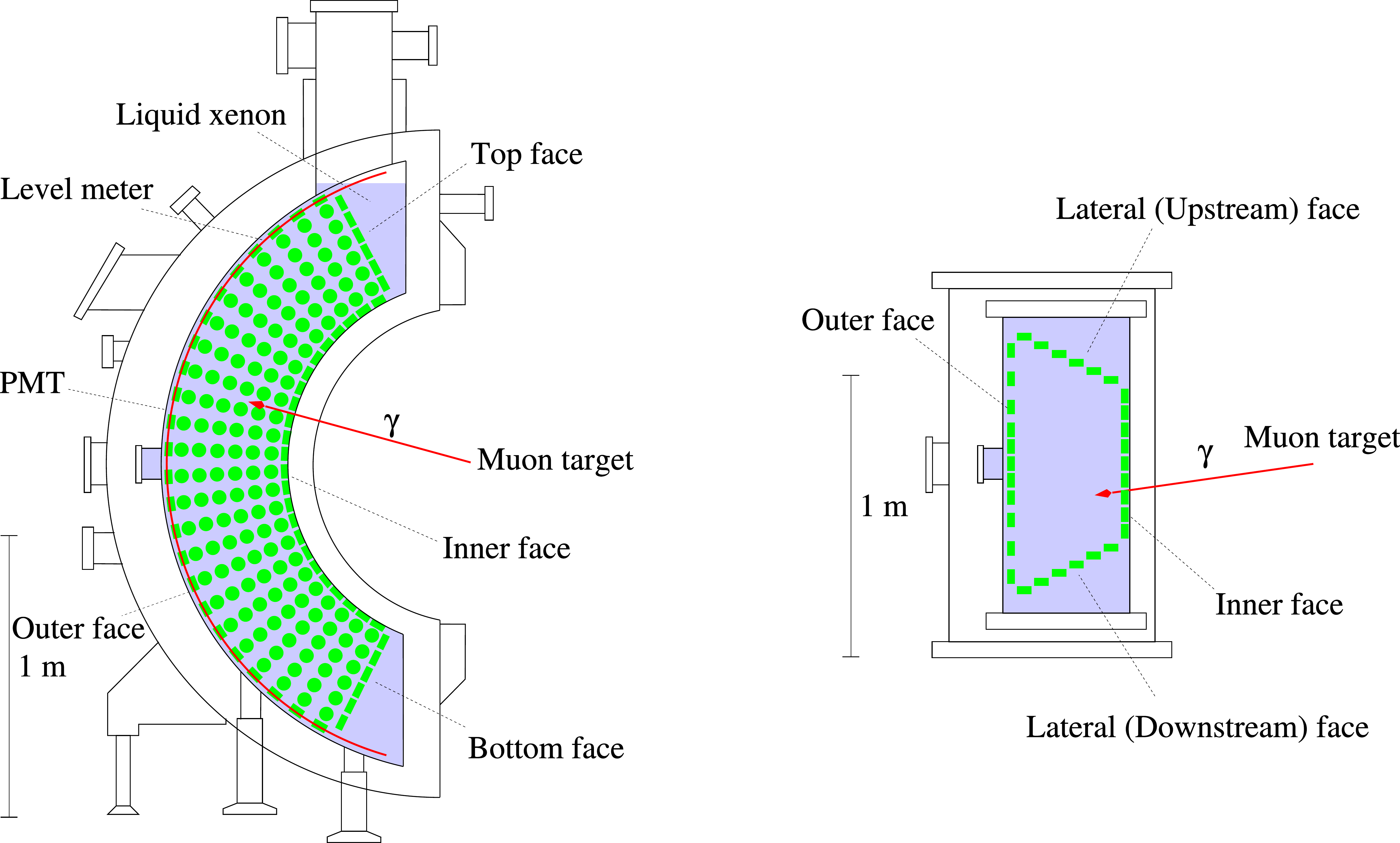}
 \caption{
Schematic view of the LXe detector: from the downstream side (left), from the top (right).
}
 \label{XEC:xecdetector}
\end{figure}

A schematic view of the
LXe detector is shown in Fig.~\ref{XEC:xecdetector}. It has a 
C-shaped structure fitting the outer radius of COBRA.
The fiducial volume is $\approx$ 800~$\ell$, covering 11\% 
of the solid angle viewed from the centre of the
stopping target. Scintillation light is detected in
846 PMTs submerged directly in the liquid xenon.
They are placed on all six faces of the detector, 
with different PMT coverage on different faces.
The detector's depth is 38.5~cm, corresponding to $\approx 14$~X$_0$.

\subsection{Calibration}
\label{sec:calibration}

Multiple calibration and monitoring tools are integrated into the experiment 
\cite{papa_2010} in order to continuously 
check the operation of single sub-detectors (e.g. LXe photodetector
gain equalisation, TC bar cross-timing, LXe and spectrometer energy scale) 
and multiple-detector comparisons simultaneously
(e.g. relative positron-photon timing).

Data for some of the monitoring and calibration tasks are recorded during normal data taking,
making use of particles coming from muon decays, for example the end-points of the positron and 
photon spectra to check the energy scale, or the positron-photon timing in RMD to check 
the LXe--TC relative timing. Additional calibrations required the installation of
new tools, devices or detectors. A list of these 
methods is presented in Table~\ref{tab:calibrations} and they are briefly discussed below.

\begin{table*}
\caption{\label{tab:calibrations} 
The calibration tools of the MEG experiment.}
{\begin{tabular*}{\textwidth}{@{\extracolsep{\fill}}ccccc@{}}
\hline
\multicolumn{2}{c}{\bf Process}  & {\bf Energy} & {\bf Main Purpose} & {\bf Frequency} \\
\hline
Cosmic rays & $\mu^{\pm}$ from atmospheric showers & Wide spectrum $\cal O$(GeV) & LXe-DCH relative position & annually \\[1.2mm]
& & & DCH alignment & \\[1.2mm]
& & & TC energy and time offset calibration & \\[1.2mm]
& & & LXe purity & on demand\\[1.2mm]
Charge exchange & $\pi^- \mathrm{p}  \to \pi^0 \mathrm{n} $  & $55, 83, 129$~MeV photons & LXe energy scale/resolution & annually \\
& $ \pi^0 \to \gamma \gamma$ & & & \\[1.2mm]
Radiative $\mu-$decay  & $\radiative$ & photons $> 40$~MeV,   & LXe-TC relative timing & continuously \\[1.2mm]
& & positrons $> 45$~MeV & Normalisation & \\[1.2mm]
Normal $\mu-$decay  & $\michel$ & 52.83 MeV end-point positrons & DCH energy scale/resolution & continuously \\[1.2mm]
& & & DCH and target alignment & \\[1.2mm]
& & & Normalisation & \\[1.2mm]
Mott positrons  & $\pos$ target $\to \pos$ target & $\approx 50$~MeV positrons & DCH energy scale/resolution & annually \\[1.2mm]
& & & DCH alignment & \\[1.2mm]
Proton accelerator & $^7 {\rm Li} (\mathrm{p}, \gamma) ^8 {\rm Be}$ & 14.8, 17.6~MeV photons & LXe uniformity/purity& weekly \\[1mm]
& $^{11} {\rm B} (\mathrm{p}, \gamma) ^{12} {\rm C}$ & 4.4, 11.6, 16.1~MeV photons & TC interbar/ LXe--TC timing & weekly \\[1.2mm]
Neutron generator & $^{58} {\rm Ni}(\mathrm{n},\gamma) ^{59}{\rm Ni}$ & 9~MeV photons & LXe energy scale & weekly \\[1.2mm]
Radioactive source & $^{241}{\rm Am}(\alpha,\gamma)^{237}{\rm Np}$ & 5.5~MeV $\alpha$'s, $56~{\rm keV}$ photons & LXe PMT calibration/purity & weekly \\[1.2mm]
Radioactive source & $^9\mathrm{Be}(\alpha_{^{241} {\rm Am}}, \mathrm{n})^{12}\mathrm{C}^{\star}$ & 4.4~MeV photons & LXe energy scale & on demand \\
                   & $^{12}\mathrm{C}^{\star}(\gamma)^{12}\mathrm{C}$ & & & \\
LED & & & LXe PMT calibration & continuously \\
\hline
\end{tabular*}}
\end{table*}

Various processes can affect the LXe detector response: xenon purity, long-term PMT gain or quantum efficiency drifts from ageing, HV variations, etc.
PMT gains are tracked using 44 blue LEDs immersed in the LXe at different positions. 
Dedicated runs for gain measurements in which LEDs are flashed at different intensities are taken every two days.
In order to monitor the PMT long-term gain and efficiency variations, flashing LED events are constantly taken (1~Hz) during physics runs.
Thin tungsten wires
with point-like $^{241}\mathrm{Am}$ $\alpha$-sources are also installed in precisely known positions in the detector fiducial volume. 
They are used for monitoring the xenon purity and 
measuring the PMT quantum efficiencies~\cite{Baldini:2006}.

A dedicated Cockcroft--Walton accelerator~\cite{calibration_cw} placed downstream of the muon beam line is 
installed to produce photons of known energy by impinging sub-MeV~protons
on a lithium tetraborate target. The accelerator was operated twice per week to generate
single photons of relatively high energy (17.6~MeV from lithium) 
to monitor the LXe detector energy scale, and 
coincident photons (4.4~MeV and 11.6~MeV from boron) 
to monitor the TC scintillator bar relative timing and 
the TC--LXe detectors' relative timing (see Table~\ref{tab:calibrations} for the relevant reactions).

A dedicated calibration run is performed annually by stopping $\pi^-$ in 
a liquid hydrogen target placed at the centre of COBRA \cite{Signorelli:2015}. Coincident photons
from $\pi^0$ decays produced in the charge exchange (CEX) reaction $\pi^- {\rm p} \to \pi^0 {\rm n}$ 
are detected simultaneously in the LXe detector and a dedicated BGO crystal detector. By appropriate 
relative LXe and BGO geometrical selection and BGO energy selection, a nearly monochromatic sample 
of 55~MeV (and 83~MeV) photons incident on the LXe are used to 
measure the response of the LXe detector at these energies and set the absolute energy scale at the signal photon energy.

% suggest skipping this
A low-energy calibration point is provided by 4.4~MeV photons from an $^{241}$Am/Be source that is moved periodically 
in front of the LXe detector during beam-off periods.

Finally, a neutron generator exploiting the $\left( {\rm n},\gamma \right)$ 
reaction on nickel shown in Table~\ref{tab:calibrations} allows an energy 
calibration under various detector rate conditions, 
in particular normal MEG and CEX data taking.

% this is out of place and i suggest removing it
Data with Mott-scattered positrons are also acquired annually to monitor and calibrate the spectrometer 
with all the benefits associated with the usage of a quasi-monochromatic energy line at $\approx 53$~MeV \cite{Rutar:2016}. 

\subsection{Front-end electronics }

The digitisation and data acquisition system for MEG uses a custom, high 
frequency digitiser
based on the switched capacitor array technique, the
Domino Ring Sampler~4 (DRS4) \cite{Ritt2010486}.
For each of the $\approx 3000$ read-out channels with a signal above some threshold, it records a waveform of 1024 samples.
The sampling rate is 1.6~GHz for the TC 
and LXe detectors, matched to the precise time measurements in these detectors, and 0.8~GHz for the DCH, matched 
to the drift velocity and intrinsic drift resolution. 

Each waveform is processed offline by applying baseline subtraction, spectral analysis, 
noise filtering, digital constant fraction discrimination etc. so as to optimise the
extraction of the variables relevant for the measurement.
Saving the full waveform provides the advantage of being able to reprocess the full waveform information offline 
with improved algorithms.

\subsection{Trigger}
% This could use some substantial worik

An experiment to search for ultra-rare events within a huge background due to a high muon stopping rate needs a quick and 
efficient event selection, which demands the combined use of high-resolution detection techniques
with fast front-end, digitising electronics and trigger. The trigger system plays an essential role in
processing the detector signals in order to find the signature of $\meg$ events in a high-background environment
\cite{trigger2013,Galli:2014uga}.
The trigger must strike a compromise between a high efficiency for signal event selection, high live-time
and a very high background rejection rate. 
The trigger rate should be kept below 10~Hz so as not to overload the data acquisition (DAQ) system.

The set of observables to be reconstructed at trigger level includes:
\begin{itemize}
\item the photon energy;
\item the relative $\egammapair$ direction;
\item the relative $\egammapair$ timing.
\end{itemize}
The stringent limit due to the latency of the read-out electronics prevents the use of any information 
from the DCH, since the electron drift time toward the anode wires is too long.
Therefore a reconstruction of the positron momentum
cannot be obtained at the trigger level
even if the requirement of a TC hit
is equivalent to the requirement of positron momentum $\gtrsim 45$~MeV.
The photon energy is the most important observable to be reconstructed,
due to the steep decrease in the spectrum at the end-point.
For this reason the calibration factors for the PMT signals of the LXe detector
(such as PMT gains and quantum efficiencies) are continuously monitored and periodically
updated. The energy deposited in the LXe detector is estimated
by the weighted linear sum of the PMT pulse amplitudes. 

The amplitudes of the inner-face PMT pulses are also sent to comparator
stages to extract the index of the PMT collecting the highest charge,
which provides a robust estimator of the photon interaction vertex 
in the LXe detector.
The line connecting this vertex and the target centre provides an estimate of the photon direction.

On the positron side, the coordinates of the TC interaction point are the 
only information available online. The radial coordinate is given simply by the 
radial location of the TC, while, due to its segmentation along $\phi$,
this coordinate is  
identified by the bar index of the first hit (first bar encountered moving
along the positron trajectory). The local $z$-coordinate on the hit bar is measured 
by the ratio of charges on the PMTs on opposite sides of the bar with a resolution
$\approx 5$~cm.

On the assumption of the momentum being that
of a signal event and the direction opposite to that of the photon,
by means of Monte Carlo (MC) simulations, each PMT index is associated
with a region of the TC.
If the online TC coordinates fall into this region, the relative
$\egammapair$ direction is compatible with the back-to-back condition.

The interaction time of the photon in the LXe detector is extracted by a fit of the
leading edge of PMT pulses with a $\approx 2$ ns resolution.
The same procedure allows the estimation of the time of the positron
hit on the TC with a comparable resolution. The relative time
is obtained from
their difference; fluctuations due to the time-of-flight of each particle
are within the resolutions.

\subsection{DAQ system}
%{\it Editor's comments: \\
%Section coordinator: Luca G., Stefan R.\\
%Text:  0.5\\
%Figure: 1.
%}

The DAQ challenge is to perform the complete read-out of all detector waveforms while maintaining the system efficiency, 
defined as the product of the online efficiency ($\epsilon_{\rm trg}$) and the DAQ live-time fraction ($f_\mathrm{LT}$), as high as possible.

At the beginning of data taking, with the help of a MC simulation, a trigger configuration which maximised the DAQ efficiency was found to have 
$\epsilon_{\rm trg}\approx 90\%$ and $f_\mathrm{LT}\approx 85\%$ and an associated event rate $R_{\rm daq}\approx 7~\mathrm{Hz}$, almost seven orders of magnitude lower than the muon stopping rate.

The system bottleneck was found in the waveform read-out time from the VME boards to the online disks, lasting as much as 
$t_\mathrm{ro}\approx$~24~$\mathrm{ms/event}$; the irreducible contribution to the dead-time is the DRS4 read-out time and 
accounts for $625~\mathrm{\mu s}$. This limitation has been overcome, starting from the 2011 run, thanks to a multiple buffer 
read-out scheme, in our case consisting of three buffers. In this scheme, in case of a new trigger during the event read-out from a buffer, 
new waveforms are written in the following one; the system experiences dead-time only when there are no empty buffers left. 
This happens when three events occur within a time interval equal to the read-out time $t_\mathrm{ro}$. The associated live-time is
\[
f_\mathrm{LT} = \exp^{-R_\mathrm{daq}\cdot t_\mathrm{ro}} \cdot [1+R_\mathrm{daq}\cdot t_\mathrm{ro} + (R_\mathrm{daq}\cdot t_\mathrm{ro})^2/2! ],
\]
and is $\geq99\%$ for event rates up to $\approx 13$~Hz. 

\begin{figure}[!ht]
\begin{center}
	\includegraphics[width=0.5\textwidth]{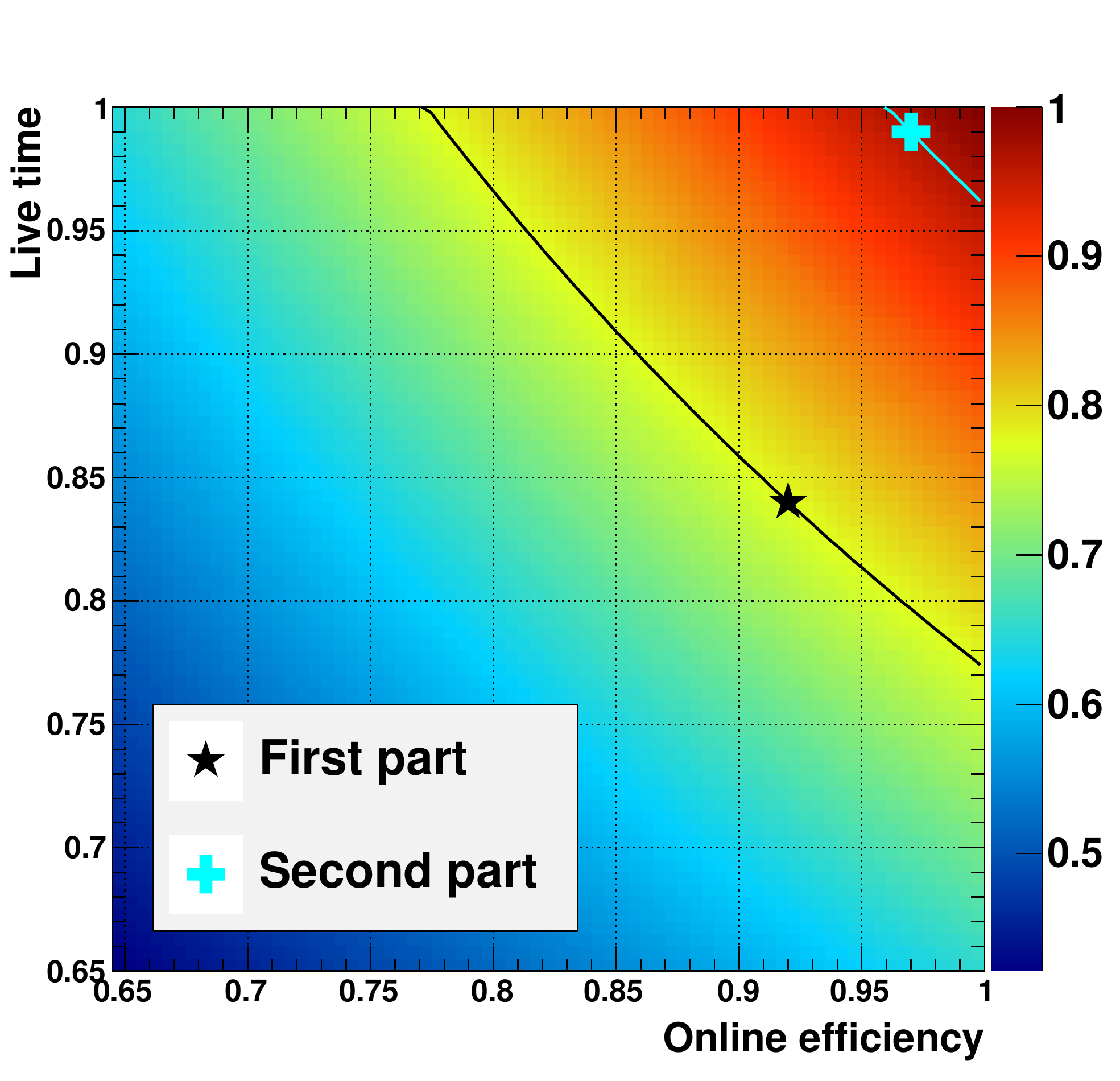}
\end{center}
\caption{Contour lines for DAQ efficiency during different run periods: without (First part: 2009-2010) and with (Second part: 2011-2013) the multiple buffer read-out scheme.}
\label{fig:daqeff}
\end{figure}

The multiple buffer scheme allows relaxation of the trigger conditions, in particular for what concerns the relative 
$\egammapair$ direction, leading to a much more efficient DAQ system, from 75\%\ in the 2009-2010 runs to 97\%\ in the 2011-2013 runs.
Figure~\ref{fig:daqeff} shows the two described working points, the first part refers to the 2009-2010 runs, 
while the second refers to the 2011-2013 runs.

%% file: reconstruction.tex
\section{Reconstruction}
\label{sec:reconstruction}
%
%{\it Editor's comments: \\
%Section coordinator: Toshiyuki Iwamoto \\
%Text:  5.5\\
%Figure: 12.
%} 

In this section the reconstruction of high-level objects is presented. More information about low-level objects 
(e.g. waveform analysis, hit reconstruction) and calibration issues are available in \cite{megdet}.

\subsection{Photon reconstruction}
%{\it Editor's comments: \\
%Section coordinator: Toshiyuki, Ryu, Yusuke\\
%Text:  1.5\\
%Figure: 2.
%} 
\label{sec:xec_reco}
A 52.83 MeV photon interacts with LXe predominantly via the pair production process, followed by an electromagnetic shower. 
The major uncertainty in the reconstruction stems from the event-by-event fluctuations in the shower development.
A series of algorithms provide the best estimates of the energy, the interaction vertex, and the interaction time 
of the incident photon and to identify and eliminate events with multiple photons in the same event.

For reconstruction inside the LXe detector, a special coordinate system $(u, v, w)$ is used: $u$ coincides with $z$ in the MEG coordinate system; $v$ is directed along the negative $\phi$-direction at the radius of the fiducial volume inner face ($r_\mathrm{in} = 67.85$~cm); $w=r-r_\mathrm{in}$, measures the depth. The fiducial volume of the LXe detector is defined as $|u|<25$~cm, $|v|<71$~cm, and 0 $<w<38.5$~cm ($|\cos\theta|<0.342$ and $120^{\circ}<\phi<240^\circ$) in order to ensure high resolutions, especially for energy and position measurements. 

The reconstruction starts with a waveform analysis that extracts charge and time for each of the PMT waveforms.
The digital-constant-fraction method is used to determine an (almost) amplitude independent pulse time, defined as the time when the signal reaches a predefined fraction (20\%) of the maximum pulse height. 
To minimise the effect of noise on the determination of the charge, a digital high-pass filter\footnote{The high-pass filter is written: 
$$
y[i] = x[i] - \frac{1}{M}\sum_{j=1}^{M}x[i-M+j],
$$
where $x[]$ is the waveform amplitude in waveform time-bins, $y[]$ is the output signal in the same time-bins, and $M=105$ is the number of points used in the average. This filter is based on the moving average, which is a simple and fast algorithm with a good response in time domain.
}
with a cutoff frequency of $\approx 10$~MHz, is applied.

The charge on each PMT ($Q_i$) is converted into the number of photoelectrons ($N_{\mathrm{pe},i}$) and into the number of scintillation photons impinging on the PMT ($N_{\mathrm{pho},i}$) as follows:
\begin{eqnarray}
&&N_{\mathrm{pe},i} = Q_i /  eG_i(t), \nonumber \\
&&N_{\mathrm{pho},i} = N_{\mathrm{pe},i} / \mathcal{E}_i(t), \nonumber
\end{eqnarray}
where $G_i(t)$ is the PMT gain and $\mathcal{E}_i(t)$ is the product of the quantum efficiency of the photocathode and the collection efficiency to the first dynode.
These quantities vary with time\footnote{Two kinds of instability in the PMT response are observed: one is a long-term gain decrease due to decreased secondary emission mainly at the last dynode with collected charge  and the other is a rate-dependent gain shift due to charge build-up on the dynodes.} and, thus, are continuously monitored and calibrated using the calibration sources instrumented in the LXe detector (see Sect.~\ref{sec:calibration}).

The PMT gain is measured using blue LEDs, flashed at different intensities by exploiting the statistical relation between the mean and variance of the observed charge,
\[
\sigma^2_{Q_i} = eG_i\bar{Q}_i + \sigma^2_\mathrm{noise}.
\] 
The time variation of the gain is tracked by using the LED events collected at $\approx 1$~Hz during physics data taking.

The quantity $\mathcal{E}_i(t)$ is evaluated using $\alpha$-particles produced by $^{241}\mathrm{Am}$ sources within the LXe volume and monochromatic (17.6-MeV) photons from a p-Li interaction (see Table~\ref{tab:calibrations}) by comparing the observed number of photoelectrons with the expected number of scintillation photons evaluated with a MC simulation,
\[
\mathcal{E}_i = \bar{N}_{\mathrm{pe},i}/\bar{N}_{\mathrm{pho},i}^\mathrm{MC}.
\]
This calibration is performed two or three times per week to monitor the time dependence of this factor. The absolute energy scale is not sensitive to the absolute magnitude of this efficiency, and this calibration serves primarily to equalise the relative PMT responses and to remove time-dependent drifts, possibly different from PMT to PMT.

\subsubsection{Photon position}
\label{sec:xec_position}
The 3D position of the photon interaction vertex $\posgamma=(\ugamma,\vgamma,\wgamma)$ is determined by a $\chi^2$-fit of the distribution 
of the numbers of scintillation photons in the PMTs ($N_\mathrm{pho}$), taking into account the solid angle subtended by each PMT 
photocathode assuming an interaction vertex, to the observed $N_\mathrm{pho}$ distribution.
To minimise the effect of shower fluctuations, only PMTs inside a radius of 3.5 times the PMT spacing for the initial estimate of 
the position of the interaction vertex 
are used in the fit. The initial estimate of the position is calculated as the amplitude weighted mean position around the PMT with 
the maximum signal. For events resulting in $\wgamma < 12$~cm, the fit is repeated with a further reduced number of PMTs, inside 
a radius of twice the PMT spacing from the first fit result.
The remaining bias on the result, due to the inclined incidence of the photon onto the inner face, is corrected using results from a MC simulation.
The performance of the position reconstruction is evaluated by a MC simulation and has been verified in dedicated CEX runs by placing 
lead collimators in front of the LXe detector. 
The average position resolutions along the two orthogonal inner-face coordinates ($u, v$) and the depth direction ($w$) are estimated 
to be $\approx 5$~mm and $\approx 6$~mm, respectively. 

The position is reconstructed in the LXe detector local coordinate system. The conversion to the MEG coordinate system relies on the alignment of the LXe detector with the rest of the MEG subsystems. 
The LXe detector position relative to the MEG coordinate system is precisely surveyed 
using a laser survey device at room temperature.
After the thermal shrinkage of the cryostat and of the PMT support structures at LXe temperature are taken into account, 
the PMT positions are calculated based on the above information. 
The final alignment of the LXe detector with respect to the spectrometer is described in Sect.~\ref{sec:relative_angle}.

\subsubsection{Photon timing}
\label{sec:tgamma}
The determination of the photon emission time from the target $\tgamma$ starts from the 
determination of the arrival time of the scintillation photons on the $i$-th PMT $t_{\gamma,i}^\mathrm{PMT}$ as described in Sect.~\ref{sec:xec_reco}.
To relate this time to the photon conversion time, the propagation time of the scintillation photons 
must be subtracted as well as any hardware-induced time offset (e.g. due to cable length).

%% Re-orginzed this paragraph by YU
The propagation time of the scintillation photons is evaluated using the $\pi^0 \to \gamma\gamma$ events 
produced in CEX runs in which the time of one of the photons is measured by two plastic scintillator counters with a lead shower converter
as a reference time. 
The primary contribution is expressed as a linear relation with the distance; the coefficient, i.e., the effective light velocity, is measured to be $\approx 8$~cm/ns.
A remaining non-linear dependence is observed and an empirical function (2D function of the distance and incident angle) is calibrated from the data. This secondary effect comes from the fact that the fraction of indirect (scattered of reflected) scintillation photons increases with a larger incident angle and a larger distance.
PMTs that do not directly view the interaction vertex $\posgamma$, shaded by the inner face wall, are not used in the following timing reconstruction.
After correcting for the scintillation photon propagation times, the remaining (constant) time offset is extracted for each PMT from the same $\pi^0 \to \gamma\gamma$ events by comparing the PMT hit time with the reference time.

After correcting for these effects, the photon conversion time $\tgamma^\mathrm{LXe}$ is obtained by combining 
the timings of those PMTs $t_{\gamma,i}^\mathrm{PMT}$ which observe more than 50 $N_\mathrm{pe}$ 
%and are not in the shadow of the walls 
by a fit that minimises 
\[%begin{equation}
\chi^2 = \sum_{i}\frac{\left(t_{\gamma,i}^\mathrm{PMT} - \tgamma^\mathrm{LXe}\right)^2}{\left(\sigma_{\tgamma}^\mathrm{1\mathchar`-PMT}(N_{\mathrm{pe},i})\right)^2}.
\]%end{equation}
PMTs with a large contribution to the $\chi^2$ are rejected during this fitting procedure to remove pile-up effects.
The single-PMT time resolution is measured in the CEX runs to be $\sigma_{\tgamma}^\mathrm{1\mathchar`-PMT}(N_\mathrm{pe}=500) = 400$--540~ps, depending on the location of the PMT, and approximately proportional to $1/\sqrt{N_\mathrm{pe}}$.
Typically 150 PMTs with $\approx 70\,000$ $N_\mathrm{pe}$ in total are used to reconstruct 50-MeV photon times. 

Finally, the photon emission time from the target $t_\gamma$ is obtained by 
subtracting the time-of-flight between the point on the stopping target 
defined by the intercept of the positron trajectory at the stopping target
and the reconstructed interaction vertex in the LXe detector from 
$t_\gamma^\mathrm{LXe}$.

The timing resolution $\sigma_{\tgamma}$ is evaluated as the dispersion of the time difference between the two photons from $\pi^0$ 
decay after subtracting contributions due to the uncertainty of the $\pi^0$ decay position and to the timing resolution of the reference counters. 
From measurements at 55 and 83~MeV, the energy dependence is estimated and corrected, resulting in $\sigma_{\tgamma} (\egamma=52.83~\mathrm{MeV}) \approx 64$~ps. 

\subsubsection{Photon energy}
\label{sec:energy_gamma}
The reconstruction of the photon energy $\egamma$ is based on the sum of scintillation photons collected by all PMTs. 
A summed waveform with the following coefficients over all the PMTs is formed and the energy is determined by integrating it:
%\[
\begin{equation}
F_i = \frac{A_i \cdot W_i(\posgamma)}{eG_i(t)\cdot \mathcal{E}_i(t)} \cdot \Omega(\posgamma)  \cdot U(\posgamma) \cdot H(t) \cdot S,
\label{eq:egamma_coeff}
%\]
\end{equation}
where 
$A_i$ is a correction factor for the fraction of photocathode coverage, which is dependent on the PMT location\footnote{The coverage on the outer face is, for example, 2.6 times less dense than that on the inner face};
$W_i(\posgamma)$ is a weighting factor for the PMT that is common to all PMTs on a given face and is determined by minimising the resolution in response to 55-MeV photons from CEX. $\Omega(\posgamma)$ is a correction factor for the solid angle subtended by photocathodes for scintillating photons emitted at the interaction vertex; it is applied only for shallow events ($\wgamma<3$~cm) for which the light collection efficiency is very sensitive to the relative position of each PMT and the interaction vertex. $U(\posgamma)$ is a position dependent non-uniformity correction factor determined by the responses to the 17.6- and 55-MeV photons. $H(t)$ is a correction factor for the time-varying LXe scintillation light yield and $S$ is a constant conversion factor of the energy scale, determined by the 55- and 83-MeV photons with a precision of 0.3\%.

A potential significant background is due to pile-up events with more than one photon in the detector nearly coincident in time.  
Approximately 15\% of triggered events suffer from pile-up at the nominal beam rate. The analysis identifies pile-up events and corrects the measured energy, thereby reducing background and increasing detection efficiency.
Three methods are used to identify and extract the primary photon energy in pile-up events.  

The first method identifies multiple photons with different timing using the $\chi^2/\mathrm{NDF}$ value in the time fit.
In contrast to the time reconstruction, all the PMTs with more than 50 $N_\mathrm{pe}$ are used to identify pile-up events.

The second method identifies pile-up events with photons at different positions by searching for spatially separated peaks in the inner and outer faces.
If the event has two or more peaks whose energies cannot be determined using the third method below, a pile-up removal algorithm is applied to the PMT charge distribution.  
It uses a position dependent table containing the average charge of each PMT in response to 17.6-MeV photons.
Once a pile-up event is identified, the energy of the primary photon is estimated by fitting the PMT charges to the table without using PMTs around the secondary photon. 
Then, the PMT charges around the secondary photon are replaced with the charges estimated by the fit.
Finally, the energy is reconstructed as a sum of the individual PMT charges with the coefficients $F_i$ (Eq.~\ref{eq:egamma_coeff}), instead of integrating the summed waveform.

\begin{figure}[tbp]
\centering
\includegraphics[width=1\columnwidth]{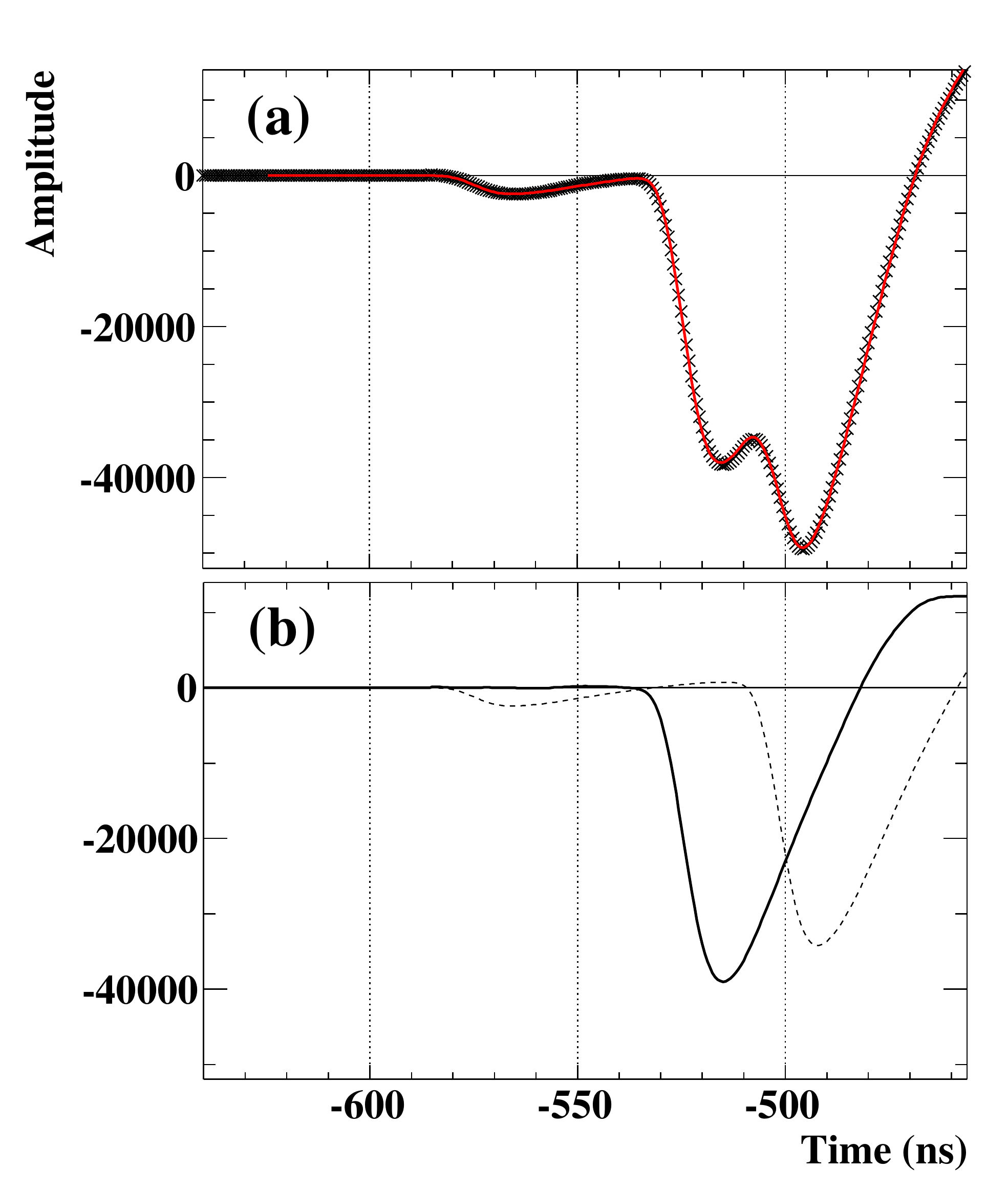}
\caption{\label{fig:xecwf}(a) Example of a LXe detector waveform for an event with three photons (2.5, 40.1 and 36.1~MeV). The cross markers show the 
waveform (with the digital high-pass filter) summed over all PMTs with the coefficients defined in the text, and the red line shows the fitted superposition 
of three template waveforms. (b) The unfolded main pulse (solid line) and the pile-up pulses (dashed).}
\end{figure}

The third method identifies multiple photons and unfolds them by combining the information from summed waveforms and the two methods above.
First, the total summed waveform is searched for temporally separated pulses.
Next, if the event is identified as a pile-up event by either of the two methods above, a summed waveform over PMTs near the secondary photon 
is formed to search for multiple pulses. 
The pulse found in the partial summed waveform is added to the list of pulses if the time is more than 5~ns apart from the other pulse times.
Then, a superimposition of $N$ template waveforms is fitted to the total summed waveform, where $N$ is the number of pulses detected in this event. 
Figure~\ref{fig:xecwf} shows an example of the fitting, where three pulses are detected. 
Finally, the contributions of pile-up photons are subtracted and the remaining waveform is used for the primary energy estimation. 

\begin{figure}[tbp]
\centering
\includegraphics[width=1\columnwidth]{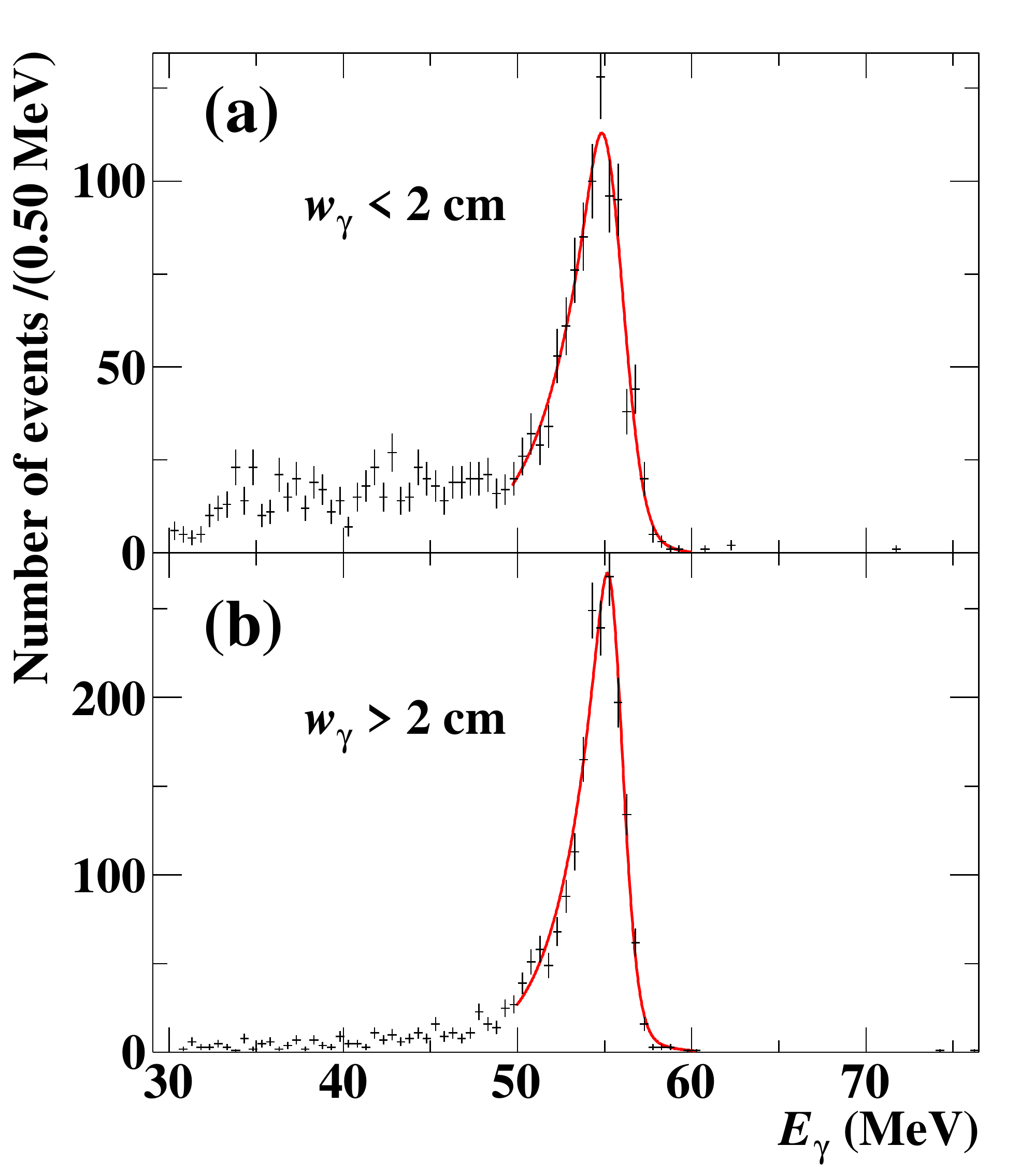}
\caption{\label{fig:xeclineshape}Energy response of the LXe detector to 54.9-MeV photons in a restricted range of ($\ugamma$, $\vgamma$) 
for two groups of events with different~$\wgamma$: (a) $0<\wgamma<2~\mathrm{cm}$ (event fraction 42\%) and (b) $\wgamma > 2~\mathrm{cm}$ (58\%).}
\end{figure}

The energy response of the LXe detector is studied in the CEX runs using $\pi^0$ decays with an opening angle between the two photons $> 170^\circ$, for which each of the photons has an intrinsic line width small compared to the detector resolution.
The measured line shape is shown in Fig.~\ref{fig:xeclineshape} at two different conversion depth ($\wgamma$) regions.
The line shape is asymmetric with a low energy tail mainly for two reasons: the interaction of the photon in the 
material in front of the LXe detector fiducial volume, and the albedo shower leakage from the inner face.
The energy resolution is evaluated from the width of the line shape on the right-hand (high-energy) side ($\sigma_{\egamma}$) by unfolding the finite width of 
the incident photon energy distribution due to the imperfect back-to-back selection and a small correction for the different background conditions between the muon and pion beams.
Since the response of the detector depends on the position of the photon conversion, 
the fitted parameters of the line shape are functions of the 3D coordinates, mainly of $\wgamma$.
The average resolution is measured to be
$\sigma_{\egamma} = 2.3\%$ ($0 < \wgamma < 2~\mathrm{cm}$, event fraction 42\%) and 1.6\% ($\wgamma > 2~\mathrm{cm}$, 58\%).

The energy resolutions and energy scale are cross-checked by fitting the background spectra measured in the muon decay data with the MC spectra folded with the detector resolutions.

\subsection{Positron reconstruction}
%{\it Editor's comments: \\
%Section coordinator: Francesco R., Gordon L., Luca G., Matteo D.\\
%Text:  3.0\\
%Figure: 8.
%} 
\label{sec:positron_rec}

\subsubsection{DCH reconstruction}
%{\it Editor's comments:\\
%Section coordinator: Francesco 
%}
\label{sec:dch_position}

The reconstruction of positron trajectories in the DCH is performed in four steps: hit reconstruction in
each single cell, clustering of hits within the same chamber, track finding in the spectrometer, and track fitting.

In step one, raw waveforms from anodes and cathodes are filtered in order to remove known noise contributions of fixed
frequencies. A hit is defined as a negative signal appearing in the waveform collected at each end of the anode wire,
with an amplitude of at least $-5$~mV below the baseline. This level and its uncertainty $\sigma_\mathrm{B}$ are estimated from the waveform itself in the region around
625~ns before the trigger time. The hit time is taken from the anode signal with larger amplitude as the
 time of the first sample more than $-3\sigma_\mathrm{B}$ below the baseline.

The samples with amplitude below $-2\sigma_\mathrm{B}$ from the baseline and in a range of $[-24,+56]$~ns
around the peak, are used for charge integration. The range is optimised to minimise the uncertainty produced by
the electronic noise. A first estimate of the $z$-coordinate, with a resolution of about 1~cm, is obtained
from charge division on the anode wire, and it allows the determination of the Vernier cycle 
(see Sect.~\ref{sec:dch-system}) in which the hit occurred.
If one or more of the four cathode pad channels is known to be defective, the $z$-coordinate
from charge division is used and is assigned a 1~cm uncertainty. Otherwise, charge integration
is performed on the cathode pad waveforms and the resulting charges are
combined to refine the $z$-measurement, exploiting the Vernier pattern. The charge asymmetries
between the upper and lower sections of the inner and outer cathodes are given by
\[
A_{\mathrm{in,out}} = \frac{Q^{\mathrm{UP}}_{\mathrm{in,out}} - Q^{\mathrm{DOWN}}_{\mathrm{in,out}}}{Q^{\mathrm{UP}}_{\mathrm{in,out}} + Q^{\mathrm{DOWN}}_{\mathrm{in,out}}} \; ,
\]
the position within the $\lambda = 5$ cm Vernier cycle is given by:
\[
\delta_z = \arctan(A_{\mathrm{in}}/A_{\mathrm{out}}) \times \frac{\lambda}{2\pi}\; .
\]
At this stage, a first estimate of the position of the hit in the ($x,y$) plane is given by the wire position.

Once reconstructed, hits from nearby cells with similar $z$ are grouped into clusters, taking 
into account that the $z$-measurement can be shifted by $\lambda$ if the wrong Vernier cycle has been selected
via charge division. These clusters are then used to build track seeds.

A seed is defined as a group of three clusters in four adjacent chambers, 
at large radius ($r>24$~cm) where the chamber occupancy is lower and only particles with large momentum are found. The clusters are required to satisfy appropriate proximity criteria
on their $r$ and $z$ values.
A first estimate of the track curvature and total momentum is obtained from the coordinates of the hit wires, and is 
used to extend the track and search for other clusters, taking advantage of the adiabatic invariant $p_T^2/B_z$, 
where $p_T$ is the positron transverse momentum, for slowly varying 
axial magnetic fields. Having determined the approximate trajectory, the left/right ambiguity of the hits on each wire can be resolved in most cases.
A first estimate of the track time (and hence the precise
position of the hit within a cell) and further improvement of the left/right solutions can be obtained by minimising the 
$\chi^2$ of a circle fit of the hit positions in the ($x,y$) plane.

At this stage, in order to retain high efficiency, the same hit can belong to different 
clusters and the same cluster to different track candidates, which can result in duplicated tracks. Only after the track fit,
when the best information on the track is available, independent tracks are defined.

A precise estimate of the ($x,y$) positions of the hits associated with the track
candidate is then extracted from the drift time, defined as the difference between the hit and track times.
The position is taken from tables relating ($x,y$) position to drift time. These are track-angle dependent and are derived using
GARFIELD software~\cite{Veenhof:1993hz}.
The reconstructed ($x,y$) position is continuously updated during the tracking process, as the
track information improves.

A track fit is finally performed with the Kalman filter technique~\cite{Billoir:1983mz,Fruhwirth:1987fm}. The GEANE
software~\cite{Fontana:2008zz} is used to account for the effect of materials in the spectrometer during
the propagation of the track and to estimate the error matrix. Exploiting the results of the first track fit,
hits not initially included in the track candidate are added if appropriate and hits which are inconsistent with the fitted track are removed.
The track is then propagated to the TC and matched to the hits in the bars (see Sect.~\ref{sec:tc-dc} for details).
The time of the matched TC hit (corrected for propagation delay) is used to provide a more accurate estimate of the track time, and hence the drift times.
A final fit is then done with this refined information. Following the fit, the track is propagated backwards to the target. The decay vertex ($\xpos$, $\ypos$, $\zpos$) and the
positron decay direction ($\phie$, $\thetae$) are defined as the point of intersection of the track with the target foil and the track direction at the decay vertex.
%is taken as the decay vertex  while the extrapolated
%direction  is taken as the positron direction at the decay vertex.
The error matrix of the track parameters at the decay vertex is computed and used in the subsequent analysis.

Among tracks sharing at least one hit, a ranking is performed based on a linear combination of
five variables denoting the quality of the track (the momentum, $\thetae$ and $\phie$ errors at the target, the number of hits and the
reduced $\chi^2$). In order to optimise the performance of the ranking procedure, the linear combination is taken as the first component
of a principal component analysis of the five variables. 
The ranking variables are also used to select tracks, along with other 
quality criteria as (for instance) the request that the backward track 
extrapolation intercepts the target within its fiducial volume. 
Since the subsequent analysis uses the
errors associated with the track parameters event by event, the selection criteria are kept loose in order to preserve high efficiency while removing badly reconstructed tracks
for which the fit and the associated errors might be unreliable. After the selection criteria are applied, the track quality ranking is used to select only one track among the surviving
duplicate candidates.

\subsubsection{DCH missing turn recovery}
%{\it Editor's comments:\\
%Section coordinator: Gordon 
%}
\label{sec:dch_missingturnrecovery}

A positron can traverse the DCH system multiple times before it exits the spectrometer. An individual crossing of the DCH system is referred to as a positron `turn'. 
An intermediate merging step in the Kalman fit procedure, described previously, attempts to identify multi-turn positrons by combining 
and refitting individually reconstructed turns into a multi-turn track. However, it is possible that not all turns of a multi-turn 
positron are correctly reconstructed or merged into a multi-turn track. If this involves the first turn, i.e. the turn closest to the 
muon stopping target, this will lead to an incorrect determination of the muon decay point and time as well as an incorrect determination 
of the positron momentum and direction at the muon decay point, and therefore a loss of signal efficiency. 

\begin{figure}[t]
\centering
\includegraphics[width=0.5\textwidth,angle=0] {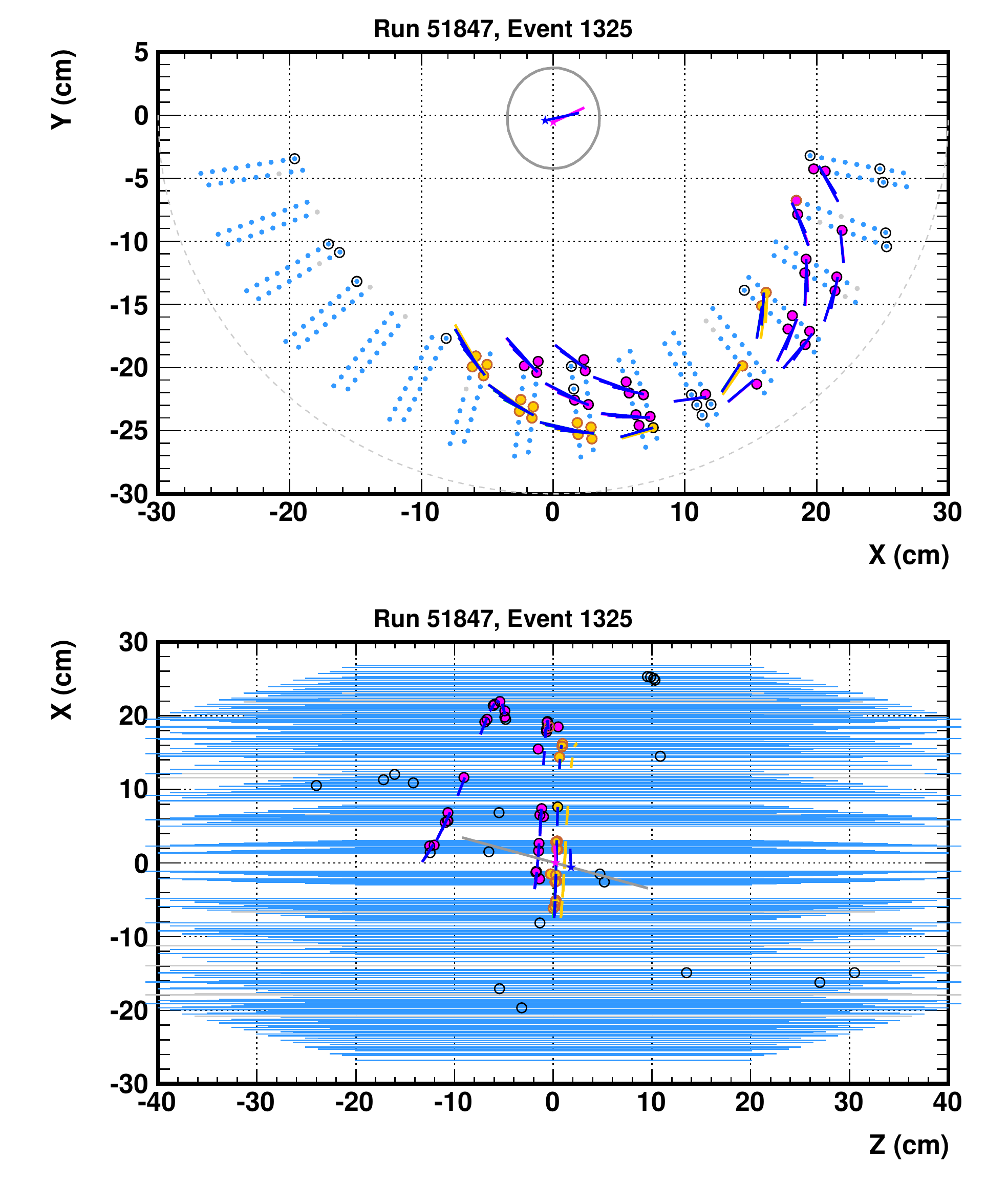}
\caption{Example of a triple-turn positron in a year 2009 event. The positron was 
originally reconstructed as a double-turn track, formed by magenta hits, but 
the MFT recovery algorithm found a missing first track formed by the brown hits. 
The track was then refitted as a triple-turn one; the corresponding positron vector 
extrapolated at the target is shown as a blue arrow and compared with that coming 
from the original double-turn fitted track, shown as a magenta arrow.}
\label{fig:Missingturnexample}
\end{figure}

After the track reconstruction is completed, a missing first turn (MFT) recovery algorithm, developed and incorporated in the DCH reconstruction software
expressly for this analysis, is used to identify and refit positron tracks with an MFT.
Firstly, for each track in an event, the algorithm identifies all hits that may potentially be part of an MFT,
based on the compatibility of their $z$-coordinates and wire locations in the DCH system with regard to the positron track.
% based on their $z$-coordinates and DCH module
%identifiers with respect to the positron track. 
The vertex state vector of the track is propagated backwards to the point of closest approach with each
potential MFT hit, and the hit selection is refined based on the $r$ and $z$ residuals between the potential MFT hits and their propagated state vector
positions. 
%Potential MFT candidates are subsequently selected based on the distribution properties and 
%number of remaining selected MFT hits.
Potential MFT candidates are subsequently selected if there are MFT hits in at least four DCH modules 
of which three are adjacent to one another, and the average signed $z$-difference between the hits and 
their propagated state vector positions as well as the standard deviation of the corresponding 
unsigned $z$-difference are smaller than 2.5~cm.
A new MFT track is reconstructed using the Kalman filter technique based on the selected MFT hits and correspondingly propagated state vectors.
Finally, the original positron and MFT tracks are combined and refitted using the Kalman filter technique, followed by a recalculation of the track
quality ranking and the positron variables and their uncertainties at the target. An example of a multi-turn positron with a recovered MFT is shown in Fig.~\ref{fig:Missingturnexample}.

\begin{figure}[t]
\centering
\includegraphics[width=0.5\textwidth,angle=0] {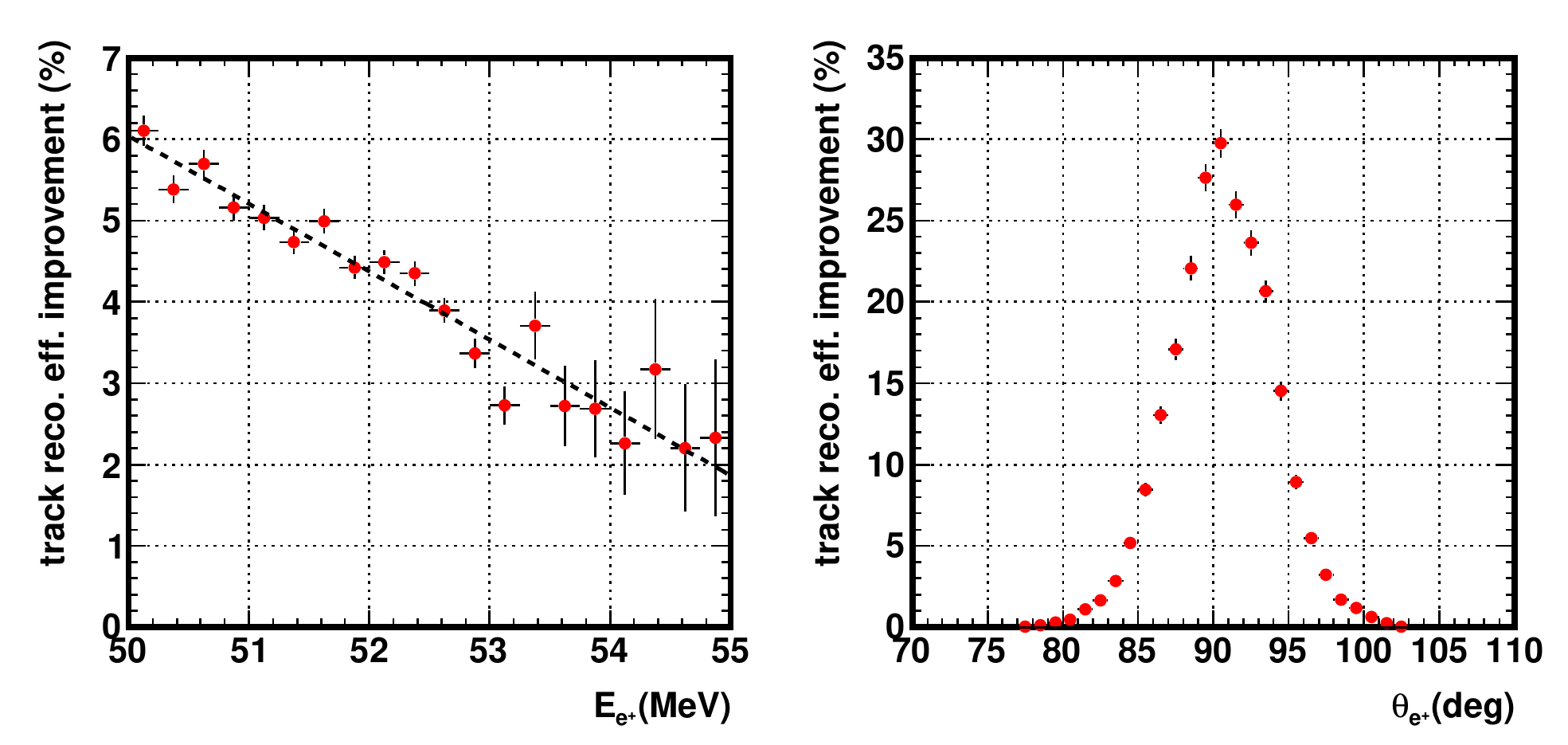}
\caption{The improvement of the overall track reconstruction efficiency due to the use of the MFT recovery algorithm as a function of $\epositron$ (left) and $\thetae$ (right).}
\label{fig:MFTefficiency}
\end{figure}

The improvement of the overall track reconstruction efficiency due to the use of the MFT recovery algorithm, defined as the ratio of the number of reconstructed Michel positrons with a recovered MFT to the total number of reconstructed Michel positrons, is measured using data and is shown as a function of $\epositron$ and $\thetae$ in Fig.~\ref{fig:MFTefficiency}. As can be seen from the left figure, the improvement of the track reconstruction efficiency at the signal energy due to the use of the MFT recovery algorithm, averaged over all angles, is $\approx 4\%$. The efficiency improvement decreases with increasing energy because the nominal track reconstruction is more efficient at higher energy. The right figure shows that the efficiency improvement is maximal for positrons emitted perpendicular to the beam direction, as expected, since these positrons are more likely to have multiple turns and cross the target twice.

\subsubsection{DCH alignment}
%{\it Editor's comments:\\
%Section coordinator: Gordon, Luca
%}
\label{sec:dch_align}

Accurate positron track reconstruction requires precise knowledge of the location and orientation of the anode wires and cathode pads in the DCH system. This is achieved by an
alignment procedure that consists of two parts: an optical survey alignment based on reference markers, and a software alignment based on reconstructed tracks.

Each DCH module is equipped with cross hair marks on the upstream and downstream sides of the module. Each module is fastened to carbon-fibre support structures on the upstream and
downstream sides of the DCH system, which accommodate individual alignment pins with an optically detectable centre. Before the start of each data-taking period an
optical survey of the cross hairs and pins is performed using a theodolite. The optical survey technique was improved in 2011 by adding corner cube reflectors next to the cross
hairs, which were used in conjunction with a laser tracker system. The resolution of the laser method is $\approx 0.2$~mm for each coordinate.

Two independent software alignment methods are used to cross-check and further improve the alignment precision of the DCH system.
The first method is based on the Millepede algorithm \cite{millipede} and uses cosmic-rays reconstructed without magnetic field. During COBRA shutdown periods, cosmic-rays are triggered using dedicated
scintillation counters located around the magnet cryostat. The alignment procedure utilises the reconstructed hit positions on the DCH modules to minimise the residuals with
respect to straight tracks according to the Millepede algorithm. The global alignment parameters, three positional and three rotational degrees of freedom per module, are
determined with an accuracy of better than 150~$\mu$m for each coordinate.

The second method is based on an iterative algorithm using reconstructed Michel positrons and aims to improve the relative radial and longitudinal alignment of the DCH modules. 
The radial and longitudinal differences between the track position and the corresponding hit position at each module are recorded for a large number of tracks. The average hit-track 
residuals of each module are used to correct the radial and longitudinal position of the modules, while keeping the average correction over all modules equal to zero. This process is 
repeated several times while refitting the tracks after each iteration, until the alignment corrections converge and an accuracy of better than 50~$\mu$m for each coordinate is reached.
The method is cross-checked by using reconstructed Mott-scattered positrons (see Sect.~\ref{sec:calibration}), resulting in very similar alignment corrections.

The exact resolution reached by each approach depends on the resolution of the optical survey used as a starting position. For a low-resolution survey, the Millepede method obtains a
better resolution, while the iterative method obtains a better resolution for a high-resolution survey.
Based on these points, the Millepede method is adopted for the years 2009-2011 and the iterative method is used for the years 2012-2013 for which the novel optical survey data are
available; in 2011, the first year with the novel optical survey data, the resulting resolution of both approaches is comparable.

\subsubsection{Target alignment}
%{\it Editor's comments:\\
%Section coordinator: Gordon
%}
\label{sec:tar_align}

Precise knowledge of the position of the target foil relative to the DCH system is crucial for an accurate
determination of the muon decay vertex and positron direction at the vertex, which are calculated
by propagating the reconstructed track back to the target,
particularly when the trajectory of the track is far from the direction normal to the plane of the target.

\begin{figure}[t]
\centering
\includegraphics[width=0.5\textwidth,angle=0] {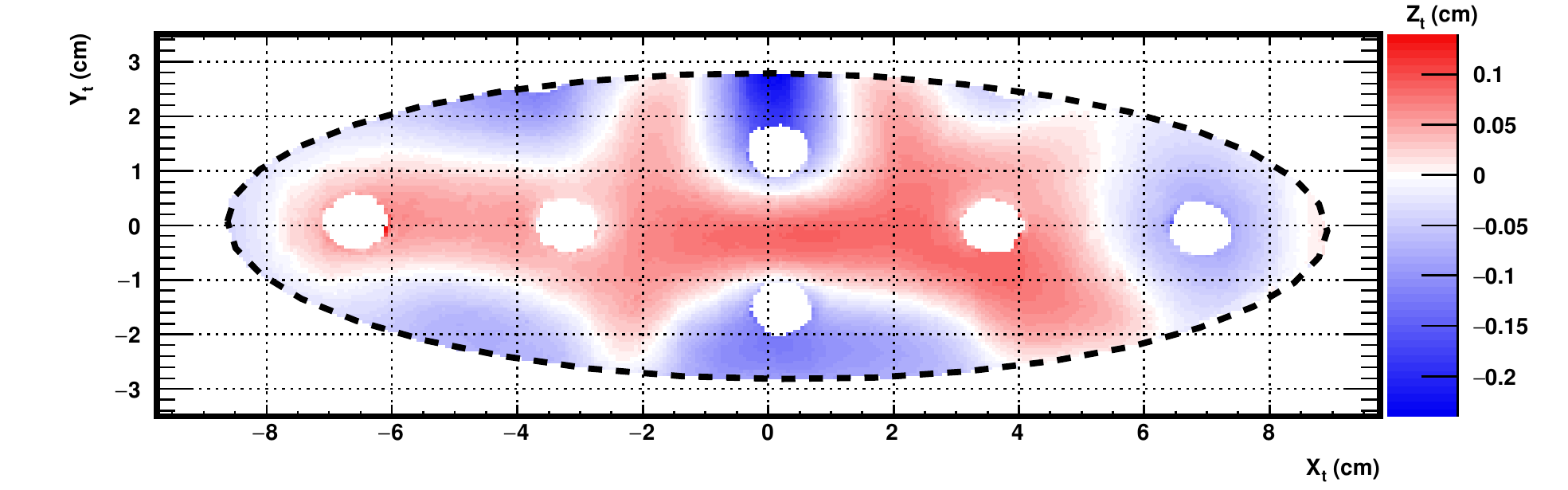}
\includegraphics[width=0.5\textwidth,angle=0] {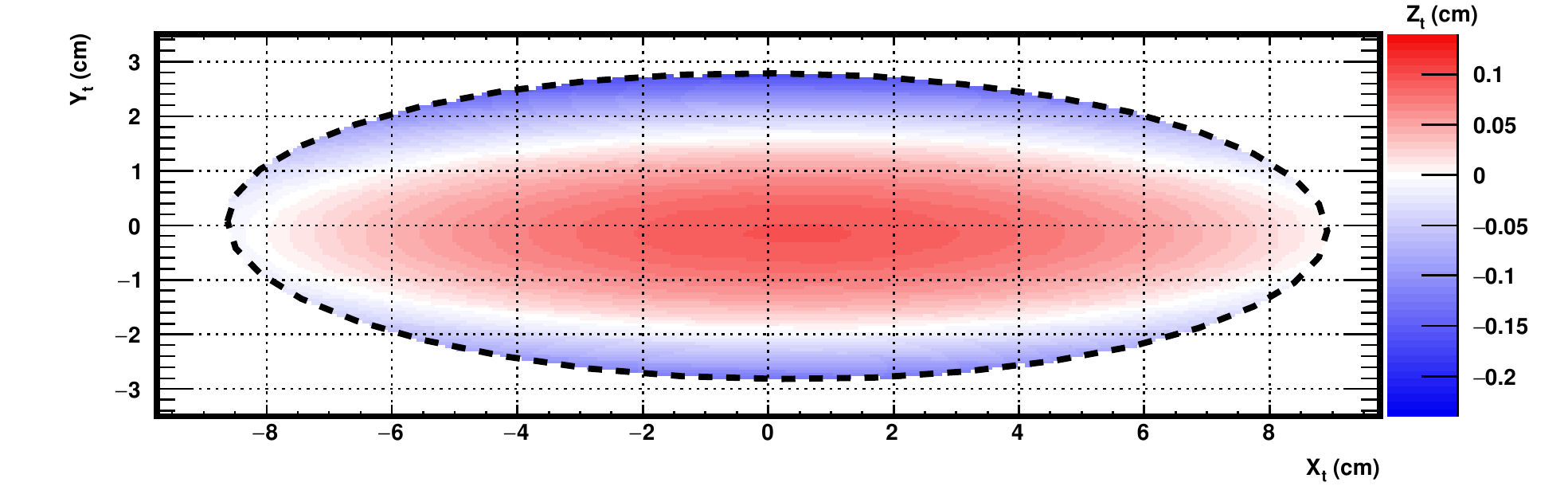}
\caption{{\it Top}: FARO scan measurements of the target aplanarity in the local target reference frame, in which $x_\mathrm{t}$ ($y_\mathrm{t}$) is the coordinate along the 
semi-major (semi-minor) axis of the target, and $z_\mathrm{t}$ as indicated by the colour axis is the coordinate perpendicular to the target plane. {\it Bottom}: 
the paraboloidal fit of the 2013 cross mark measurements. 
The paraboloidal approximation is valid since the vertices 
are concentrated at the centre of the target, as 
shown in Fig.~\ref{fig:vertices2012}.
%the paraboloidal fit of the 2013 cross mark measurements. 
}
\label{fig:faroscan}
\end{figure}

Both optical alignment techniques and software tools using physics data are used to measure and cross-check the target position.
The positions of the cross marks on the target foil (see Fig.~\ref{fig:target}) are surveyed each year using a 
theodolite, with an estimated accuracy of \mbox{$\pm$(0.5, 0.5, 1.5)~mm} in the \mbox{($x$, $y$, $z$)} directions. 
For each year, a plane fit of the cross mark measurements is used in the propagation of tracks back to the target
as a first approximation of the target foil position.
However, the residuals between the cross mark measurements and the plane fits indicate that the target foil has
developed a gradual aplanarity over time.
This is confirmed by 
measurements of the target aplanarity performed with a high-precision FARO 3D laser scanner~\cite{FARO} at the end of 2013, as shown in the top panel of 
Fig.~\ref{fig:faroscan}. 
Therefore, the propagation of tracks back to the target is improved by using a paraboloidal approximation 
\mbox{$z_\mathrm{t} - z_0$} $=$ \mbox{$c_x(x_\mathrm{t}-x_0)^2$} $+$ \mbox{$c_y(y_\mathrm{t}-y_0)^2$}
of the target foil obtained by fitting 
separately the cross mark measurements for each year.
%In this function, \mbox{$(x_\mathrm{t}, y_\mathrm{t}, z_\mathrm{t})$} is the
%local target reference frame in which 
In this function, \mbox{$(x_\mathrm{t}, y_\mathrm{t}, z_\mathrm{t})$} is the local target 
coordinate system (i.e. not the nominal MEG coordinate system) in which 
$x_\mathrm{t}$ ($y_\mathrm{t}$) is the coordinate 
along the semi-major (semi-minor) axis of the target, and 
$z_\mathrm{t}$ is the coordinate perpendicular to the target plane. 
The fit parameters are \mbox{$(x_0, y_0, z_0)$} for the position of the paraboloid 
extremum, and $c_x$ and $c_y$ for the 
paraboloid curvatures in the $x_\mathrm{t}$ and $y_\mathrm{t}$ directions. The paraboloidal fit of the 2013 cross mark measurements, shown in the bottom panel of Fig.~\ref{fig:faroscan}, exhibits
the largest aplanarity among all years. In this fit $c_y=-0.03$~${\rm cm}^{-1}$, which corresponds to a focal length of $\approx 8$~cm for the semi-minor axis of the target in 2013.

\begin{figure}[t]
\centering
\includegraphics[width=0.5\textwidth,angle=0] {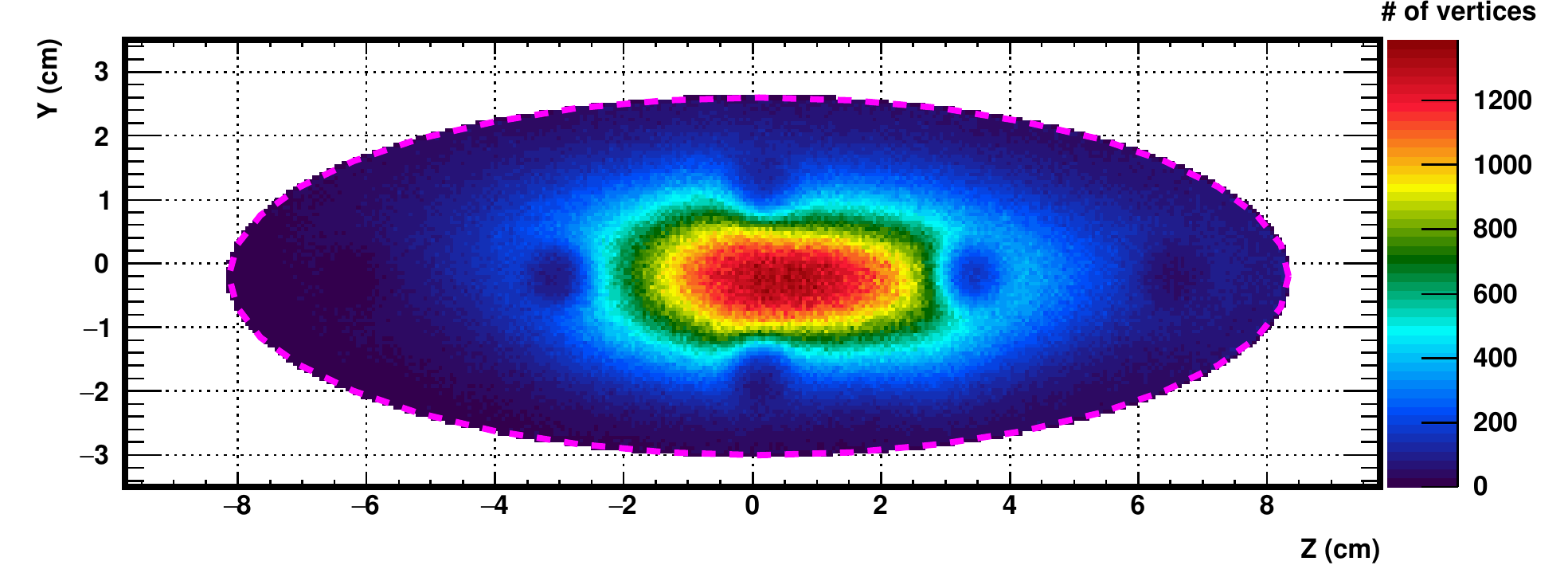}
\caption{Vertex distribution in 2012 projected on the $z-y$ plane. The four central target holes appear as dips around the beam spot in the centre of the plot.}
\label{fig:vertices2012}
\end{figure}

\begin{figure}[t]
\centering
\includegraphics[width=0.4\textwidth,angle=0] {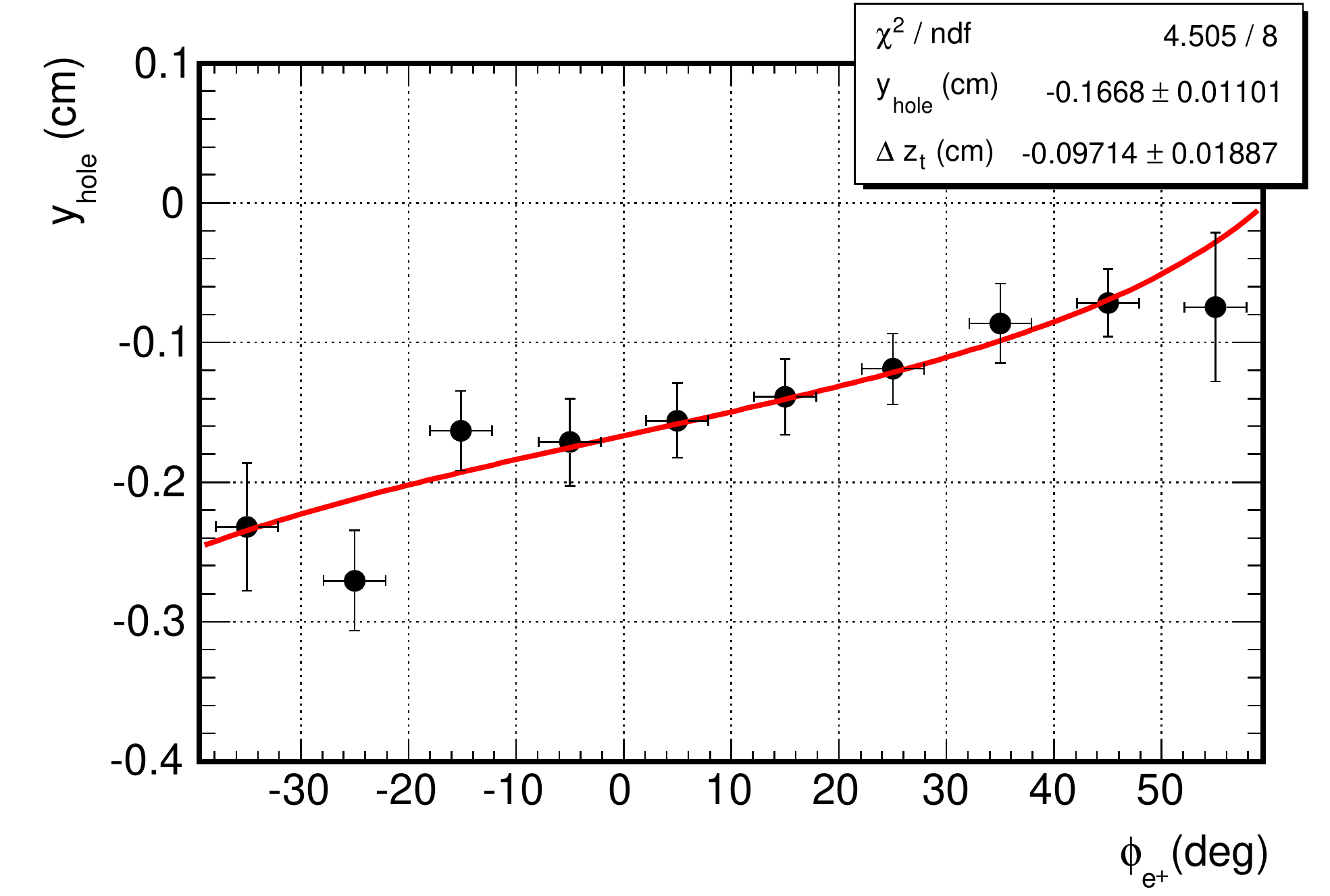}
\caption{The reconstructed $y_\mathrm{t}$ position of the left-central target hole in 2011. The fit indicates that the true hole position is shifted by 1~mm in the negative 
$z_\mathrm{t}$ direction (i.e. towards the LXe detector) with respect to its position according to the fitted optical survey.}
\label{fig:leftcentraltargethole2011}
\end{figure}

The alignment of the target foil in the $z_\mathrm{t}$ direction and the corresponding systematic uncertainty have a significant 
effect on the analysis.
In the paraboloidal approximation of the target foil, the value and uncertainty of $z_0$ are the most relevant.
The fitted $z_0$-values that are used in the track propagation are validated and
corrected by imaging
the holes in the target foil (see Fig.~\ref{fig:target}) using reconstructed
Michel positrons. The target holes appear as dips in projections
of the vertex distribution, as shown in Fig.~\ref{fig:vertices2012}.
For each year, the $z_0$-value of the paraboloidal fit is checked 
by determining the reconstructed $y_\mathrm{t}$ position of the four central target holes
as a function of the positron angle $\phie$. Ideally the target hole
positions should be independent of the track direction, while
a $z_0$-displacement with respect to the fitted value would
induce a dependence of $y_\mathrm{t}$ on
$\tan\phie$, to first order linear. Figure~\ref{fig:leftcentraltargethole2011} shows the reconstructed $y_\mathrm{t}$
position of the left-central target hole in 2011 as a function of $\phie$, fitted with a tangent function; the fit indicates a
$z_0$-displacement of 1~mm towards the LXe detector. By imaging all four central holes for each year, the systematic uncertainty 
of $z_0$ is estimated
as $\sigma^\mathrm{sys}_{z_0} \approx 0.3$~mm for 2009-2012 data and $\sigma^\mathrm{sys}_{z_0} \approx 0.5$~mm
for 2013 data.

The effect of the non-paraboloidal deformation on the analysis and its systematic uncertainty
are estimated by using a 2D map of the $z_\mathrm{t}$-difference between
the 2013 paraboloidal fit and the FARO measurements, as a function of
$x_\mathrm{t}$ and $y_\mathrm{t}$ (i.e. the difference between the top and bottom panels of Fig.~\ref{fig:faroscan}).
As discussed in detail in Sec.~\ref{sec:maximum likelihood analysis}, 
this map is scaled by a factor $k_\mathrm{t}$  for each year, to represent 
the increase of the non-paraboloidal deformation of the target
over time.

%Therefore the $z_\mathrm{t}$-coordinate of the intersection point between a track and the target foil is estimated as
%\begin{eqnarray}
%z_{t,\pos} (x_{\mathrm{t},\pos}, y_{\mathrm{t},\pos}, z_0, k_\mathrm{t}) &&= \nonumber \\
%z_0 + k_\mathrm{t} \cdot z_\mathrm{t,FARO}&&(x_{\mathrm{t},\pos}, y_{\mathrm{t},\pos}) + (1 - k_\mathrm{t}) \cdot z_\mathrm{t,par}(x_{\mathrm{t},\pos}, y_{\mathrm{t},\pos}), \nonumber
%\end{eqnarray}
%where $x_{t,\pos}$ and $y_{t,\pos}$ are the local target coordinates of the intersection of the track with the paraboloidal fit,
%while $z_\mathrm{t,FARO}$ and $z_\mathrm{t,par}$ are the functions defining the coordinates from the FARO measurement and the paraboloid, respectively.

\subsubsection{DCH performance}
%{\it Editor's comments:\\
%Section coordinator: Francesco Renga
%}
\label{sec:dch-perf}

We developed a series of methods to extract, from data, an estimate of the resolution functions, defined for 
a generic observable $q$ as the distribution of the errors, $q - q_\mathrm{true}$.

A complete overview of the performance of the spectrometer can be found in~\cite{megdet}, where the methods 
used to evaluate it are also described in detail. Two methods are used to extract the resolution functions for 
the positron parameters. The energy resolution function, including the absolute energy scale, is extracted with good 
accuracy from a fit to the energy spectrum of positrons from Michel decay. A core resolution of 
$\sigma^\mathrm{core}_{\epositron} \approx 330$~keV is found, with a $\approx 18$\% tail component with 
$\sigma^\mathrm{tail}_{\epositron} \approx 1.1$~MeV, with exact values depending on the data subset. The resolution functions 
for the positron angles and production vertex are extracted exploiting tracks that make two turns inside the 
spectrometer. The two turns are treated as independent tracks, and extrapolated to a prolongation of the target 
plane at a position between the two turns. The resulting differences in the position and direction of the two turns are then used to extract the position 
and angle resolutions. The same method is used to study the correlations among the variables and to cross-check the 
energy resolution. However, since the two-turn tracks are a biased sample with respect to the whole dataset, 
substantial MC-based corrections are necessary. 
These corrections are introduced as multiplicative
factors to the width of the resolution functions, ranging from 0.75 to 1.20.
Moreover, no information can be extracted about a possible 
reconstruction bias, which needs to be estimated from the detector alignment procedures described later in this paper.
After applying the corrections, the following average resolutions are found: $\sigma_{\thetae}=9.4$ mrad;
$\sigma_{\phie}=8.4$~mrad; $\sigma_{\ypos}= 1.1$~mm and $\sigma_{\zpos}=2.5$~mm at the target.

In order to maximise the sensitivity of the analysis for the search for $\meg$ 
(discussed in detail in Sec.~\ref{sec:maximum likelihood analysis}), instead of using 
these average resolutions we use the per-event estimate 
of the uncertainties, as provided by the track fit. It is done by replacing 
the resolution function of a generic observable $q$ with the PDF of the 
corresponding \emph{pull}: 
\[
\mathrm{pull} = \frac{q - q_{\mathrm{true}}}{\sigma^{\prime}_q}
\]
where $\sigma^{\prime}_q$ is the uncertainty provided by the track fit. 
Following a well established procedure (see for instance 
\cite{Aubert:2002rg}) this PDF is a gaussian function whose width 
$\sigma_{q}$ accounts for the bias in the determination of 
$\sigma^{\prime}_q$. The correlations between variables in the signal PDF 
are treated as correlations between pulls.
%
%The reliability of these 
%uncertainties is verified by 
%estimating the distribution of the \emph{pulls} on data :
%\[
%\mathrm{pull} = \frac{q - q_{\mathrm{true}}}{\sigma^{\prime}_q}
%\]
%where is the uncertainty provided by the track fit. In case of 
%Gaussian errors, if the uncertainty is correctly estimated and there is no bias in the reconstruction of $q$, the 
%pulls are expected to follow a Normal distribution. 
%Nevertheless, non-Gaussian effects 
%(Coulomb multiple scattering with thin layers of material at large angles, tails in DCH 
%position measurement errors, etc.) lead to expected deviations
%from the Normal distribution, which are taken into account by the \emph{pull} paramater for the observable
%$q$ ($s_{q}$) defined as the 
%standard deviation of the pull distribution. This parameter is extracted from data with the same methods used for the average resolution functions and discussed above.
%Practically, the pull 
%distribution is used instead of the error distribution. Additionally, correlations 
%among variables are also described as correlations among pulls. In this case, we 
%do not extract the per-event correlation factors from the fit matrix.
%Instead, we prefer to use analytic formulae for the correlation factors as a function of the track parameters, thanks
%to our solid understanding of per-event correlations based on geometrical models.

\subsubsection{TC reconstruction}
%{\it Editor's comments:\\
%Section coordinator: Matteo
%}
\label{sec:tc-rec}

Each of the timing counter (TC) bars acts as an independent detector. It exploits the 
fast scintillating photons released by the passage of a positron to infer
the time and longitudinal position of the hit. A fraction of the scintillating photons 
reaches the bar ends where they are read-out by PMTs.

The signal from each TC PMT is processed with a Double Threshold Discriminator (DTD) 
to extract the arrival time of the scintillating photons minimising the time walk effect. 
A TC hit is formed when both PMTs on a single bar have signals above the higher DTD threshold.
The times $t^\mathrm{TC,in}_\pos$ and $t^\mathrm{TC,out}_\pos$, measured by the two 
PMTs belonging to the same bar, are extracted by a template fit to a 
NIM waveform (square wave at level -0.8 V) fired at the lower DTD threshold and digitised by a DRS.

The hit position along the bar is derived by the following technique.
A positron impinging on a TC bar at time $t^{TC}_\pos$ has a relationship with the measured PMT times given by:
\begin{eqnarray}\label{tcinouttime}
t^\mathrm{TC,in}_\pos &=&t^\mathrm{TC}_\pos+b_\mathrm{in}+W_\mathrm{in}+\frac{\frac{L}{2}+z_\pos^\mathrm{TC}}{v_\mathrm{eff}} \nonumber \\
t^\mathrm{TC,out}_\pos &=& t^\mathrm{TC}_\pos+b_\mathrm{out}+W_\mathrm{out}+\frac{\frac{L}{2}-z_\pos^{\mathrm{TC}}}{v_\mathrm{eff}}
\end{eqnarray}
where $b_{\mathrm{in,out}}$ are the offsets and $W_\mathrm{in,out}$ are the contributions from the time walk effect from the inner and outer PMT, respectively, 
$v_\mathrm{eff}$ is the effective velocity of light in the bar and 
$L$ is the bar length; the $z$-axis points along the main axis of the bar and its origin is taken in the middle of the bar.
Adding the two parts of Eq.~\ref{tcinouttime} the result is:
\[
t^\mathrm{TC}_\pos =\frac {t^\mathrm{TC,in}_\pos+t^\mathrm{TC,out}_\pos}{2} -  \frac{b_\mathrm{in}+b_\mathrm{out}}{2} - \frac{W_\mathrm{in}+W_\mathrm{out}}{2} - \frac{L}{2v_\mathrm{eff}}.
\]

Subtracting the two parts of Eq.~\ref{tcinouttime} the longitudinal coordinate of the impact point along the bar is given by:
\[
z^\mathrm{TC}_{\pos}=\frac{v_\mathrm{eff}}{2} \left((t^\mathrm{TC,in}_\pos-t^{\mathrm{TC,out}}_\pos) - (b_\mathrm{in}-b_\mathrm{out}) - (W_\mathrm{in}-W_\mathrm{out})\right).
\]
The time (longitudinal positions) resolution of TC is determined using tracks hitting multiple bars from the distribution 
of the time (longitudinal position) difference between hits on neighbouring bars corrected for the path length.
The radial and azimuthal coordinates are taken as the corresponding coordinates of the centre of each bar.

The longitudinal position resolution is $\sigma_{z^{\mathrm{TC}}_{\pos}} \approx 1.0$ cm and
the time resolution is $\sigma_{t^{\mathrm{TC}}_{\pos}} \approx 65$ ps.

The TC, therefore, provides the information required to reconstruct all positron variables
necessary to match a DCH track (see Sect.~\ref{sec:tc-dc}) and recover the muon decay time 
by extrapolating the $t^\mathrm{TC}_\pos$ along the track trajectory back to the target to obtain 
the positron emission time $t_\pos$.

\subsubsection{DCH-TC matching}
%{\it Editor's comments:\\
%Section coordinator: Francesco
%}
\label{sec:tc-dc}

The matching of DCH tracks with hits in the TC is performed as an intermediate step in the track fit procedure, 
in order to exploit the information from the TC in the track reconstruction.

After being reconstructed within the DCH system, a track is propagated to the 
first bar volume it encounters 
(reference bar). If no bar volume is crossed, the procedure is repeated with an 
extended volume to account for extrapolation uncertainties. 
Then, for each TC hit within $\pm~5$ bars from the reference one, the track is 
propagated to the corresponding bar volume and the hit is matched with the track according 
to the following ranking :
\begin{enumerate}
\item the TC hit belongs to the reference bar, with the longitudinal distance between the 
track and the hit $\left| \Delta z^{\mathrm{TC}} \right| < 12$~cm (the track position 
defined as the entrance point of the track in the bar volume);
\item the TC hit belongs to another bar whose extended volume is also crossed by the track, and 
$\left| \Delta z^{\mathrm{TC}} \right| < 12$~cm (the track position defined as the entrance 
point of the track in the extended bar volume);
\item the TC hit belongs to a bar whose extended volume is not crossed by the 
track, but where the distance of closest approach of the track to the bar axis 
is less than 5~cm, and $\left| \Delta z^{\mathrm{TC}} \right| < 12$~cm (the track position 
defined as the point of closest approach of the track to the bar axis).
\end{enumerate}
Among all successful matching candidates, those with the lowest ranking are chosen. Among them,
the one with the smallest $\Delta z^{\mathrm{TC}}$ is used.

The time of the matched TC hit is assigned to the track, which is then back-propagated to the chambers in order 
to correct the drift time of the hits for the track length timing contribution. 
The Kalman filter procedure is also applied to propagate the track
back to the target to get the best estimate
of the decay vertex parameters at the target, including the time $\tpositron$.

\subsubsection{Positron AIF reconstruction}
%{\it Editor's comments:\\
%Section coordinator: Gordon
%}
\label{sec:aif_rec}

The photon background in an energy region very close to the signal is dominated by positron AIF in the detector (see Sect.~\ref{sec:singlegammabackground}). If the positron crosses part of the DCH before it annihilates, it can leave a trace of hits which are correlated to the subsequent photon signal. A pattern recognition algorithm has been developed that can identify these types of positron AIF events. Since positron AIF contributes to the accidental background, this algorithm can help to distinguish accidental background events from signal and RMD events. The algorithm is summarised in the following.

\begin{figure}[t]
\centering
\includegraphics[width=0.5\textwidth,angle=0] {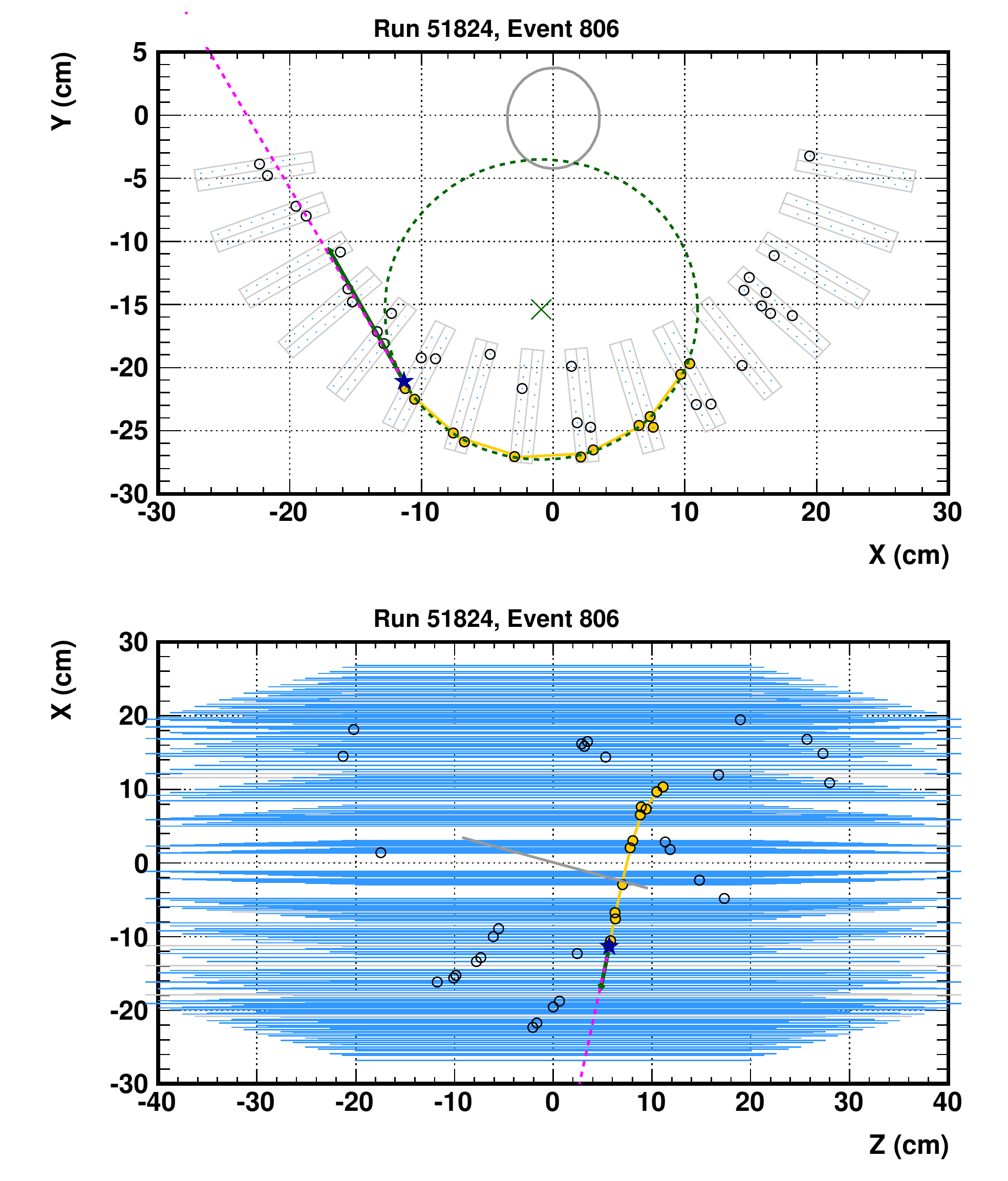}
\caption{Example of a reconstructed positron AIF candidate in a 2009 event due 
to a downstream muon decay. The reconstructed AIF vertex is indicated as a blue star, 
visible in the upper plot at $\left( x,y \right) = \left(-12,-21 \right)$ approximately. 
The AIF direction is indicated as a green arrow, originating from this star and 
pointing towards lower $x$ and higher $y$ coordinates. The vector connecting the 
AIF vertex and the photon conversion vertex in the LXe detector is indicated as 
a magenta dashed line. Note that the green arrow and the magenta line nearly 
overlap, as expected for a true AIF event.}
\label{fig:AIFcandidate}
\end{figure}

The procedure starts by building positron AIF seeds from all reconstructed clusters. An AIF seed is defined as a set of 
clusters on adjacent DCH modules which satisfy a number of minimum proximity criteria. A positron AIF candidate ($\posAIF$) is reconstructed 
from each seed by performing a circle fit based on the $xy$-coordinates of all clusters in the seed. The circle fit is 
improved by considering the individual hits in all clusters. The $xy$-coordinates of hits in multi-hit clusters are refined 
and left/right solutions based on the initial circle fit are determined by taking into account the timing information of 
the individual hits, which also results in an estimate of the AIF time. The $xy$-coordinates of the AIF vertex 
are determined by the intersection point of the circle fit with the first DCH cathode plane after the last cluster hit. 
If the circle fit does not cross the next DCH cathode plane, the intersection point of the circle fit with the support 
structure of the next DCH module or the inner wall of COBRA is used. The $z$-coordinate of the AIF vertex is calculated 
by extrapolating the quadratic polynomial fit of the $xz$-positions of the last three clusters of the AIF candidate to the $x$-coordinate of the AIF vertex. 
The AIF candidate direction is taken as the direction tangent to both the circle fit and the quadratic polynomial fit at the AIF vertex. Fig.~\ref{fig:AIFcandidate} shows an example of a 
reconstructed AIF candidate.

\subsection{Combined reconstruction}

This section deals with variables requiring signals both in the spectrometer and in the LXe detector.

\subsubsection{Relative photon--positron angles}
%{\it Editor's comments: \\
%Section coordinator: Ryu, Yusuke \\
%Text:  0.5\\
%Figure: 1.
%} 
\label{sec:relative_angle}

Since the LXe detector is not capable of reconstructing the direction of the incoming photons, 
this direction is determined by connecting the reconstructed interaction vertex of the 
photon in the LXe detector to the reconstructed decay vertex on target: it is defined through
its azimuthal and polar angles ($\phigamma$, $\thetagamma$).

The degree to which the photon and positron are not back-to-back is quantified in terms 
of the angle between the photon direction and the positron direction reversed at the target 
in terms of azimuthal and polar angle differences:
\begin{eqnarray}
\thetaegamma &=& (\pi - \thetae) - \thetagamma, \nonumber \\
\phiegamma &=& (\pi + \phie) - \phigamma \nonumber.
\end{eqnarray}
There are no direct calibration source for measuring the resolutions of the measurements of these relative angles.
Hence, they are obtained by combining 1) the position resolution of the LXe detector and 2) the position and angular resolutions of the spectrometer,
taking into account the relative alignment of the spectrometer and the calorimeter.

There are correlations among the errors in measurements of the positron observables at the target both due to the fit and also
introduced by the extrapolation
to the target. Additionally, the errors in the photon angles contain a contribution from the positron position error at the target.
Due to the correlations, the relative angle resolutions are not the quadratic sum of the 
photon and positron angular resolutions.

The $\thetaegamma$ resolution is evaluated as $\sigma_{\thetaegamma} = (15.0 - 16.2)$~mrad depending on the
year of data taking 
by taking into account the correlation between $z_e$ and $\thetae$.
Since the true positron momentum and $\thetaegamma$ of the $\meg$ signal are known,
$\phie$ and $\ypos$ can be corrected using the reconstructed energy of the
positron and $\thetaegamma$. 
The $\phiegamma$ resolution after correcting these
correlations is evaluated as $\sigma_{\phiegamma}= (8.9-9.0)$~mrad depending on the year.

The systematic uncertainty of the positron emission angle relies on the accuracy of the relative alignment among the magnetic field, the DCH modules, and the target (see Sect.~\ref{sec:dch_align} and \ref{sec:tar_align} for the alignment methods and the uncertainties).
The position of the target (in particular the error in the position and orientation of the target plane)  and any distortion of the target plane directly affect the emission angle measurement and are found to be one of the dominant sources of systematic uncertainty on the relative angles. 
%For instance, an uncertainty of $0.5$~mm in $x$ corresponds to \mbox{$\approx 4$~mrad} in contribution to the uncertainty in $\phiegamma$.

The measurement of the photon direction depends on the relative alignment between the spectrometer and the LXe detector.
They are aligned separately using optical alignment techniques and calculations of LXe detector distortions and motion during filling and cool-down. Additionally,
the relative alignment is cross-checked by directly measuring two types of physics processes that produce hits in both detectors:
\begin{itemize}
\item Positron AIF events,
\item Cosmic rays without the COBRA magnetic field.
\end{itemize}
Each of the two measurements provides independent information of the displacement with a precision 
better than 1.0~mm in the longitudinal direction ($\delta z$) relative to the survey; however they 
are subject to systematic uncertainties due to the non-uniform distribution of both positron AIF 
and cosmic ray events and to the different shower development because those events are not coming 
from the target. The two results in $\delta z$ ($\delta z = 2.1 \pm 0.5$~mm for AIF and $\delta z =1.8 \pm 0.9$~mm 
for cosmic-rays) agree well, resulting in an average $\delta z = 2.0\pm0.4$~mm. 
On the other hand, the survey data may have some unknown systematic uncertainties 
because the survey can be done only at room temperature and the effects of shrinkage 
and the detector distortions at LXe temperature are only taken into account by the calculation. 
The difference of $\delta z=2.0$~mm between the survey and the two measurements suggests the 
existence of these possible uncertainties and we take the difference into account as the 
systematic uncertainties. The nominal displacement is taken as the average of the survey 
and the average of the measurements, 
$\delta z = 1.0 \pm 0.6$ mm where the uncertainty is the systematic. 
This uncertainty corresponds to 0.85~mrad at the centre value of $\thetaegamma$ (converted by the radial position of the LXe detector $r_\mathrm{in} = 67.85$~cm).
There is no cosmic ray measurement available in other degrees of freedom and we can not extract the systematic uncertainty of $\phiegamma$.   
Therefore, we regard the observed value of $\delta z$ 
as an estimate of the systematic uncertainty of the survey while keeping all other survey results for the alignment.
Finally, we assign the same systematic uncertainty estimated for $\thetaegamma$ to $\phiegamma$. %, namely $\delta\phiegamma = \delta\thetaegamma$.

\subsubsection{Relative photon--positron time}
%{\it Editor's comments: \\
%Section coordinator: Cecilia Voena \\
%Text:  0.5\\
%Figure: 1.
%}
\label{sec:teg_time}

\begin{figure}[t]
\centering
\includegraphics[width=0.5\textwidth,angle=0] {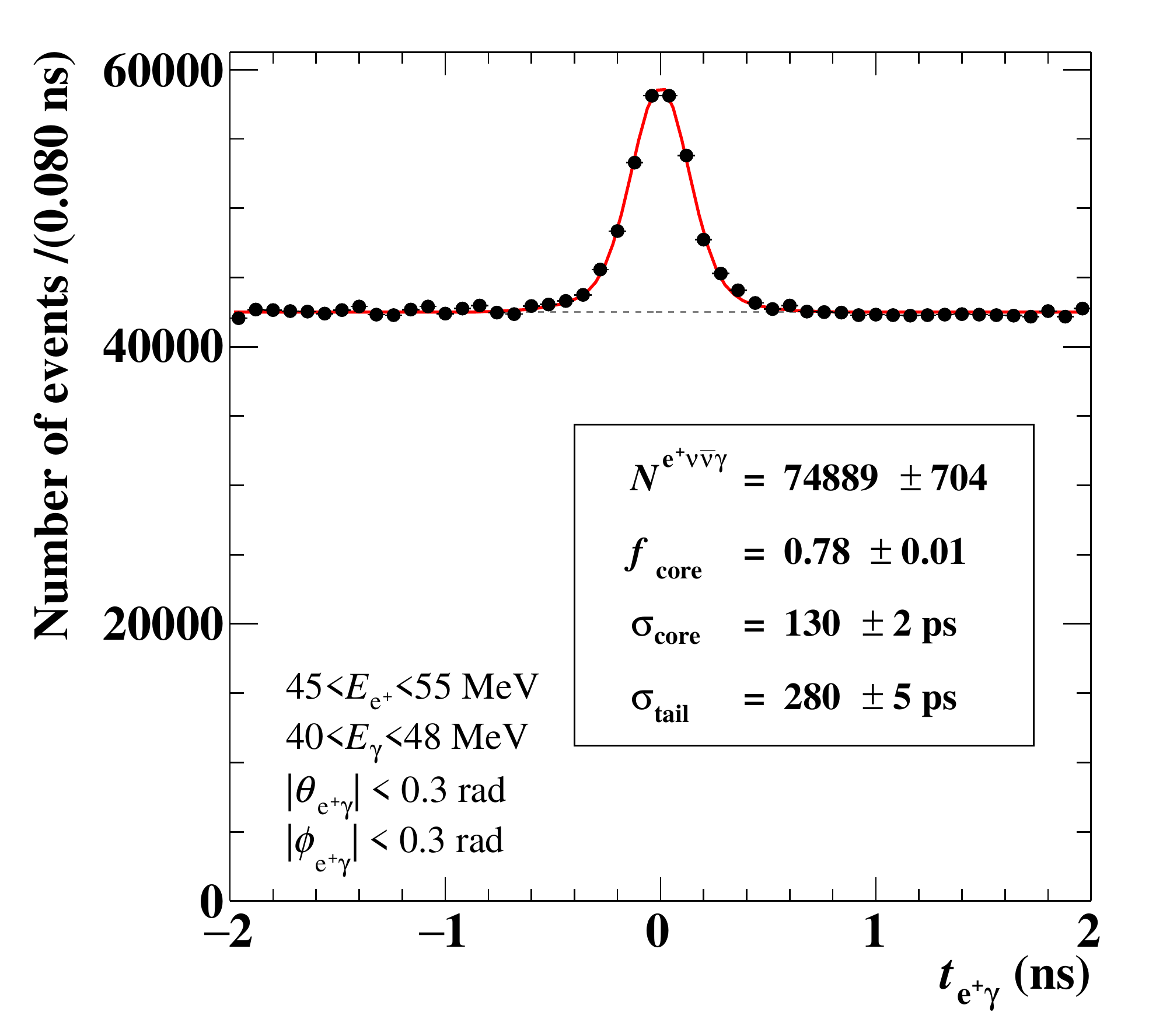}
\caption{Distribution of $\tegamma$ for MEG standard trigger. The peak is from RMD, the flat component is from accidental 
coincidences.}
\label{fig:teg}
\end{figure}

The relative time $\tegamma = \tgamma-\tpositron$ is defined as the difference between the 
photon time (see Sect.~\ref{sec:tgamma}) and the positron time (see Sect.~\ref{sec:tc-rec}) calculated at the target.
The relative time is calibrated using the RMD peak
observed in the energy side-band\footnote{Side-bands are defined in Sect.~\ref{sec:blinding}} and shown in Fig.~\ref{fig:teg}.
The centre of this distribution is used to correct the time offset
between the TC and LXe detectors.
The position of the RMD-peak corresponding to $\tegamma = 0$ is
monitored constantly during the physics data-taking period and found
to be stable to within $15$~ps.
In order to obtain the resolution on $\tegamma$ for signal events,
the resolution of Fig.~\ref{fig:teg} must be corrected for the photon
energy dependence as measured in the CEX calibration run and for the positron energy
dependence (from a MC simulation), resulting in $\sigma_{\tegamma}=122\pm 4$~ps.

The dominant contributions to the  $\tegamma$ resolution are
the positron track length uncertainty (in timing units 75~ps), the 
TC intrinsic time resolution (65~ps), and the LXe detector time 
resolution (64~ps).

\subsubsection{Photon-AIF analysis}
%{\it Editor's comments: \\
%Section coordinator: Gordon \\
%Text:  0.5\\
%Figure: 1.
%}
\label{sec:aif_gamma}

In order to determine if a photon originates from positron AIF, the following three quantities are calculated for each possible $\posAIF$$\gamma$-pair 
from all reconstructed $\posAIF$ candidates and photons in the event: the angular differences between the AIF candidate direction 
and the vector connecting the photon and the AIF vertex ($\theta_{\rm AIF}$ and $\phi_{\rm AIF}$), and the time difference between the photon and 
the AIF candidate ($t_{\rm AIF}$). If there are multiple $\posAIF$ candidates per event, a ranking of $\posAIF$$\gamma$-pairs is performed by minimising 
the $\chi^2$ based on these three observables. 

\begin{figure}[t]
\centering
\includegraphics[width=0.45\textwidth,angle=0] {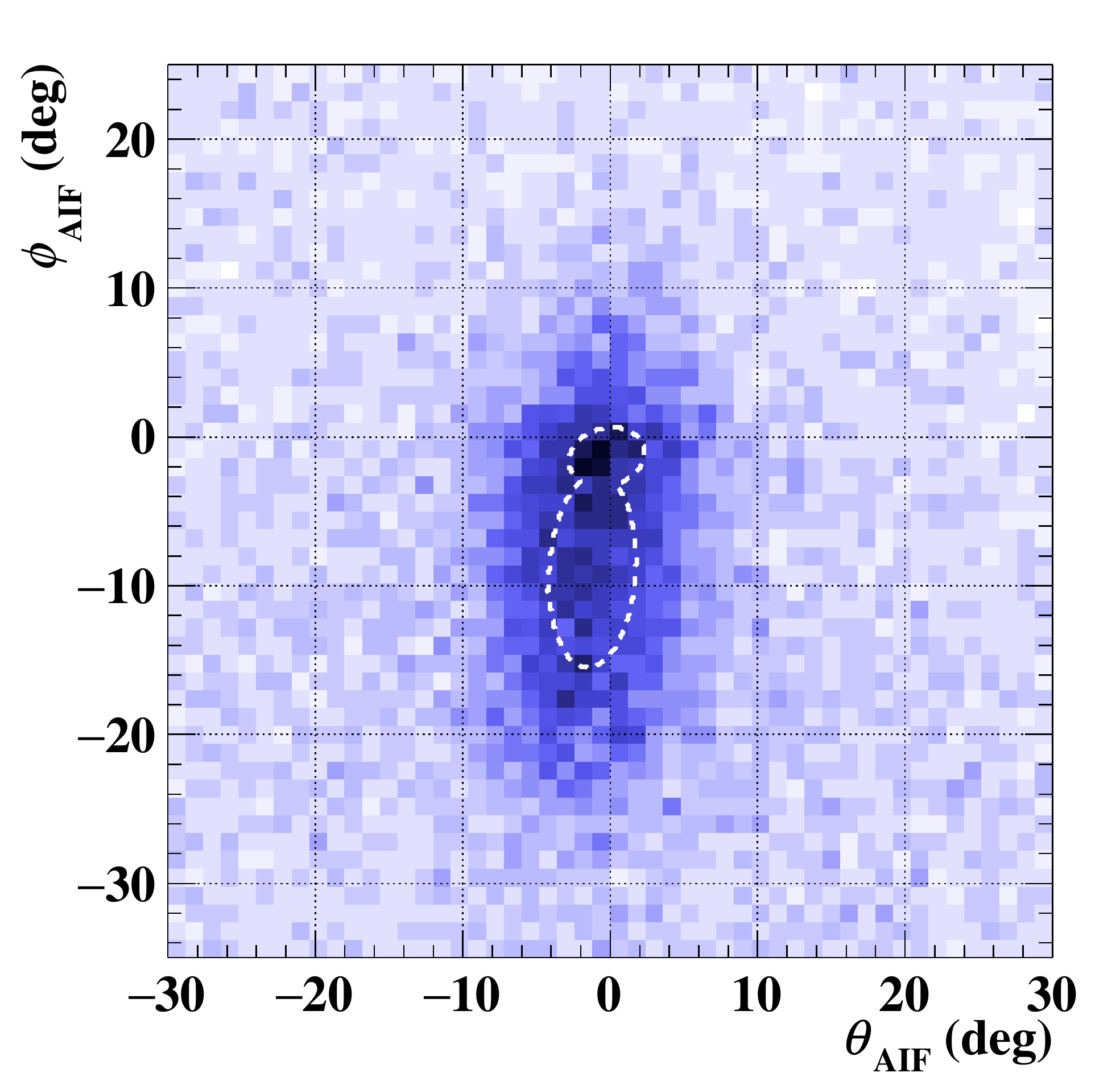}
\caption{The $\phi_{\rm AIF}$ vs. $\theta_{\rm AIF}$ distribution of the highest ranked $\posAIF$$\gamma$-pairs per event in a sample of year 2009 events. 
The peak in the centre of the plot is caused by photons originating from positron-AIF in the DCH. The events located inside the dashed line contour are removed by the AIF cut.}
\label{fig:deltathetaAIF_vs_deltaphiAIF}
\end{figure}

A plot of $\phi_{\rm AIF}$ vs. $\theta_{\rm AIF}$ for the highest ranked $\posAIF$$\gamma$-pairs 
per event in a random sample of year 2011 events is shown in Fig.~\ref{fig:deltathetaAIF_vs_deltaphiAIF}. The peak at the centre is caused by photons 
originating from positron AIF in the DCH. The peak has a tail in the negative $\phi_{\rm AIF}$ direction since the AIF vertex is reconstructed at the first DCH 
cathode foil immediately after the last hit in the $\posAIF$ candidate. However, if the last hit is located in the left plane of a DCH module, it is equally 
likely (to first order) that the AIF occurred in the first cathode foil of the next DCH module.

The observables $\theta_{\rm AIF}$ and $\phi_{\rm AIF}$ are combined into a 1D \lq\lq{}distance\rq\rq{} from the peak where the correlation between $\theta_{\rm AIF}$ and 
$\phi_{\rm AIF}$ and the structure of the two peaks are taken into account. The smaller the distance, the more likely the event is a true AIF background.
Events falling within 0.7$\sigma_\mathrm{AIF}$ of either of the two peaks are cut.
The estimation for the fraction of rejected AIF background events is 1.9\%, while losing 1.1\% of signal events. The fraction of rejected RMD events is the same as that of the signal.
The cut based on the AIF analysis is employed in the physics analysis to remove outlier events which happen to be signal-like with an AIF photon.

%% file: analysisresults.tex
\section{Analysis}
\label{sec:analysis}
%{Section coordinator: Fabrizio, Wataru \\
%Text:  8.\\
%Figure: 14.
%}
\subsection{Analysis strategy}
\label{sec:Analysis strategy}
%{\it Editor's comments: \\
%Section coordinator: Wataru\\
%}
The MEG analysis strategy is a combination of blind and maximum likelihood
analysis. The blind analysis is chosen to prevent any bias in the evaluation
of the expected background in the analysis region and the maximum likelihood
analysis is preferred to the simpler box analysis in order to avoid boundary
effects at the borders of the analysis region and to improve the sensitivity 
by correctly taking into account the probabilities of events
being due to signal, RMD or accidental background.

The ${\meg}$ event is characterised by an $\egammapair$-pair,
simultaneously emitted 
with equal momentum magnitude and opposite directions, and with
energy of $m_{\mu}/2 = 52.83~{\rm MeV}$ each. 
The ${\meg}$ event signature is therefore very simple and the sensitivity 
of the experiment is limited by the ability to reject 
background $\egammapair$-pairs, of various origins. 
Positron and photon energies 
($\epositron$ and $\egamma$), $\egammapair$ relative time 
($\tegamma$), and relative azimuthal and polar angles $\thetaegamma$ and $\phiegamma$
are the observables available to distinguish possible ${\meg}$ candidates from background pairs.
In the maximum likelihood analysis presented here, $\thetaegamma$ and $\phiegamma$ are treated 
separately, with independent distributions, since these variables can
have different experimental resolutions. 

This maximum likelihood analysis is thoroughly cross-checked by an alternative independent maximum likelihood analysis 
where some of the methods are simplified; for example, 
the relative stereo angle $\Thetaegamma$ is used
instead of the relative polar and azimuthal angles. 

\subsection{Dataset}
%{\it Editor's comments: \\
%Section coordinator: Toshiyuki\\
%Text:  0.5\\
%Figure: 0.
%}

\begin{figure}[!ht]
\centering
	\includegraphics[width=1\columnwidth]{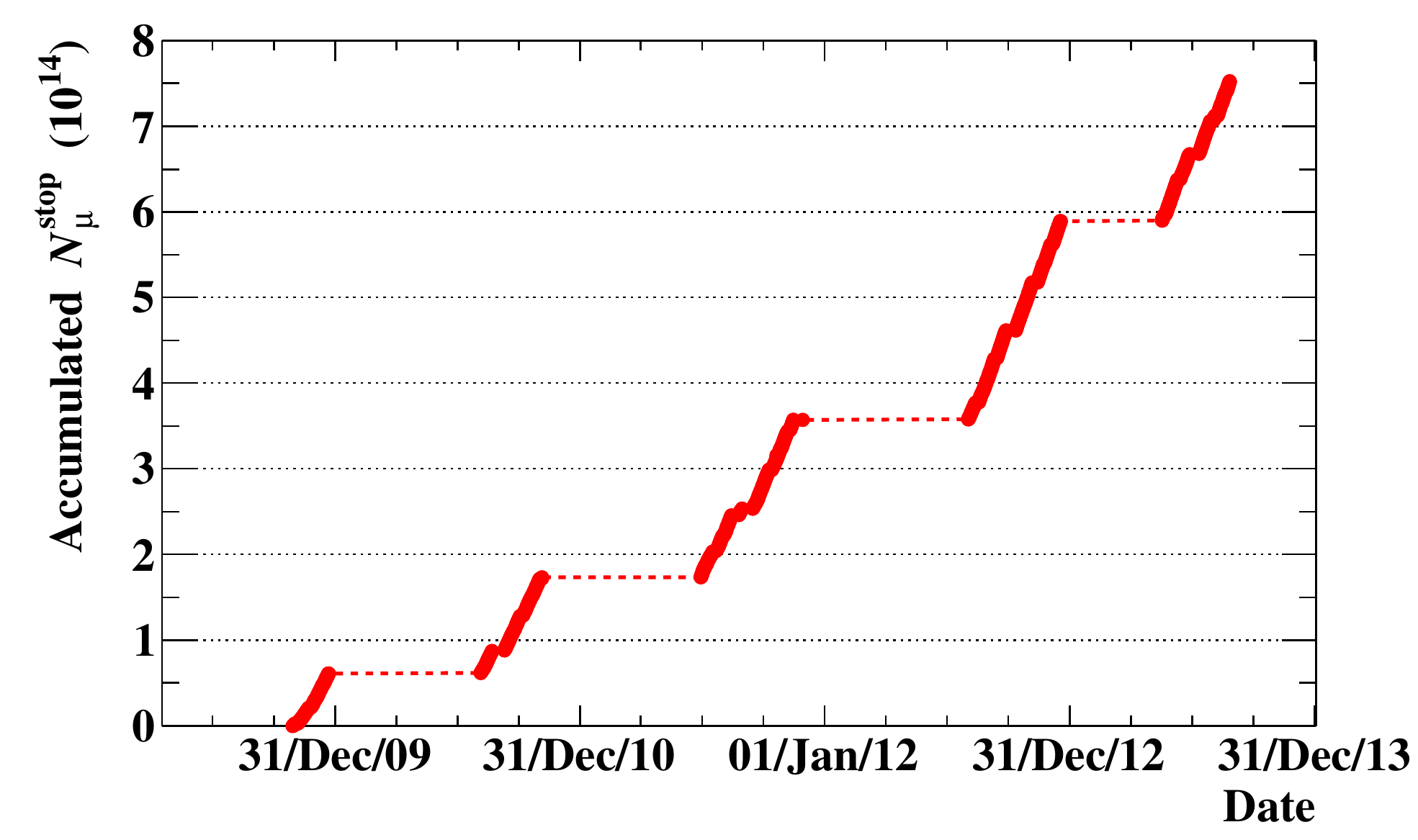}
\caption{The accumulated number of stopped muons on target as function of time.}
\label{fig:numberofmuons}
\end{figure}

Data were accumulated intermittently in the years 2008--2013. 
Figure~\ref{fig:numberofmuons} shows the
data collection period divided into each calendar year by the planned PSI winter accelerator shutdown 
periods of 4--5 months. 
Shutdown periods are used for detector maintenance, modification and repair work. 
The data accumulated in 2008 were presented in \cite{meg2009},
but the quality of those data was degraded by problems with the tracking system and therefore 
they are not considered in this analysis.

In total, $7.5 \times 10^{14}$ muons were stopped on target in 2009--2013.
The analysis based on the $3.6\times10^{14}$ muons stopped on target in 2009--2011
has already been published~\cite{meg2013}. The data from the remaining $2.3\times10^{14}$ muons stopped
on target in 2012, and from $1.6\times10^{14}$ muons stopped on target in 2013 are included in this analysis, thus completing the full dataset.
%\subsubsection{Preselection}
%%

In the first stage of the MEG analysis, events are pre-selected with loose requirements, 
requiring the presence of (at least) one positron track candidate and
%$\left| \tegamma \right| < 4~{\rm ns}$, a relative timing window
%more than $25$ times larger than the experimental resolution. 
a time match given by $-6.9 < t_{\rm LXe-TC} < 4.4 ~{\rm ns}$, where $t_{\rm LXe-TC}$ is the relative difference 
between the LXe time and the TC time associated with the positron candidate.
The window is asymmetric to include multiple turn events.
This procedure reduces our data size to $\approx 16$\% of the recorded events. 
No requirements are made on photon and positron energies or relative 
directions. Such loose cuts ensure that even in the presence of not yet optimised 
calibration constants the possibility of losing a good ${\meg}$ event is negligible. 

\subsection{Blinding}
\label{sec:blinding}
%{\it Editor's comments: \\
%Section coordinator: Fabrizio, Wataru\\
%Text:  0.5\\
%Figure: 1.
%}

\begin{figure}[htb]
\centering
  \includegraphics[width=0.5\textwidth,angle=0] {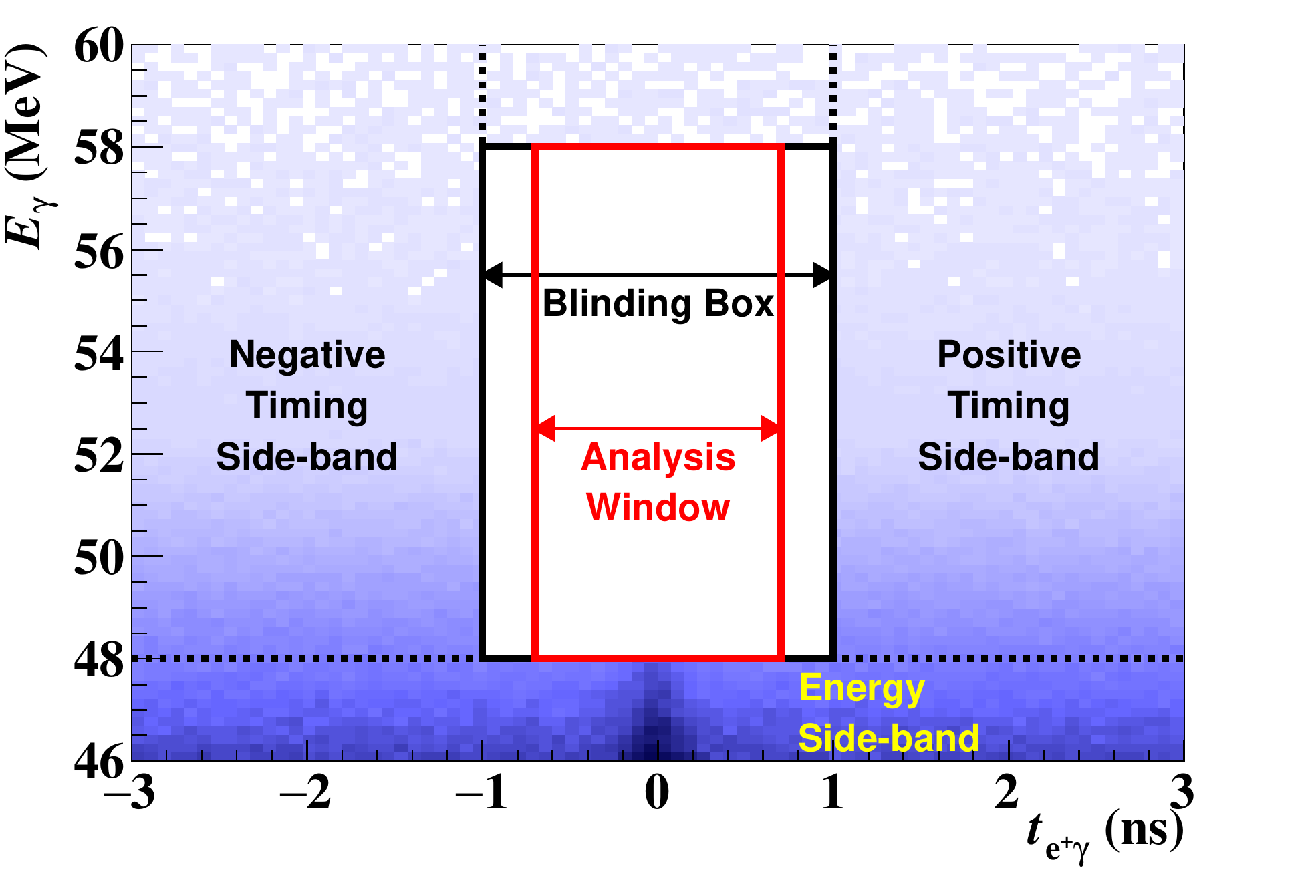}
 \caption{
The MEG blinding box and a possible definition of side-bands.
}
 \label{bb}
\end{figure}

Every time the pre-selected events are processed, events falling in the window in
the $(\tegamma, \egamma)$
plane defined by $| \tegamma | < 1~{\rm ns}$ and
$48.0 < \egamma < 58.0~{\rm MeV}$ (\lq\lq Blinding Box\rq\rq) are hidden 
and written to a data stream not accessible by the collaboration.
The MEG blinding box is shown in Fig.~\ref{bb}.

For purposes of various studies, a number of side-band regions were defined. Events with $| \tegamma | > 1~{\rm ns}$
fall in the \lq\lq timing side-bands\rq\rq, the left side-band corresponding to
$\tegamma < -1~{\rm ns}$ and the right side-band to
$\tegamma > 1~{\rm ns}$, while events with arbitrary
relative timing and with $\egamma < 48.0~{\rm MeV}$ fall into the
\lq\lq energy side-band\rq\rq. 
Different photon energy windows are used for different timing side-band studies.
For example, events with $48.0 < \egamma < 58.0~{\rm MeV}$ are used
when the timing side-band data are compared with the data in the analysis window, 
and events with $ \egamma > 40.0~{\rm MeV}$ are used for the single photon background study.
RMD events, with zero relative timing, belong to the
energy side-band and, as stated in Sect.~\ref{sec:teg_time}, are used to 
accurately calibrate the timing difference between LXe detector and TC. Events in 
the timing side-bands are very likely to be accidental events; hence, their positron and 
photon energy spectra and relative angle distributions are uncorrelated. 
We also define \lq\lq angle side-bands\rq\rq~the regions corresponding to
$50 < | \thetaegamma | < 150~{\rm mrad}$
or $75 < | \phiegamma | < 225~{\rm mrad}$,
which are used for self-consistency checks of the analysis procedure.

Side-band events are studied in detail to optimise the algorithms and
analysis quality, to estimate the background in the analysis window, and to
evaluate the experimental sensitivity by using toy MC simulations.
%, as will be explained later
At the end of the optimisation procedure, the events in the blinding box are analysed and
a maximum likelihood fit is performed to extract the number of signal ($\nsig$), RMD
($\nrd$) and accidental background ($\nacc$) events. The likelihood fit is performed on
events falling in the \lq\lq Analysis Window\rq\rq~defined
by $48.0 < \egamma < 58.0~{\rm MeV}$, $50.0 < \epositron < 56.0~{\rm MeV}$,
$| \tegamma | < 0.7~{\rm ns}$,
$| \thetaegamma | < 50~{\rm mrad}$ and
$| \phiegamma | < 75~{\rm mrad}$. 
The projection of the analysis window in the $(\tegamma, \egamma)$ plane is also shown in Fig.~\ref{bb}.
The size of the analysis window is chosen 
%on the basis of experimental resolutions, 
to be between five
and twenty times the experimental resolutions of all observables in order to prevent any risk of losing good
events and to restrict the number of events to be fitted at a reasonable level.
The same fitting procedure is preliminarily applied to equal size regions in the
timing and angle side-bands (with appropriate shifts on relative timings or angles)
to verify the consistency of the calculation.
\subsection{ Background study }
%{\it Editor's comments: \\
%Section coordinator:  Toshiyuki, Yusuke\\
%Text:  1.5\\
%Figure: 3.
%}

The background in the search for the $\megsign$ decay comes either 
from RMD or from an accidental overlap between a Michel positron 
and a photon from RMD or AIF.
All types of background are thoroughly studied 
in the side-bands prior to analysing events in the analysis window.

\subsubsection{Accidental background}
%{\it Editor's comments: \\
%Section coordinator:  Wataru
%}
\label{sec:accbackg}

The accidental overlap between a positron with energy close to the kinematic edge 
of the Michel decay and an energetic photon from RMD or positron AIF is the 
leading source of the background.

 \paragraph{Single photon background}
\label{sec:singlegammabackground}
\hspace{0cm} \\
High energy single photon background events are mainly produced by two 
processes: RMD and AIF of positrons. 
The contribution from external Brems\-strahlung is negligibly small 
in our analysis window. RMD is the Michel decay with the emission of a 
photon, also called inner Bremsstrahlung. The integrated fraction of 
the spectrum of photons from RMD is roughly 
proportional to the square of the integration window size near the signal energy, 
which is usually determined by the energy resolution \cite{fronsdal_1959_pr,eckstein_1959_ap}. AIF photon background 
events are produced when a positron from Michel decay annihilates with an 
electron in the material along the positron trajectory into two photons 
and the most energetic photon enters the LXe detector. The emission direction 
of the most energetic photon is closely aligned to that of the original positron and the cross section is peaked with one photon carrying most of the energy.
The total number of AIF background events depends on the layout and the material 
budget of the detector along the positron trajectory.

\begin{figure}[tbp]
\centering
\includegraphics[width=18pc]{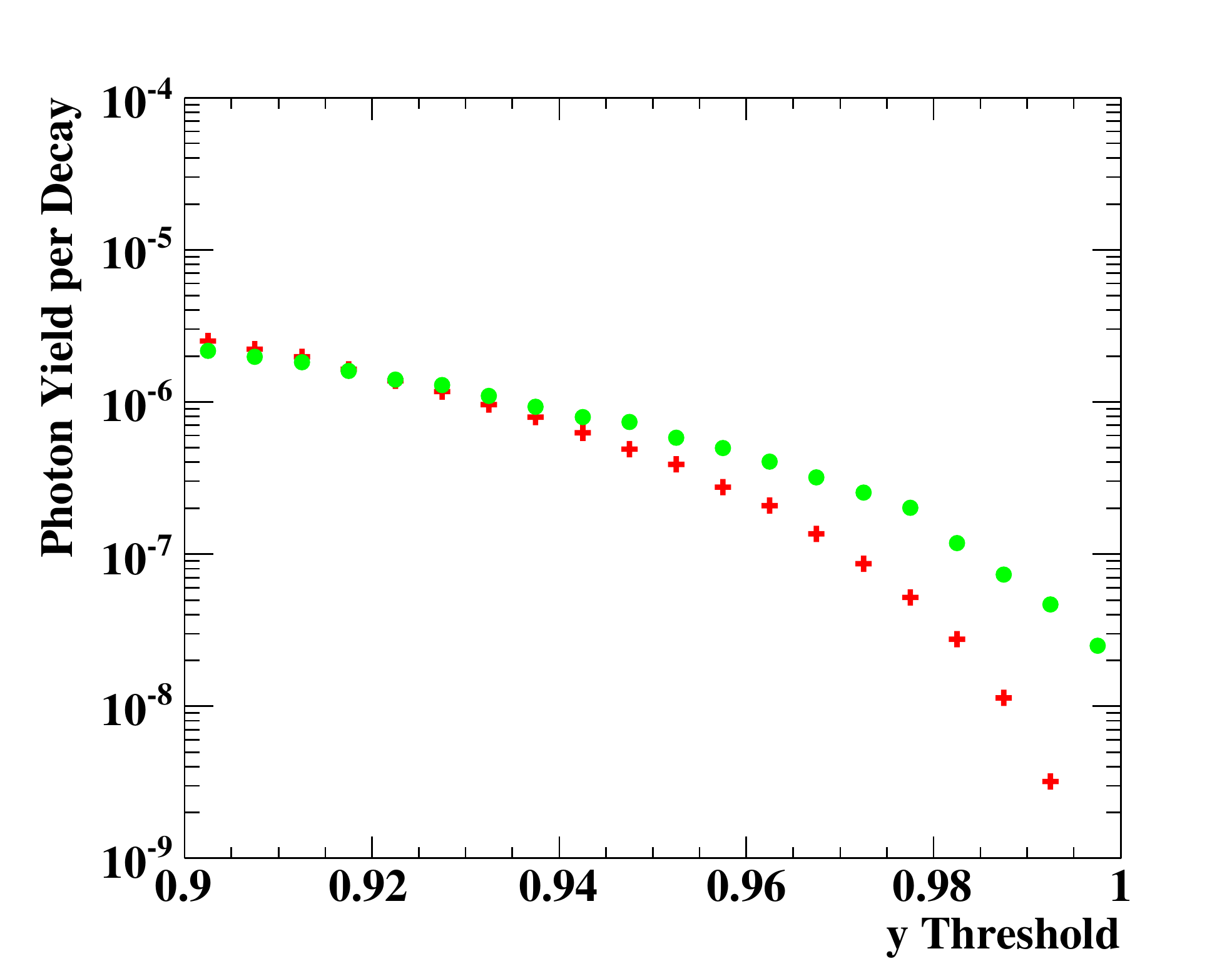}
\caption{\label{fig:gamma_single_bg}The RMD (red crosses) and AIF (green circles) 
photon background spectra in the MEG detector estimated 
by a MC simulation around the kinematic end-point. The variable on the horizontal axis is $y=2\egamma/m_{\mu}$ where $\egamma$ is the 
photon energy and $m_\mu$ is the muon mass.}
\end{figure}

Figure~\ref{fig:gamma_single_bg} shows the single photon background spectra 
calculated from a MC simulation of the MEG detector as a function of the 
normalized photon energy $y = 2 E_{\gamma}/m_{\mu}$. The green circles show 
the AIF photon background spectrum and the red crosses show that due to RMD.
The integrated photon yield 
per decay above $y$ is plotted on the vertical 
axis (the maximum allowed value for 
$y$ is slightly smaller than one for RMD and slightly larger than one for AIF, 
due to the electron mass). The RMD photon fraction is 55\%, and the 
AIF photon fraction is 45\% in the $y>0.9$ region. From Fig.~\ref{fig:gamma_single_bg}, 
AIF becomes dominant in the $y>0.92$ region. 
Since the energy spectra decrease 
rapidly as a function of $y$ near the kinematic end-point, 
a good energy resolution 
reduces steeply the single photon background.

In addition to the RMD and AIF components in the analysis window, there are contributions from pile-up photons 
and cosmic-ray components, totalling at most 4--6\%. The pile-up rejection methods are 
discussed in Sect.~\ref{sec:energy_gamma}. The cosmic-ray events are rejected by using 
topological cuts based on the deposited charge ratio of the inner to outer face and 
the reconstructed depth ($w$) because these events mostly come from the outer face 
of the LXe detector while signal events are expected from the inner face. After applying 
these cuts, photon background spectra are measured directly from the timing side-band 
data, and the measured shape is used in the analysis window.

 \paragraph{Single positron background}
\hspace{0cm} \\
The single positron background in the analysis window results from the Michel decay positrons.
Although the theoretical positron energy spectrum of the Michel decay is well known \cite{kinoshita_1959}, 
the measured positron spectrum is severely distorted by the design of the spectrometer 
which tracks only high momentum positrons, 
and therefore introduces a strong momentum dependence in the tracking efficiency. 
The resolution in the momentum reconstruction also influences the measured spectrum.
The positron spectrum obtained by our detector with the resolution function and the acceptance curve are shown in ~\cite{megdet}. 
There is a plateau region near the signal energy where the measurement rate of the positrons reaches its maximum,
which allows us to extract the shape of the positron background precisely from the data with high statistics.

\begin{figure}[tbp]
\centering
\includegraphics[width=20pc]{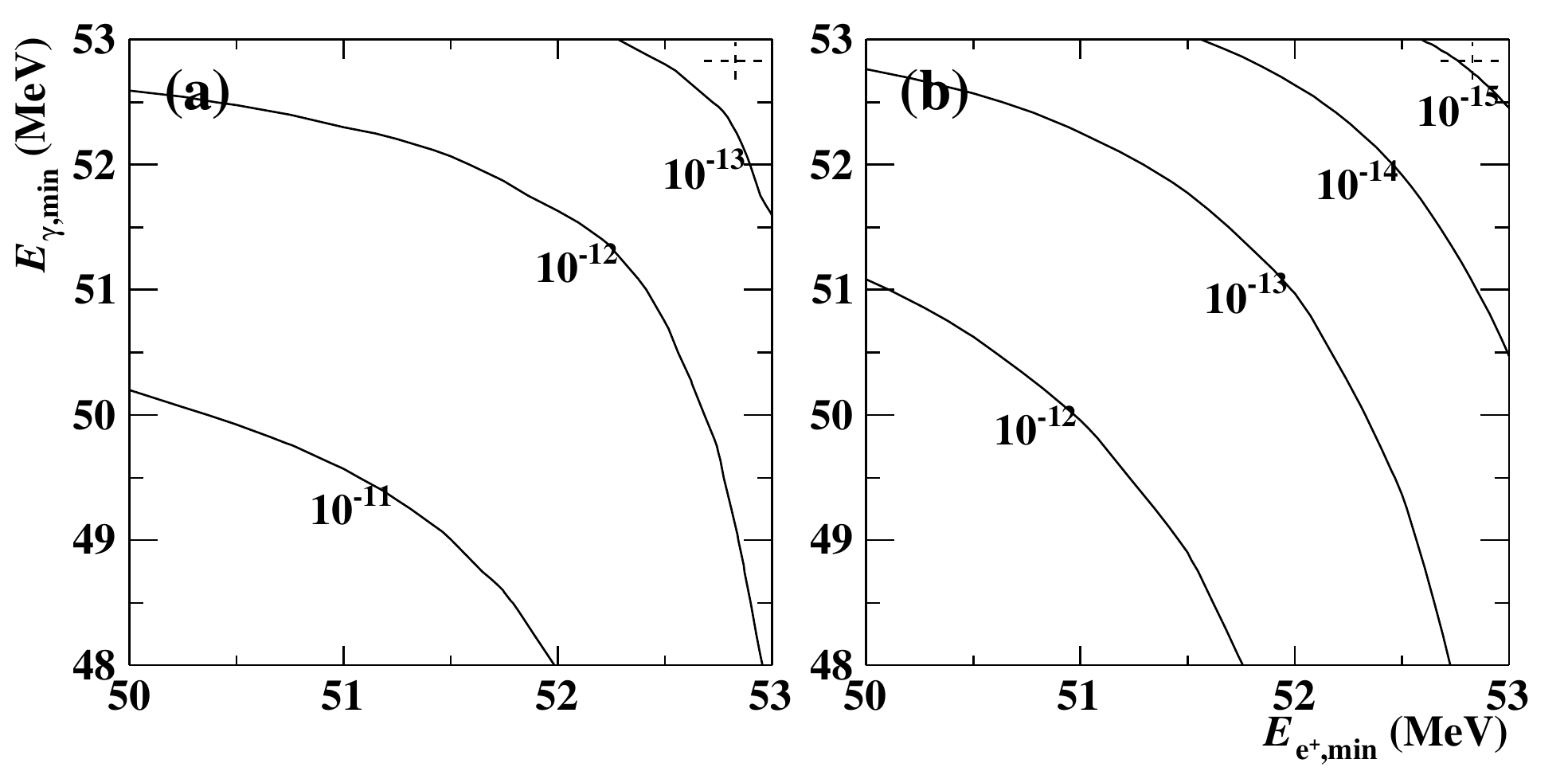}
\caption{\label{fig:bg_contour}Effective branching ratios of the two types of background 
into the kinematic window defined by $E_\mathrm{e^+,min}<\epositron<53.5$~MeV, 
$E_{\gamma,\mathrm{min}}<\egamma<53.5$~MeV, $|\tegamma|<0.24$~ns and $\cosThetaegamma<-0.9996$. 
(a) Accidental background evaluated from the timing side-band. (b) RMD background 
from $\radiative$ calculated with theoretical formula folded with detector responses.}
\end{figure}

 \paragraph{Effective branching ratio}
\hspace{0cm} \\
The effective branching ratio of the accidental background, defined by the 
background rate normalised to the muon stopping rate,
can be approximately expressed by \cite{kuno_2001}
\[
B_\mathrm{acc} \propto R_\mu \, \delta \epositron \, (\delta \egamma)^2 \, \delta \tegamma \, \delta \thetaegamma \, \delta \phiegamma,
\]
where $R_\mu$ is the muon stopping rate and $\delta q$ 
is the width of the integration region 
defined by the detector resolution for the observable $q$.
Figure~\ref{fig:bg_contour} (a) shows the effective branching ratio 
for the accidental background as a function of the lower edges of the integration 
regions of $\epositron$ and $\egamma$.
The same plot for the RMD background is shown 
in Fig.~\ref{fig:bg_contour} (b), which is described in detail 
in the Sect.~\ref{sec:rmdbackground}. It can be seen that the accidental 
background is much more severe than the RMD background.

The rate of the accidental background expected in the analysis window is evaluated
using the data from a wider time window in the side-bands with larger statistics. 
The background rate measured in the side-bands is used as a statistical constraint
in the maximum likelihood analysis.
The distributions of the observables relevant for the physics analysis are also precisely 
measured in the timing side-bands and used 
%as the accidental background PDFs 
in the maximum likelihood analysis (Sect.~\ref{sec:maximum likelihood analysis}).

\subsubsection{RMD background}
%{\it Editor's comments: \\
%Section coordinator:  Yusuke\\
%}
\label{sec:rmdbackground}
A second background source consists of the $\radiative$ RMD process, producing 
a time-coincident $\egammapair$-pair. The RMD events fall into 
the analysis window when the two neutrinos have small momentum and are 
identical to the signal in the limit of neutrino energies equal to zero.
Observation of the RMD events provides a strong internal 
consistency check for the $\meg$ analysis since it is a source of time-coincident $\egammapair$-pairs.
 
\begin{figure}[tbp]
\centering
\includegraphics[width=20pc]{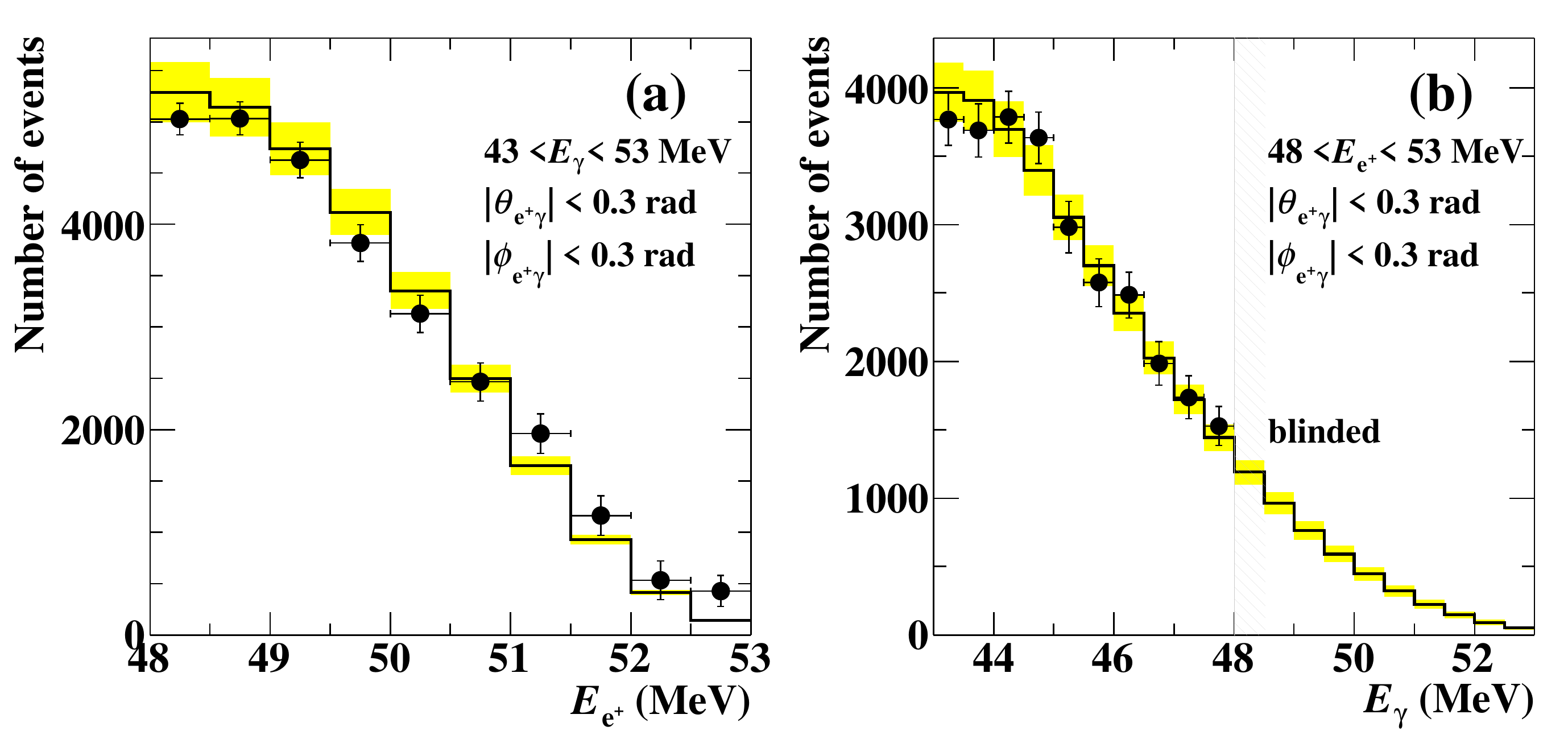}
\caption{\label{fig:rmd_dist}Projected distributions of $\radiative$ events measured 
in the energy side-band (dots with error bars) compared with the expectations 
(histograms with the uncertainty specified by the yellow bands). 
The expectations are calculated with the theoretical formula folded 
with the detector responses and a normalisation based on Michel events.
}
\end{figure}

The RMD in the energy side-band defined by
$43.0 < \egamma < 48.0$~MeV, $48.0<\epositron<53.0$~MeV, 
$|\phiegamma| < 0.3$~rad, 
and $|\thetaegamma| < 0.3$~rad are studied.
The RMD events are identified by a peak 
in the $\tegamma$ distribution as shown in Fig.~\ref{fig:teg}.
The distribution of RMD in terms of energy and angle is measured 
by fitting the $\tegamma$-distribution divided into energy 
and angle bins. Figure~\ref{fig:rmd_dist} shows the measured projected distributions.
The rates and shapes are compared with the Standard Model calculation 
(in lowest order) \cite{kuno_2001} and found to be consistent.
The measured branching ratio within the energy side-band agrees with 
the expectation to within 5\%.

The estimated number of RMD events in the $\meg$ analysis window is calculated 
by extrapolating the energy side-band distribution to the analysis window, 
giving an estimate of $\langle \nrd \rangle = 614 \pm 34$, which is used as 
a statistical constraint in the likelihood analysis.

The RMD branching ratio is highly suppressed when the integration region is 
close to the limit of $\meg$ kinematics. The effective branching ratio, 
which is calculated by considering the detector resolution, is plotted in 
Fig.~\ref{fig:bg_contour}~(b) as a function of the lower edges of the integration regions
on $\epositron$ and $\egamma$. For example, the effective branching ratio 
for $52.0 < \egamma < 53.5$~MeV and $52.0<\epositron<53.5$~MeV is $3\times 10^{-14}$, 
more than twenty times lower than that due to the accidental background.

\subsection{ Maximum likelihood analysis}
%{\it Editor's comments: \\
%Section coordinator: Fabrizio, Wataru, Ryu \\
%Text:  2.\\
%Figure: 4.
%}
\label{sec:maximum likelihood analysis}

\subsubsection{Likelihood function}
\label{sec:likelihood}
%{\it Editor's comments: \\
%Section coordinator:  Wataru
%}

The numbers of signal, RMD and accidental background events 
 in the analysis window, ($\nsig$, $\nrd$, $\nacc$),
are determined by a maximum likelihood analysis. 
In addition, two target parameters $\vector{t}$ for each year, representing
the position ($z_0$) and deformation ($k_\mathrm{t}$) of the muon stopping target
are also included 
as fitting parameters in the likelihood function (see Sect.~\ref{sec:tar_align}).
Of particular interest is $\nsig$, 
   while the other parameters ($\nrd$, $\nacc$, $\vector{t}$) are treated 
   as nuisance parameters which are profiled 
   in the calculation of the confidence intervals, as discussed 
   in Sect.~\ref{sec:feldman-cousins}.
The extended likelihood function is thus defined as
\begin{eqnarray}
\label{eq:likelihood}
\lefteqn{{\cal L}\left(\nsig, \nrd, \nacc, \vector{t}\right) = } \nonumber \\
&&\frac {e^{- N}}{\nobs!} C(\nrd, \nacc, \vector{t})\times \nonumber  \\
&& \prod_{i=1}^{\nobs} \left( {\nsig} S(\vector{x_i}, \vector{t})
      + {\nrd} R(\vector{x_i})
+ {\nacc} A(\vector{x_i}) \right), \label{eq:Likelihood_Definition}
\end{eqnarray}
where $\vector{x_i} = \{\egamma, \epositron, \tegamma,
\thetaegamma, \phiegamma \}$ is the vector of observables for the $i$-th event.

$S$, $R$ and $A$ are
the probability density functions (PDFs) for the signal, RMD and accidental background events, respectively.
$N = \nsig + \nrd + \nacc$ is the total number of events in the fit
and $\nobs$
is the total number of detected events in the
analysis window. $C$ is a term for the constraints of nuisance parameters.

The expected numbers of
RMD and accidental background events 
with their respective uncertainties are evaluated in the side-bands and are 
     applied as Gaussian constraints on $\nrd$ and $\nacc$
     in the $C$ term in Eq.~\ref{eq:likelihood}.

The target position parameters $z_0$ are subject to
Gaussian constraints whose
widths are the year dependent systematic
uncertainties; 
the target deformation parameters $k_\mathrm{t}$ are constrained with uniform distributions in year dependent
intervals. 
%in which the maximum allowed values are 0.1, 0.1, 0.4, 0.5 and 1.0, for 
% 2009-2013 data, respectively.

\subsubsection{PDFs}
%{\it Editor's comments: \\
%Section coordinator:  Ryu
%}

\label{sec:pdfs}
\paragraph{Event-by-event PDFs}
\hspace{0cm} \\

The PDFs for signal, RMD and accidental background events are formed as a function of
the five observables ($\egamma$, $\epositron$, $\tegamma$, $\thetaegamma$, $\phiegamma$) 
taking into account the correlations between them and the
dependence of each of them and of their uncertainties on the photon interaction vertex,
the muon decay vertex and the track reconstruction quality.

Because the detector resolutions depend on the detector conditions and the hit position in the
detector, this approach uses different PDFs for each event (event-by-event PDFs).
The energy response, the position resolution and the background spectrum of the LXe detector 
are evaluated as function of the interaction vertex.
For the positron PDF, the fitting errors of the tracking variables are used to
compute the resolutions; namely the resolution on the observable $q$ ($\sigma_q$) is replaced 
by a product of the
pull parameter ($s_q$) and the fitting error ($\sigma_q^{\prime}$). The pull parameters 
are extracted from the data as described in Sect.~\ref{sec:dch-perf} and
are common to all events in a given DAQ period.
The correlations between observables are also treated on an event-by-event basis.
For example, the errors on the momentum and the angle are correlated 
because the emission angle of positrons is computed by
extrapolating the fitted tracks to the target plane. 
Since the true positron momentum of the signal is known, the
mean of the signal angle PDF can be corrected as a function of the observed momentum.

Because the energies, relative timing and angles for the signal are fixed and known, 
the signal PDFs are described by the product of the detector 
response function for each observable.
The correlations between the errors of the observables are 
implemented in the $\tegamma$, $\thetaegamma$ and $\phiegamma$ PDFs 
by shifting the centres and modifying the
resolutions.
The possible reconstruction bias due to errors on the target position and deformation is
included in the signal PDF by shifting the centre of the $\phiegamma$ PDF by an amount
computed from $\vector{t}$.
The amount of the shift is computed geometrically by shifting the target by
\mbox{$\delta z_0 + k_\mathrm{t} \cdot (z_\mathrm{t,FARO}(x_{\pos}, y_{\pos}) - z_{\mathrm{t},2013}(x_{\pos}, y_{\pos}))$} in the $z_\mathrm{t}$
direction, where $\delta z_0$, $z_\mathrm{t,FARO}$ and $z_{\mathrm{t},2013}$ are the
deviation of $z_0$ from the nominal value and the coordinates defined by the FARO measurements and the 2013 paraboloid fit, respectively (see Fig.~\ref{fig:faroscan}).
For the $\tegamma$ PDF, events are categorised by using $\qualitye$, which consists of
the track-fitting quality and the matching quality between the fitted 
track and the hit position on the TC. The resolution and the 
central value are extracted for each category from the observed RMD timing peak.
The dependence on $\egamma$ and $\epositron$ is taken into account.
Most of the parameters used to describe the correlations are extracted from data 
by using the double-turn method (see Sect.~\ref{sec:dch-perf}), while a few 
parameters (for instance, the slope parameter for the
$\delta_{\tegamma}$--$\delta_{\epositron}$ correlation, where $\delta_x$ is
the difference between the observed and the true value of the observable $x$) 
are extracted from a MC simulation.

The RMD PDF is formed by the convolution of the detector response and the kinematic distribution 
in the parameter space, ($\egamma$, $\epositron$, $\thetaegamma$, $\phiegamma$), 
expected from the Standard Model \cite{kuno_2001}. The correlations between 
the variables are included in the kinematic model. The PDF for $\tegamma$ 
is almost the same as that of the signal PDF, while the correlation between
$\delta_{\tegamma}$ and $\epositron$ is excluded.

The accidental background PDFs are extracted from the timing side-band data.
For $\epositron$, the spectrum, after applying the same event selection on the
track reconstruction quality as for the physics analysis, is fitted with a function formed by
the convolution of the Michel positron spectrum and a parameterised function describing the detector response.
For $\egamma$, the energy spectra after applying the pile-up and cosmic-ray cuts and a loose
selection on the $\egammapair$ relative angle, are fitted with a function to represent
background photon, remaining cosmic-ray and the pile-up components convoluted 
with the detector response.
The $\thetaegamma$ and $\phiegamma$ PDFs are represented by polynomial functions 
fitted to the data after applying the same event selection except for the $\tegamma$.
For $\tegamma$, a flat PDF is used.

\paragraph{Constant PDFs}
\hspace{0cm} \\

The event-by-event PDFs employ the entire information we have about detector 
responses and kinematic variable correlations. A slightly less 
sensitive analysis, based on an alternative set of PDFs, is used as a cross 
check; this approach was already implemented in \cite{meg2013}. 

In this alternative set of PDFs the events are characterised by 
\lq\lq categories\rq\rq, mainly determined by the tracking quality of positrons 
and by the reconstructed depth of the interaction vertex in the LXe detector 
for photons. A constant group of PDFs is determined year by year, one for each 
of the categories mentioned above; the relative stereo angle $\Thetaegamma$ 
is treated as an observable instead of $\thetaegamma$ and $\phiegamma$ separately, 
while the three other kinematic variables ($\epositron$, $\egamma$ 
and $\tegamma$) are common to the two sets of PDFs. Correlations 
between kinematic variables are also taken into account with a simpler 
approach and the systematic uncertainties associated with the target position are 
included by shifting $\Thetaegamma$ of each event by an appropriate 
amount, computed by a combination of the corresponding shifts of $\thetaegamma$ 
and $\phiegamma$. Signal and RMD PDFs are modelled as in the event-by-event analysis by using 
calibration data and theoretical distributions, folded with detector response. 
This likelihood function is analogous to Eq.~\ref{eq:likelihood} with the inclusion 
of the Gaussian constraints on the expected number of RMD and accidental background events 
and of the Poissonian constraint on the expected total number of events.  
In what follows we refer to this set of PDFs as \lq\lq constant PDFs\rq\rq~and 
to the analysis based on it as \lq\lq constant PDFs' analysis\rq\rq. 
\subsubsection{Confidence interval}\label{sec:feldman-cousins}
%{\it Editor's comments: \\
%Section coordinator:  Ryu
%}

The confidence interval of $\nsig$ is calculated following the Feldman-Cousins approach
\cite{feldman_1998} with the profile-likelihood ratio ordering \cite{PDBook_2014}.
The test statistic $\lambda_p$ for sorting experiments is defined by
% Any better way for hat and double hat ?
\begin{eqnarray}\label{eq:test_statistic}
\lambda_{p}(\nsig) &=&
\left\{
\begin{array}{lll}
\frac{{\mathcal L}(\nsig      ,  \hat{\hat{\vector{\theta}}}(\nsig))}
     {{\mathcal L}(0          ,  \hat{\hat{\vector{\theta}}}(0    ))}
 & {\rm if } & \hat{N}_{\rm sig} <   0\\
\frac{{\mathcal L}(\nsig      ,  \hat{\hat{\vector{\theta}}}(\nsig))}
     {{\mathcal L}(\hat{N}_{\rm sig},       \hat{\vector{\theta}}        )}
 & {\rm if } & \hat{N}_{\rm sig} \ge 0,
\end{array} \right.\nonumber 
\end{eqnarray}
where ${\vector{\theta}}$ is a vector of nuisance parameters 
($\nacc$, $\nrd$ and $\vector{t}$), $\hat{N}_{\rm sig}$ and $\hat{\vector{\theta}}$
are the values of
$\nsig$ and $\vector{\theta}$ which maximise the likelihood, $\hat{\hat{\vector{\theta}}}(\nsig)$
is the value of $\vector{\theta}$ which maximises the likelihood for the specified $\nsig$.
The confidence interval is calculated using the
distribution of the likelihood ratio for an ensemble of 
pseudo experiments simulated based on the PDFs.

The following systematic uncertainties are included in the calculation of the 
confidence interval:
the normalisation (defined in Sect.~\ref{sec:normalisation}),
the alignment of the photon and the positron detectors,
the alignment (position and deformation) of the muon stopping target,
the photon energy scale,
the positron energy bias,
the centre of the signal $\tegamma$ PDF,
    the shapes of the signal and background PDFs, and
the correlations between the errors of the positron observables.
The dominant systematic uncertainty is due to the target alignment
as described in Sect.~\ref{sec:sensitivity}, 
which is included in the maximum likelihood fit by profiling the target parameters. 
The other uncertainties are included by
randomising them in the generation of the pseudo experiments 
used to construct the distribution of the likelihood ratio.

\subsection{Normalisation}
\label{sec:normalisation}
%{\it Editor's comments: \\
%Section coordinator:  Yusuke, Luca \\
%Text:  1.\\
%Figure: 1.
%} \\
The branching ratio as a function of the number of signal events ($\nsig$) is 
expressed by

\[%begin{eqnarray}
\mathcal{B}(\meg) \equiv \frac{\Gamma(\meg)}{\Gamma_{\mathrm{total}}}
				     = \frac{\nsig}{N_{\mu} },
\]%end{eqnarray}
where the normalisation factor $N_{\mu}$ is the number of muon decays effectively 
measured during the experiment.

Two independent methods are used to calculate $N_{\mu}$.
Since both methods use control samples measured simultaneously 
with a signal, they are independent of the instantaneous beam rate.

%%%%%%%%%%%%%%%%%%%%%%%%

\subsubsection{Michel positron counting}
%{\it Editor's comments: \\
%Section coordinator:  Daisuke\\
%}

The number of high momentum Michel positrons is counted using a pre-scaled %Michel positron %LU
TC based trigger enabled during the physics data taking. %and using the same analysis cuts.
Because $\mathcal{B}(\michel)\approx 1$, $N_\mu$ is calculated as follows:
\[ %begin{eqnarray}
	N_{\mu} =
	\frac{{N^{\mathrm{e}\nu\bar{\nu}}}}
		 {f^{\mathrm{e}\nu\bar{\nu}}_{\epositron}} \times
	\frac{P^{\mathrm{e}\nu\bar{\nu}}}{\epsilon^{\mathrm{e}\nu\bar{\nu}}_{\mathrm{trg}}} \times
	\frac{\epsilon^{\mathrm{e}\gamma}_{\mathrm{e}}}
		{\epsilon^{\mathrm{e}\nu\bar{\nu}}_{\mathrm{e}}} \times 
	A^{\mathrm{e}\gamma}_{\gamma} \times
	\epsilon^{\mathrm{e}\gamma}_{\gamma} \times
	\epsilon^{\mathrm{e}\gamma}_{\mathrm{trg}} \times
	\epsilon^{\mathrm{e}\gamma}_{\mathrm{sel}},
	 \label{eq:Norm_Michel}
\]%end{eqnarray}
where
$N^{\mathrm{e}\nu\bar{\nu}} = 245\,860$ is the number of Michel positrons 
detected with $50.0<E_\mathrm{e} < 56.0$~MeV; $f^{\mathrm{e}\nu\bar{\nu}}_{\epositron} 
= 0.101 \pm 0.001$ is the fraction of the Michel spectrum for this energy range 
(the uncertainty coming from the systematic uncertainty on the $\epositron$ bias);
$P^{\mathrm{e}\nu\bar{\nu}} = 10^7$ is the pre-scaling factor of the Michel 
positron trigger, which requires a correction factor 
$\epsilon^{\mathrm{e}\nu\bar{\nu}}_{\mathrm{trg}} = 0.894 \pm0.009$ %, which depends on the beam rate, 
to account for the dead-time of the trigger scaler due to pile-up in the TC;
$\epsilon^{\mathrm{e}\gamma}_{\mathrm{e}}/\epsilon^{\mathrm{e}\nu\bar{\nu}}_{\mathrm{e}}$ 
is the ratio of signal-to-Michel efficiency for detection of positrons in this energy range;
$A^{\mathrm{e}\gamma}_{\gamma}=0.985\pm0.005$ is the geometrical acceptance for 
signal photon given an accepted signal positron;
$\epsilon^{\mathrm{e}\gamma}_{\gamma}$ is the efficiency for detection and reconstruction 
of 52.83 MeV photons; $\epsilon^{\mathrm{e}\gamma}_{\mathrm{trg}}$ is the trigger 
efficiency for signal events; and $\epsilon^{\mathrm{e}\gamma}_{\mathrm{sel}}$ is the 
$\egammapair$-pair selection efficiency for signal events given a 
reconstructed positron and a photon.

The absolute values of the positron acceptance and efficiency
cancel in the ratio 
$\epsilon^{\mathrm{e}\gamma}_{\mathrm{e}}/\epsilon^{\mathrm{e}\nu\bar{\nu}}_{\mathrm{e}}$. 
Momentum dependent effects are derived from the Michel spectrum fit, resulting in 
$\epsilon^{\mathrm{e}\gamma}_{\mathrm{e}}/\epsilon^{\mathrm{e}\nu\bar{\nu}}_{\mathrm{e}}
= 1.149 \pm 0.017$.

The photon efficiency is evaluated via a MC simulation taking into account 
the observed event distribution. The average value is $\epsilon^{\mathrm{e}\gamma}_{\gamma}=0.647$. 
The main contribution to the photon inefficiency is from conversions before %%%LU
the LXe detector active volume: 14\% loss in the COBRA magnet, 7\% in the cryostat and PMTs, 
and 7\% in other materials. Another loss is due to shower escape from the inner face, resulting in a 6\% loss.
The photon efficiency is also measured in the CEX run. By tagging an 83-MeV photon from a
$\pi^0$ decay, the efficiency for detection of 55-MeV photons is measured to be 0.64--0.67, 
consistent with the evaluation from a MC simulation. 
With an additional selection efficiency of 0.97 resulting from the rejection of pile-up and cosmic-ray events,
$\epsilon^{\mathrm{e}\gamma}_{\gamma}=0.625 \pm 0.023$.

The trigger efficiency consists of three components; photon energy, time coincidence, 
and direction match. The efficiency of photon energy is estimated from the online energy 
resolution and found to be $\gtrsim 0.995$ for $\egamma>48.0$~MeV. 
The efficiency of the time coincidence is estimated from the online time 
resolution and found to be fully efficient. 
The direction match efficiency is evaluated, based on a MC simulation, to be 
$\epsilon^{\mathrm{e}\gamma}_{\mathrm{trg}} = 0.91 \pm 0.01$ and 0$.96 \pm 0.01$ 
for the data up to and after 2011, respectively (see Fig.~\ref{fig:daqeff}).

For $\egammapair$-pairs that satisfy the selection criteria for each particle, 
two kinds of further selection are imposed. One is the cut for the AIF-like events described in 
Sect.~\ref{sec:aif_rec}, resulting in 1.1\% inefficiency for the signal events.
The other is defined by the analysis window, in particular those for the relative angles and timing.
The inefficiency is evaluated via a MC simulation taking into account the pile-up and detector condition.
A loss of 3.2\% is due to the tails in the angular responses.
Additionally, about 1.5\%\ of events are outside the time window, mainly due to the erroneous reconstruction of positron trajectories when one of the turns, usually the first, %LU
is missed. 
As a result, $\epsilon^{\mathrm{e}\gamma}_{\mathrm{sel}} = 0.943\pm0.010$.

In total, the Michel positron counting method provides $N_{\mu}$ with a 4.5\% uncertainty.
%%%%%%%%%%%%%%%%%%%%%%%%

\subsubsection{RMD channel}
%{\it Editor's comments: \\
%Section coordinator:  Yusuke\\
%}
The other method for normalisation uses RMD events detected in the $\meg$ trigger data.
As in the Michel method, $N_\mu$ is expressed as,
\[%begin{equation}
\label{eq:Norm_RMD}
N_{\mu} = \frac{N^{\mathrm{e}\nu\bar{\nu}\gamma}}
				{\mathcal{B}^{\mathrm{e}\nu\bar{\nu}\gamma}} \times
		\frac{\epsilon^{\mathrm{e}\gamma}_{\mathrm{e}}}
			 {\epsilon^{\mathrm{e}\nu\bar{\nu}\gamma}_{\mathrm{e}}} \times 
		\frac{\epsilon^{\mathrm{e}\gamma}_{\gamma}}
			 {\epsilon^{\mathrm{e}\nu\bar{\nu}\gamma}_{\gamma}} \times
		\frac{\epsilon^{\mathrm{e}\gamma}_{\mathrm{trg}}}
			 {\epsilon^{\mathrm{e}\nu\bar{\nu}\gamma}_{\mathrm{trg}}} \times
		\frac{\epsilon^{\mathrm{e}\gamma}_{\mathrm{sel}}}
			 {\epsilon^{\mathrm{e}\nu\bar{\nu}\gamma}_{\mathrm{sel}}},
\]%end{equation}
where $\mathcal{B}^{\mathrm{e}\nu\bar{\nu}\gamma}$ is the partial branching ratio of 
RMD in the relevant kinematic range,
and the other factors are defined in the same way as for the Michel case.
Since the same dataset is used and the photon is also detected in this mode,
all the efficiency factors are expressed in signal-to-RMD ratios.
In contrast, the efficiency ratios need to be evaluated differentially 
as functions of the relevant kinematic variables because the kinematic range is wider 
than the $\meg$ analysis window.

We use events reconstructed in the energy side-band defined in 
Sect.~\ref{sec:rmdbackground}, corresponding to 
$\mathcal{B}^{\mathrm{e}\nu\bar{\nu}\gamma} = 4.9\times 10^{-9}$.
The number of RMD events is extracted from the fit to the $\tegamma$ 
distribution separately for each year dataset and for 12 statistically 
independent sub-windows, resulting in $N^{\mathrm{e}\nu\bar{\nu}\gamma} = 29\,950\pm527$ in total.

The momentum dependent ratio of the positron detection efficiency is extracted from the Michel 
spectrum fit. An additional correction for the momentum dependence of the missing turn 
probability is applied based on the evaluation of a MC simulation. A pre-scaled 
trigger with a lowered $\egamma$ threshold (by $\approx 4$~MeV) allows for a relative 
measurement of the energy-dependent efficiency curve of the LXe detector.
The efficiency ratio of the direction match is evaluated from the distribution of 
accidental background. The effect of muon polarisation \cite{Baldini:2015lwl}, which 
makes the background distribution non-flat (asymmetric) even in case of a fully efficient detector 
and trigger, is taken into account. 
Inefficiency due to the AIF-like event cuts and the tail in the time reconstruction 
are common to signal and RMD, and thus, only tails in the angular responses are relevant.
A more detailed description of the RMD analysis is found in \cite{megrmd}. %\cite{Adam:2014fca}.

A $\chi^2$ fit is performed to extract $N_{\mu}$ from the measured RMD spectrum. 
The systematic uncertainty on each factor, correlated among different windows, 
is accounted for in the fit. The uncertainty on $N_{\mu}$ from 
the fit to the full dataset is 5.5\%.

\subsubsection{$N_{\mu}$ summary}
\begin{figure}[tbp]
\centering
\includegraphics[width=18pc]{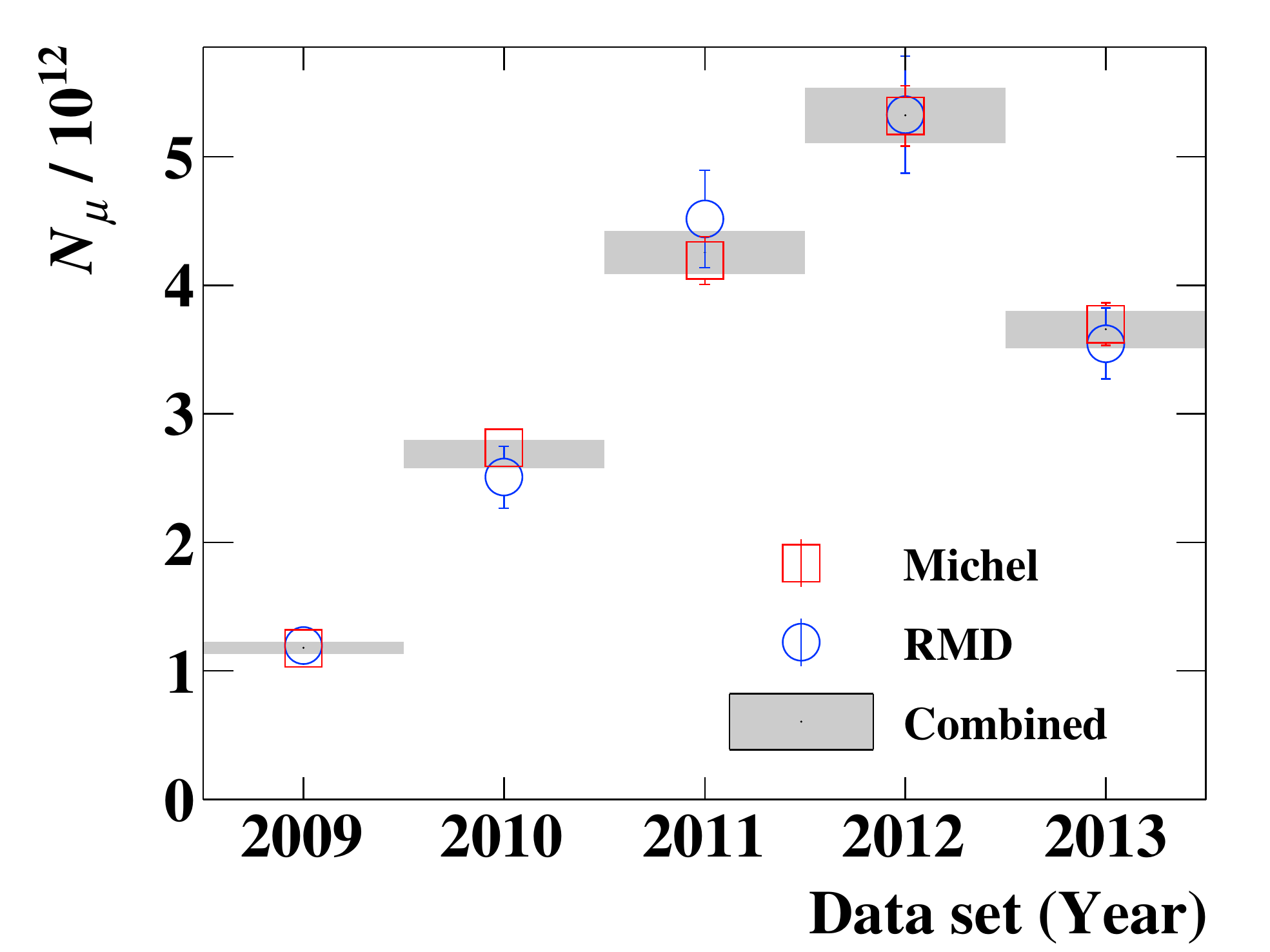}
\caption{\label{fig:norm2}$N_\mu$ calculated with the two methods and their weighted average for each year dataset.}
\end{figure}

%%%%%%%%%%%%%%%%%%%%%%%%
The normalisation factors calculated by the two methods are shown in Fig.~\ref{fig:norm2}. 
The two independent results are in good agreement and combined to give $N_{\mu}$ with
a 3.5\% uncertainty.
The single event sensitivity for the full dataset is $1/N_\mu = (5.84 \pm 0.21) \times 10^{-14}$.

The normalisation factor can also be expressed by 
\[ %begin{equation}
N_\mu = N_\mu^\mathrm{stop}\cdot \Omega \cdot \epsilon_\mathrm{tot}, \nonumber
\] %end{equation}
where $N_\mu^\mathrm{stop}$ is the total number of muons stopped in the target,
$\Omega$ is the geometrical acceptance of the apparatus and $\epsilon_\mathrm{tot}$ is the overall efficiency.
The integration of the estimated stopping rate, corrected for by the variation of the primary 
proton beam current, over the live-time gives an estimate of 
$N_\mu^\mathrm{stop}\approx 7.5\times 10^{14}$ (see Fig.~\ref{fig:numberofmuons}).
Therefore, an estimate of the overall signal acceptance of $\Omega \cdot \epsilon_\mathrm{tot} \approx 2.3$\%\ is obtained. 
This is consistent with $\Omega\approx 0.11$\ and our estimates of detector 
efficiencies, $\epsilon_\mathrm{tot} = \epsilon_\mathrm{e^{+}} \cdot \epsilon_\gamma \approx 0.30\times 0.63$.
%%%%%%%%%%%%%%%%%%%%%%%%

\subsection{Results }
%{\it Editor's comments: \\
%Section coordinator: Fabrizio, Wataru, Ryu, Daisuke\\
%Text:  3.\\
%Figure: 5.
%}

A maximum likelihood analysis is performed to extract the number of signal events from the full dataset
after the analysis tools are fully optimised and background studies in the side-bands are completed.
The sensitivity and the results in the analysis window are presented and discussed in the following sections.

\subsubsection{Sensitivity}
\label{sec:sensitivity}
\begin{figure}[htb]
\centering
 \includegraphics[width=0.5\textwidth,angle=0] {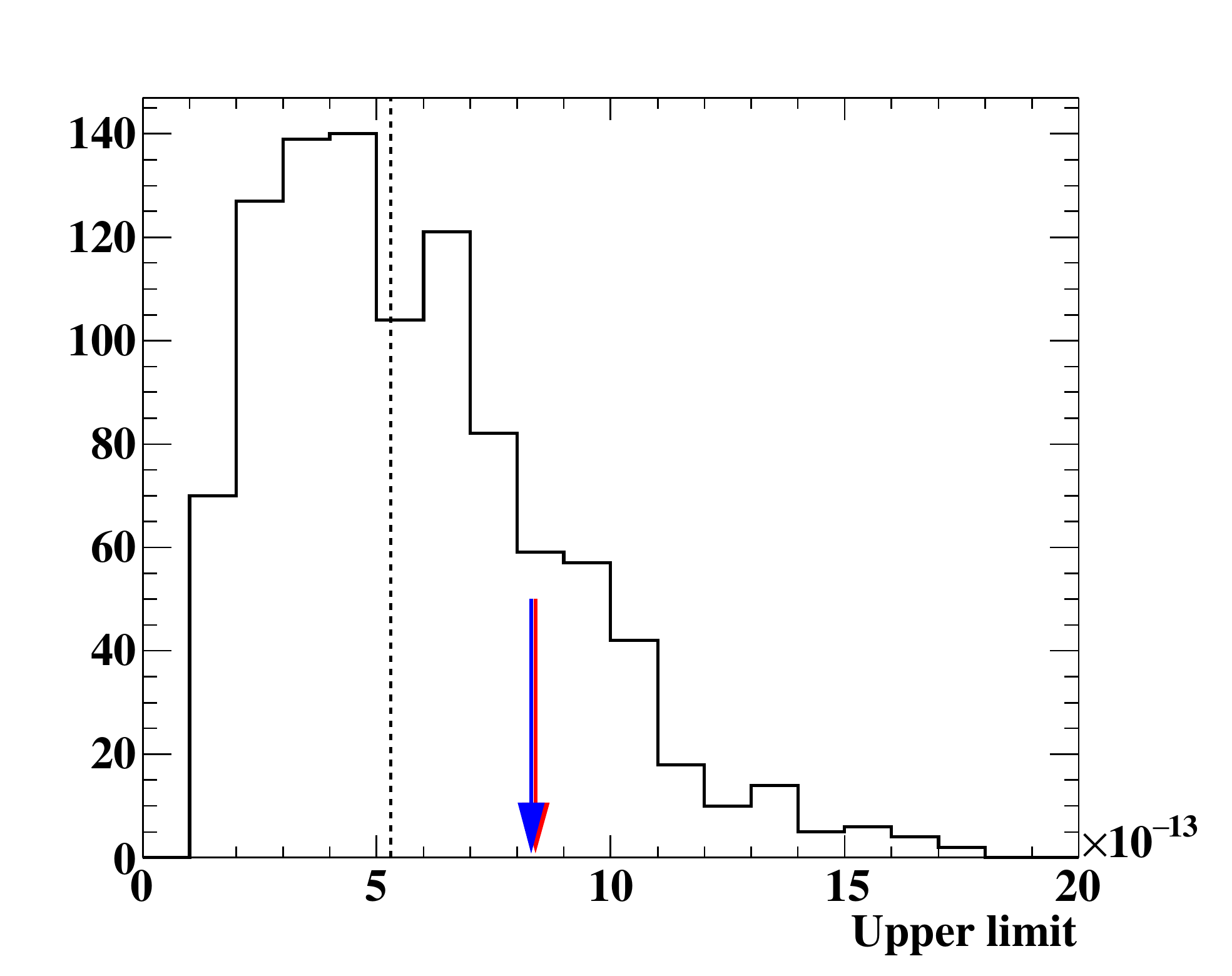}
 \caption{
Distribution of the branching ratio upper limits for pseudo experiments 
simulated for the full dataset. The sensitivity, defined as the median 
of the distribution and shown as a dashed vertical line, equals to 5.3$\times 10^{-13}$. The 
upper limits observed in the timing side-bands are indicated with arrows for comparison
(the overlap of the two arrows is accidental).
}
 \label{fig:UL distribution}
\end{figure}

The sensitivity of the analysis is evaluated 
by taking the median of the distribution of the branching ratio upper limits at
90\% C.L.~observed for an ensemble of pseudo experiments with a null signal hypothesis.
The rates of RMD and accidental background events estimated from the side-band 
studies are assumed in the pseudo experiments. All the systematic uncertainties 
as listed in Sect.~\ref{sec:feldman-cousins} are taken into account in the sensitivity evaluation.
Figure~\ref{fig:UL distribution} shows the distribution of the branching ratio upper limits
for the pseudo experiments simulated for the full dataset. The sensitivity 
is found to be 5.3$\times 10^{-13}$.
The sensitivities of the 2009--2011 and 2012--2013 datasets have also been evaluated 
separately as presented in Table~\ref{tab:BRtable}. 

The average contributions of the 
systematic uncertainties are evaluated by calculating the sensitivities without 
including them. The dominant one is found to be the uncertainty on the target
alignment; it degrades the sensitivity by 13\% on average, while the total
contribution of the other systematic uncertainties is less than 1\%.
The sensitivity for the 2009--2011 dataset is found to be slightly worse 
than previously quoted in \cite{meg2013} due to a more conservative assignment 
of the systematic uncertainty on the target alignment. 

The maximum likelihood analysis 
has also been tested in fictitious analysis windows in the timing side-bands 
centred at $\tegamma=\pm 2\,\mathrm{ns}$ without the Gaussian constraint 
on $\nrd$. The upper limits observed in the 
negative and positive timing side-bands are 8.4$\times 10^{-13}$ and 8.3$\times 10^{-13}$, respectively.
These are consistent with the upper limit distribution for 
pseudo experiments as indicated in Fig.~\ref{fig:UL distribution}.

\subsubsection{Likelihood analysis in the analysis window}
\label{sec:likelihood analysis in signal region}

\begin{figure}[htb]
\centering
  \includegraphics[width=20pc,angle=0] {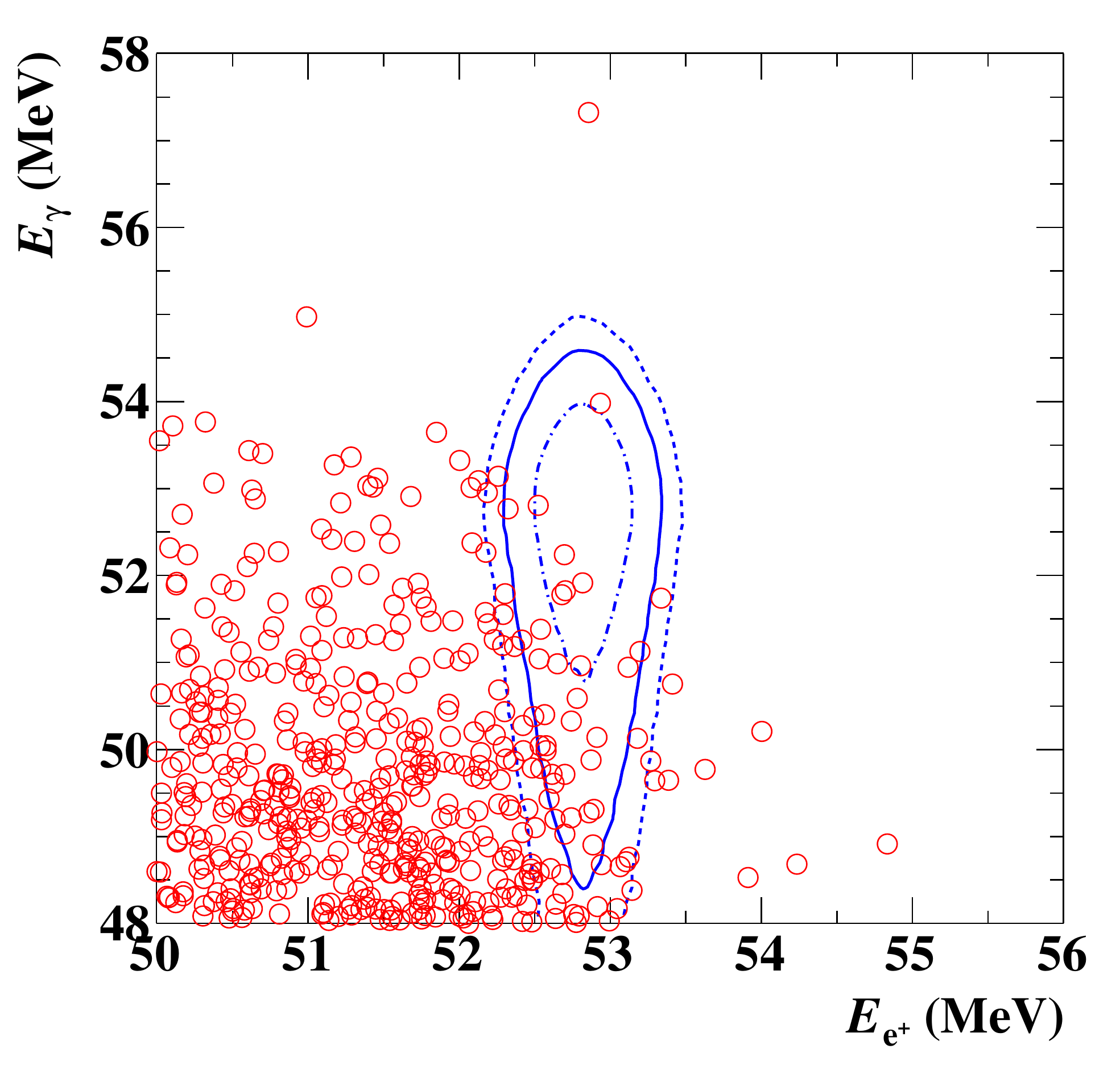}
  \includegraphics[width=20pc,angle=0] {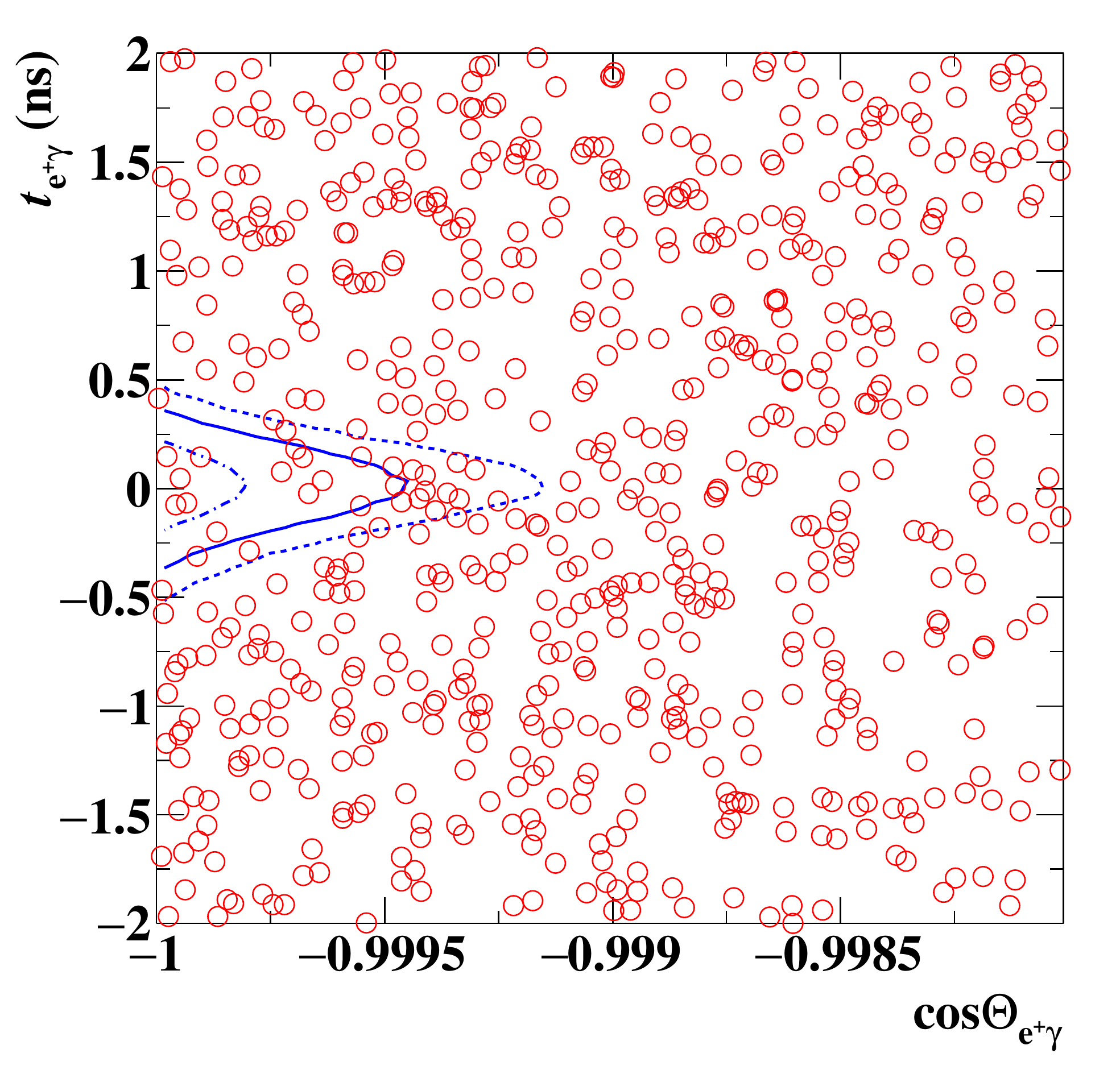}
 \caption{
Event distributions of observed events in the $(\epositron, \egamma)$- 
and $(\cos\Thetaegamma, \tegamma)$-planes. In the top figure, 
selections of $\cos\Thetaegamma < -0.99963$ and 
$|\tegamma| < 0.24$ ns are applied with 90\% efficiency for each 
variable, and in the bottom figure $51.0 < \egamma < 55.5$ MeV 
and $52.4 < \epositron < 55.0$ MeV are applied with 74\% and 90\% efficiency 
respectively. The signal PDF contours ($1\sigma$, $1.64\sigma$ 
and $2\sigma$) are also shown.
}
 \label{fig:distribution2D}
\end{figure}

Figure~\ref{fig:distribution2D} shows the event distributions for the 
2009-2013 full dataset on the $(\epositron, \egamma)$- and 
$(\cos\Thetaegamma, \tegamma)$-planes. 
The contours of the averaged signal PDFs are also shown for comparison. 
No significant correlated excess is observed within the signal contours.

\begin{figure}[htb]
\centering
  \includegraphics[width=20pc,angle=0] {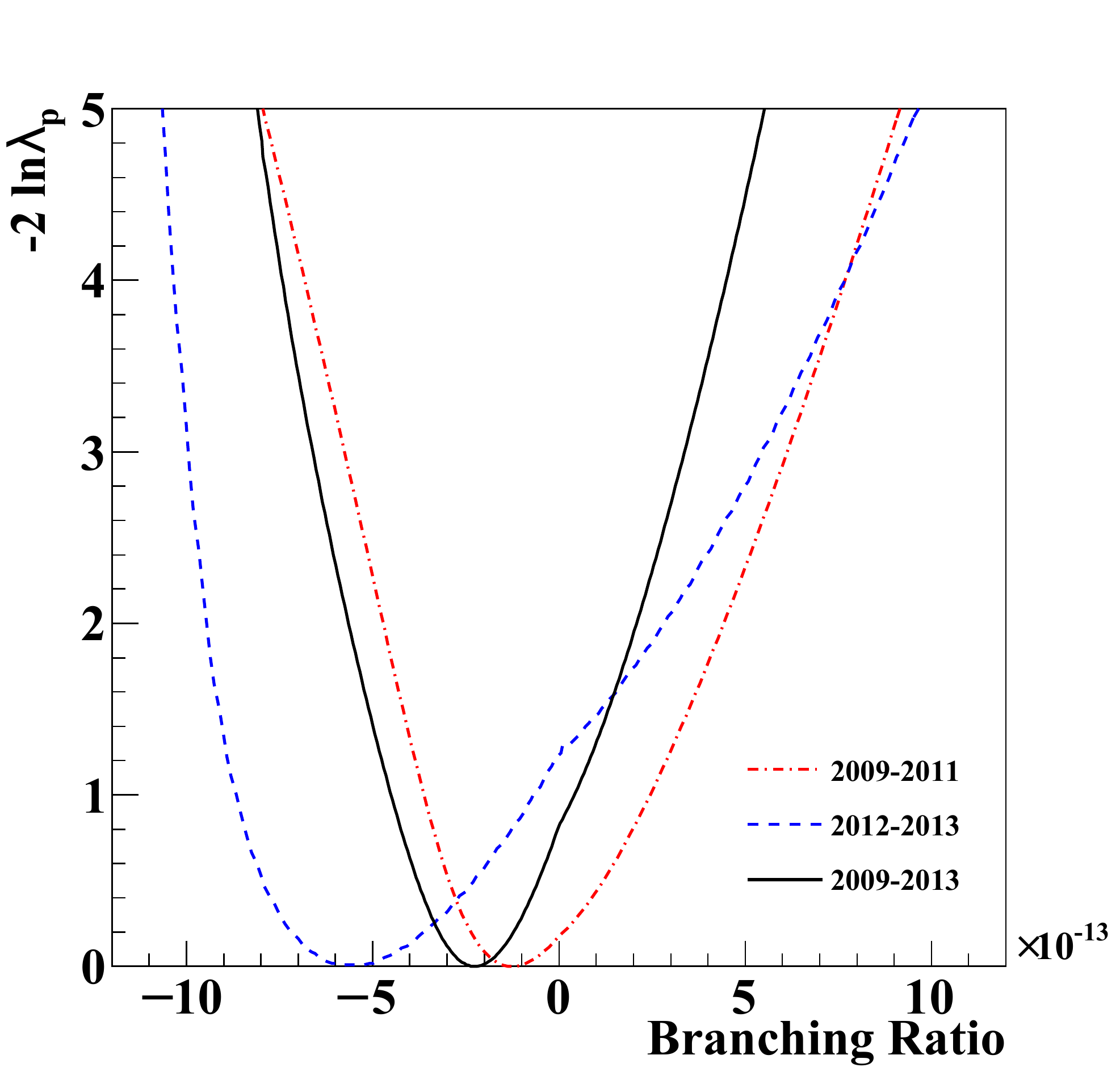}
 \caption{
 The negative log-likelihood ratio ($\lambda_{\rm p}$) as a function of 
    the branching ratio.
}
 \label{fig:NLLcurve}
\end{figure}

A maximum likelihood analysis is performed to evaluate 
the number of signal events in the analysis window by the method described 
in Sect.~\ref{sec:maximum likelihood analysis}. 
Figure~\ref{fig:NLLcurve} shows the profile-likelihood ratios as a function of the branching ratio observed for 2009--2011, 2012--2013, and 2009--2013 full dataset, 
  which are all consistent with a null-signal hypothesis. 
The kinks visible in the curves (most obvious in 2012--2013) are due to the profiling 
of the target deformation parameters (see Sect.~\ref{sec:likelihood}). 
In the positive side of the branching ratio, the estimate of the target shape 
parameters in the profiling is performed by looking for a positive excess 
of signal-like events in the $\phiegamma$ distribution. On the other hand, 
in the negative side, it is done by looking for a deficit of signal-like events. 
These parameters are therefore fitted to opposite directions (the paraboloid shape 
or the deformed shape defined by the FARO measurement) in the positive and 
the negative sides of the branching ratio. The likelihood curve shifts from 
one to another of the two shapes crossing 0 in the branching ratio. 
The best fit value on the branching ratio for the full dataset is 
  $-2.2\times10^{-13}$.
 The upper limit of the confidence interval is calculated following the frequentist approach
 described in Sect.~\ref{sec:feldman-cousins} to be $4.2\times10^{-13}$
 at 90\% C.L.

\begin{figure*}[htb]
\centering
  \includegraphics[width=36pc,angle=0] {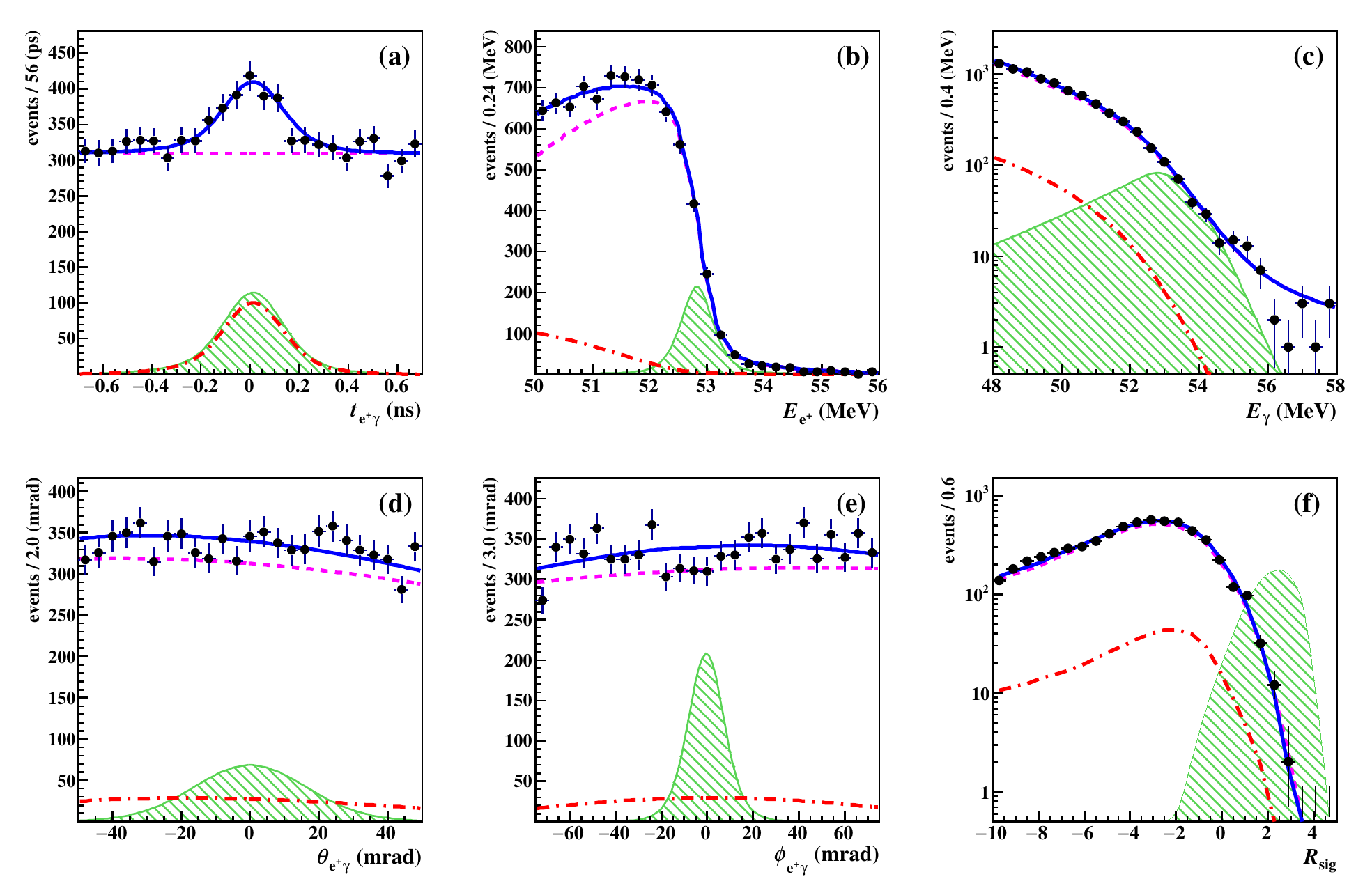}
 \caption{
The projections of the best fitted likelihood function to the five main 
   observables and $R_{\rm sig}$ together with the data spectra for the full dataset.
The magenta dash and red dot-dash lines are individual components of 
the fitted PDFs of ACC and RMD, respectively. The blue solid line is the sum 
of the best fitted PDFs. The green hatched histograms show the signal PDFs 
corresponding to 100 times magnified $\nsig$ upper limit.
}
 \label{fig:FitResult1D}
\end{figure*}

The projection of the best fitted function on each observable is shown 
in Fig.~\ref{fig:FitResult1D} (a)--(e), where all the fitted spectra are 
in good agreement with the data spectra. The agreement is also confirmed by 
the relative signal likelihood $R_{\rm sig}$ defined as
\begin{equation}
R_{\rm sig} = \log_{10} \left( \frac{S(\vector{x}_i)}{f_\mathrm{R}R(\vector{x}_i)+f_\mathrm{A}A(\vector{x}_i)} \right),
\end{equation}
where $f_\mathrm{R}$ and $f_\mathrm{A}$ are 
the expected fractions of the RMD and accidental background events which 
are estimated to be 0.07 and 0.93 in the side-bands, respectively. 
Figure~\ref{fig:FitResult1D} (f) shows the $R_{\rm sig}$ distribution observed in 
the full dataset together with the expected distribution from the fit result.

The results from the maximum likelihood analysis are summarised in Table~\ref{tab:BRtable}.
The dominant systematic uncertainty is due to the target alignment uncertainty, 
which increases the upper limit by 5\% while the other uncertainties 
increase it by less than 1\% in total.

The upper limit on the branching ratio is consistent with the sensitivity 
under the background-only hypothesis presented in Sect.~\ref{sec:sensitivity}.
This result is confirmed by following the profile of the log-likelihood 
curve as a function of the number of signal events, in parabolic approximation, 
and by independent analysis, based on a set of the constant PDFs,
which will be discussed in Sect.~\ref{sec:ConstanPDFs}.

\begin{table}
 \centering
 \caption{\label{tab:BRtable} 
   Best fit values of the branching ratios (${\cal B}_\mathrm{fit}$), upper limits at 90\% C.L. ($\ul$) and sensitivities ($\sens$)}
{\begin{tabular}{@{\extracolsep{\fill}}|l|ccc|@{}}
\hline
{\bf dataset} & 2009--2011 & 2012--2013 & 2009-2013 \\
\hline
{\bf $\bestfit\times10^{13}$} & $-1.3$\,\,\,\, & $-5.5$ \,\, & $-2.2$ \,\, \\
\hline
{\bf $\ul\times10^{13}$} & $6.1$ & $7.9$ & $4.2$ \\
\hline
{\bf $\sens\times10^{13}$} & $8.0$ & $8.2$ & $5.3$ \\
\hline
\end{tabular}}
\end{table}

A maximum likelihood fit without the constraints on $\nrd$ and $\nacc$ estimated 
in the side-bands is performed as a consistency check.
The best fit values of $\nacc$ and $\nrd$ for the combined dataset are $7684\pm103$ and $663\pm59$, respectively. 
They are consistent with the respective expectations of $7744\pm41$ and $614\pm34$ 
and also with the total number of observed events ($\nobs = 8344$) in the analysis window.

\subsubsection{Discussion}
\label{sec:Discussion}
 
\paragraph{Constant PDFs' analysis}
\label{sec:ConstanPDFs}
\hspace{0cm} \\
 
\begin{figure}[htb]
\centering
 \includegraphics[width=20pc,angle=0] {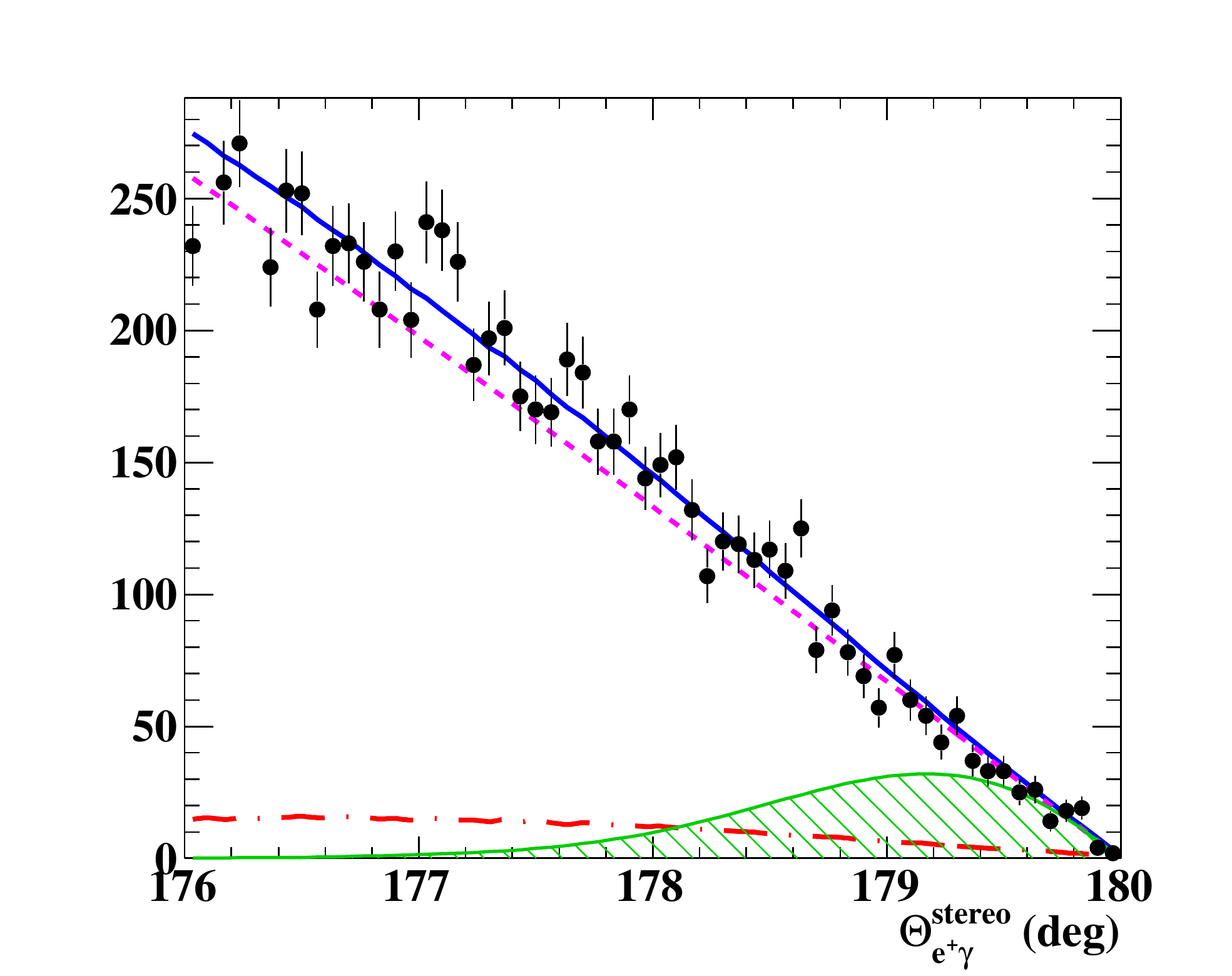}
 \caption{
   The distribution of the relative stereo angle $\Thetaegamma$ 
obtained in the constant PDFs' analysis for experimental data (black dots) and 
the fitted spectrum. The RMD and accidental background 
components and their sum are shown 
with the red dot-dashed, magenta dashed and blue solid curves, respectively; 
the green hatched histogram shows the signal PDF 
corresponding to 100 times magnified $\nsig$ upper limit.
} 
 \label{fig:FigPisa}
\end{figure}

A maximum likelihood fit is also performed by using the constant PDFs, 
obtaining results in good agreement with those of the analysis 
based on event-by-event PDFs. The best fit and upper limit at 90\% C.L.~on 
the branching ratio obtained by this analysis on the full dataset 
are $-2.5 \times 10^{-13}$ and $4.3 \times 10^{-13}$ respectively, 
in close agreement with the results of the event-by-event PDFs' analysis 
presented in Sect.~\ref{sec:likelihood analysis in signal region}. 
The fitted numbers of 
RMD and accidental events are $630 \pm 66$ and $7927 \pm 148$, in agreement 
with the expected values of $683 \pm 115$ and $7915 \pm 96$ obtained by extrapolations from the side-bands. These numbers also agree with those of 
the event-by-event PDFs' analysis when one takes into account that the angular 
selection based on the relative stereo angle ($\Thetaegamma > 176^{\circ}$) 
selects $\approx 3\,\%$ more accidental events than that based on $\thetaegamma$ 
and $\phiegamma$. Figure~\ref{fig:FigPisa} shows
an example of the results obtained with the constant PDFs' analysis 
for the projection of the best fitted function on 
$\Thetaegamma$: the fitted and the data distributions are in good agreement. 

\begin{figure}[htb]
\centering
 \includegraphics[width=20pc,angle=0] {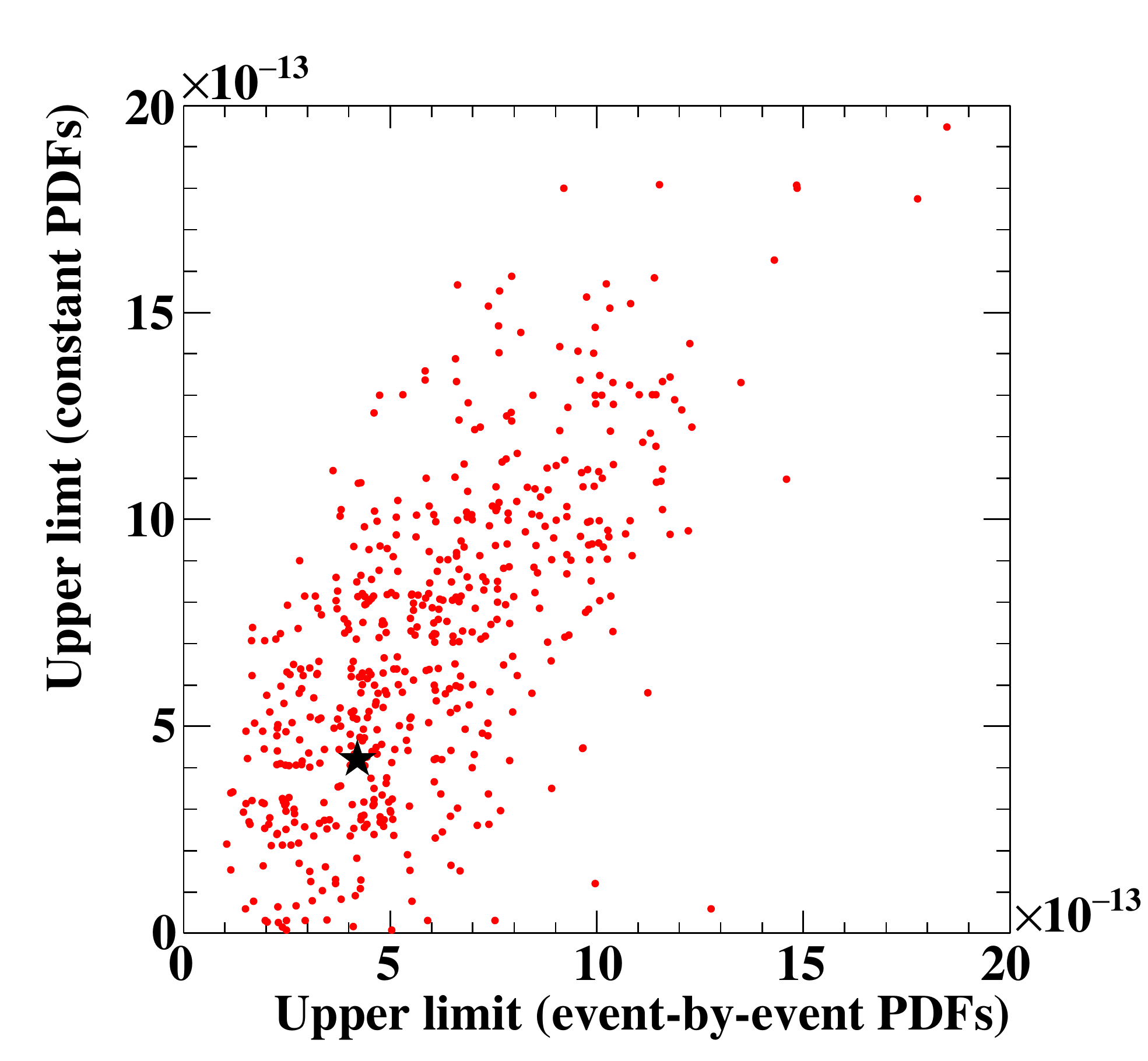}
 \caption{
   The comparison of the $90\,\%$ C.L.~upper
limits reconstructed on a sample of several hundred common toy 
MCs by the constant PDFs' and the event-by-event PDFs' analyses. 
The upper limits observed in the experimental 
data are marked by a star. There is a clear correlation between the upper limits from the two 
analyses
with a $\approx 20$\% better sensitivity on average for the 
event-by-event PDFs' analysis.}
 \label{fig:commontoy}
\end{figure}

The consistency of the two analyses is also checked by a set of 
pseudo experiments, specifically produced to be compatible with the 
structures of both the analyses (\lq\lq common toy MCs\rq\rq). 
The upper limits at 90\% C.L.~observed in the two analyses 
for a sample of several hundred common toy MCs are compared in 
Fig.~\ref{fig:commontoy}; the experimental result is marked 
by a star. There is a clear correlation between the upper limits from the two 
analyses with a $\approx 20$\% better sensitivity on average for the 
event-by-event PDFs' analysis.
By analysing the distribution of the differences between the upper limit 
reconstructed by the two analyses on this sample of common toy MCs,
we found that the probability of obtaining a difference in the upper limit 
at least equal to that measured on the real data is 70\.\%.
\paragraph{Comparison with previous analysis}
\label{sec:ComparisonWithPreviousAnalysis}
\hspace{0cm} \\
 
The previous MEG publication \cite{meg2013} reported on the analysis 
based on the 2009--2011 dataset. The analysis presented here includes 
a re-analysis of the 2009-2011 dataset with improved algorithms. 
Since the analysis algorithms are revised, the reconstructed observables 
are changed slightly, albeit within the detector resolutions. 
A change in the results of the analysis is expected due to statistical effects. 
The expected difference in the upper limit between the old and new analyses 
for the 2009--2011 dataset is evaluated by a set of toy MC simulations based 
on the expected changes in the reconstructed observables, and shows a spread of 
$4.2\times 10^{-13}$ (RMS) with a mean of nearly zero. 
The difference observed in the experimental data 
is $0.4\times 10^{-13}$ and lies well within the spread.
%

%% file: conclusions.tex
\section{Conclusions}
\label{sec:conclusions}
%
%{
%Section coordinator: Paolo \\
%Text:  1. \\
%Figure: 0. 
%}

A sensitive search for the lepton flavour violating muon decay mode 
\meg\ was performed with the MEG detector in the years 2009--2013.
A blind, maximum-likelihood analysis found no significant event excess 
compared to the expected background and established a new upper limit for the 
branching ratio of $\BR(\meg) < 4.2 \times 10^{-13}$ with 90\%\ C.L. 
This upper limit is the most stringent to date and provides important constraints 
on the existence of physics beyond the Standard Model.

The new measured upper limit improves our previous result \cite{meg2013} by a
factor $1.5$; the improvement in sensitivity amounts to a factor $1.5$.
Compared with the previous limit from 
the MEGA collaboration \cite{brooks_1999_prd}, our new upper limit represents 
a significant improvement by a factor $30$.
% and a significant constraints for BSM theories and 
%supersymmetric parameter space. For instance, in several models 
%where the CLFV is driven by a dipole-mediated mechanism, the sensitivity 
%we reached allows to explore the scale of
%supersymmetry up to several thousands of ${\rm TeV}$.

An effort to upgrade the existing MEG detector is currently underway with the goal of achieving an additional 
improvement in the sensitivity of close to an order of magnitude \cite{Baldini:2013ke}. The modifications are designed 
to increase acceptance, enable a higher muon stopping rate, and improve limiting detector resolutions. 
Tracking and timing detectors for measuring the positrons have been completely 
redesigned and other parts of the detector have been refurbished. The 
newly designed experiment, MEG II, will be able to use a muon decay rate twice that of MEG. The improved detector is expected to improve  
the branching ratio sensitivity to $5 \times 10^{-14}$ with three years of data taking 
planned for the coming years.